\documentclass[fleqn,usenatbib]{mnras}

\usepackage{newtxtext,newtxmath}
\usepackage{color}
\usepackage[T1]{fontenc}
\usepackage[symbol]{footmisc}
\usepackage{multirow}
\usepackage{graphicx}
\DeclareRobustCommand{\VAN}[3]{#2}
\let\VANthebibliography\thebibliography
\def\thebibliography{\DeclareRobustCommand{\VAN}[3]{##3}\VANthebibliography}

\usepackage{graphicx}	
\usepackage{amsmath}	
\usepackage[normalem]{ulem}
\usepackage{lineno}

\title[AICs and Multimessenger Signals]{Collapse of Rotating White Dwarfs and Multimessenger Signals}

\author[T. Kuroda, K. Kawaguchi, and M. Shibata]{
Takami Kuroda$^{1}$\thanks{E-mail: takami.kuroda@aei.mpg.de},
Kyohei Kawaguchi$^{1,2}$,
and Masaru Shibata$^{1,2}$
\\
$^{1}$Max-Planck-Institut f{\"u}r Gravitationsphysik, Am M{\"u}hlenberg 1, D-14476 Potsdam-Golm, Germany\\
$^{2}$Center for Gravitational Physics and Quantum-Information,
Yukawa Institute for Theoretical Physics, Kyoto University, Kyoto, 606-8502, Japan
}

\date{Accepted XXX. Received YYY; in original form ZZZ}

\pubyear{2025}

\begin{document}
\label{firstpage}
\pagerange{\pageref{firstpage}--\pageref{lastpage}}
\maketitle

\begin{abstract}
We present results of numerical relativity simulations for the collapse of rotating magnetized white dwarfs (WDs) in three dimension, aiming at discussing the explosion dynamics and associate multi-messenger signals: gravitational waves (GWs), neutrinos, and electromagnetic counterparts.
All WDs initiate gravitational collapse due to electron captures and then experience prompt type explosions after the proto neutron star formation.
We observe the explosions dominated by a bipolar structure and the emergence of strong spiral waves in rapidly rotating models.
The spiral waves facilitate to increase both the explosion energy and ejecta mass, though the final values still fall in the category of low explosion energy supernovae with small ejecta mass.
The spiral waves also produce strong GWs, which may expand the horizon distance of such events against GWs up to $\sim 10$\,Mpc for third-generation ground-based detectors.
Additionally as an intriguing implication, we demonstrate that such accretion or merger induced collapse of WDs might be able to explain some of the rapidly evolving optical transients, such as fast blue optical transients (FBOTs), as previously suggested.
Based on the simulation results together with several assumptions, we confirm that the magnetar may account for the brighter side of some of observed FBOTs, while a combination of ejecta-envelope interaction which can be also followed by radioactive decay of heavy elements synthesized along with the explosion might still explain the fainter branch even in the absence of magnetar formation.
\end{abstract}





\maketitle

\section{Introduction}
\label{sec:Introduction}
The nature of lower ejecta mass in accretion- or merger-induced collapse (AIC/MIC) of white dwarfs (WDs) could be the key to understanding rapidly evolving transients, whose features are hardly interpreted by standard supernovae (SNe).
WDs in a binary system may acquire mass from its non-degenerate companion star \citep{Nomoto79,Nomoto&Kondo91} or through a merger with its companion WD \citep{Saio&Nomoto85,Shen12,Schwab21}.
Depending on the mass accretion history, WDs can avoid a violent nuclear burning, i.e., Type Ia SNe, and steadily increases both their mass and central density.
Once the WD mass exceeds the Chandrasekhar mass and its central density achieves a sufficiently high value ($\sim10^{10}$\,g\,cm$^{-3}$), electron capture takes place at its center, which triggers a collapse of WDs to a neutron star (NS) and subsequent explosion, namely electron capture SN (ECSN) \citep{Nomoto&Kondo91,Yoon&Langer05}.

ECSNe of WDs--hereafter we simply term them as AICs regardless of whether the WD gains mass from a non-degenerate (AIC) or degenerate companion star (MIC)--are characterized by their relatively low ejecta mass.
This is a natural consequence of the initially {\it low} total mass of the system, i.e., the WD mass at the Chandrasekhar limit, which practically ranges from $\sim1.4$\,$M_\odot$ for a non-spinning WD to $\lesssim2$\,$M_\odot$ for a rapidly rotating one depending also on the actual evolutionary path \citep[e.g.,][]{Yoon&Langer05}.
As the ejecta mass is at most the difference between initial WD and remnant NS masses, the expected ejecta mass ($M_{\rm ej}$) is obviously quite low, typically $M_{\rm ej}\lesssim$ a few times of $0.1$\,$M_\odot$.
Such values are at least one order of magnitude smaller than the canonical values for Type II/Ibc SNe associated with the massive stellar collapse.
Furthermore, the absence of massive WD envelopes generally avoids the deceleration of ejecta and keeps the ejecta velocities nearly at their original high values, which could be another noteworthy feature of AICs.

During the evolutionary phase of WDs, they may acquire some unique properties, which result in the formation of a remnant NS having distinct features.
The most important is stemming from spinning up of WDs via the mass accretion from a companion star \citep{Yoon&Langer05,Stefano11} or merger with a companion WD \citep{Hachisu86,Segretain97,Marsh04,Dan14}.
For instance, \cite{Yoon&Langer05} proposed that some WDs can increase their angular frequencies to $\gtrsim$\,a few rad\,s$^{-1}$ by mass accretion from its companion.
In addition these WDs may avoid the path toward the Type Ia SN branch, though the detailed evolutionary paths depend on various effects, such as rotational instabilities, magnetic torque, and viscous effects between the WD and its surrounding disk \citep[e.g.,][]{Saio&Nomoto04}.
If the core of WDs gains such high angular frequencies ($\sim$\,a few rad\,s$^{-1}$) at the onset of an AIC, it is enough to produce a rapidly rotating NS, whose angular frequency reaches $\gtrsim1000$\,rad\,s$^{-1}$.

In addition, the merger of two WDs may often amplify the seed magnetic fields to strong values \citep{Tout08,Garcia-Berro12}.
Indeed, the existence of strongly magnetized WDs, whose surface magnetic fields reach in some cases $\sim10^{8\rm{-}9}$\,G, are observationally confirmed both in isolated and binary WDs \citep{Wickramasinghe00,Schmidt03}.
Numerical simulations of merging WDs also support the formation of such strongly magnetized WDs \citep{Zhu15}.
Assuming a simple dipole-like magnetic field configuration, the core magnetic fields of the observed strongly magnetized WDs might achieve $\sim10^{10\rm{-}11}$\,G.
Such magnetic fields at center can be further amplified by the compression during the WD collapse and subsequently by several other possible processes such as the magnetorotational instability \citep{Balbus91} or Tayler-Spruit dynamo \citep[e.g.,][]{Spruit02,Reboul-Salze24}, as well as by simple rotational winding.
Consequently, strong magnetic fields, whose poloidal component $B_{\rm p}$ reaches the order of $B_{\rm p}\sim10^{15}$\,G, i.e. magnetar \citep{Usov92}, might be formed in some AIC events.
Therefore AIC has recently been gaining a further attention because of its potential formation path not only to a rapidly rotating NS but also to a magnetar.

Beside such unique compact star formation channels, the properties of AIC, i.e., low ejecta mass with relatively high velocities, may explain some of the intriguing rapidly evolving transients, such as fast blue optical transients (FBOTs: \cite{Drout14}).
Indeed the work of \cite{Yu15} presented a potential connection between these rapid transients and the activities of newly formed NSs, or rather magneters, in the aftermath of AIC or binary NS (BNS) mergers.
FBOTs are characterized by their fast rise time till their peak luminosities ($t_{\rm peak}\lesssim10$\,d) and by bolometric peak luminosities ranging from $L_{\rm peak}\sim10^{42}$\,erg\,s$^{-1}$ to $\sim10^{44}$\,erg\,s$^{-1}$ \citep[][and references therein]{Drout14,Inserra19}.
Among those remarkable features, the fast rise time ($t_{\rm peak}\sim$\,a few to several days) can be explained for instance by a small ejecta mass of $M_{\rm ej}\sim0.1$\,$M_\odot$ and relatively fast ejecta velocity of $v_{\rm ej}\sim0.1c$ \citep[e.g.,][]{Margutti19}, which is certainly compatible with the typical AIC models.
To explain their peak luminosities, we can apply the most standard energy source, namely the radioactive decay of $^{56}$Ni, for fainter events with $L_{\rm peak}\sim10^{42}$\,erg\,s$^{-1}$, assuming $^{56}$Ni mass of $0.1$\,M$_\odot$.
On the other hand, this energy source can be ruled out in some events exhibiting higher bolometric luminosities $L_{\rm peak}\sim10^{44}$\,erg\,s$^{-1}$ \citep[e.g., represented for instance by AT\,2018cow,][]{Prentice18,Smartt18}, as the required amount of $^{56}$Ni is several\,$M_\odot$, far exceeding the expected ejecta mass of AICs.
Furthermore, the post-peak light curves of these rapid transients are rich in variety and not described by a single power low, inferring that their emission mechanisms may not be explained only by a single scenario \citep{Pursiainen18}.
This fact led us to consider alternative central engine models of FBOTs.

In the investigation of the alternative models, we have to pay attention to both the peak luminosity and the post-peak decline features.
To account for the former peak luminosities, a rapidly rotating nascent magnetar or a fall back model onto the black hole (BH) are proposed \citep{Yu15,Margutti19,Lyutikov19,Pasham22}, where the latter BH model can be potentially excluded in the context of AIC scenarios, because of the expected low total mass in the system.
Therefore the magnetar engine might be more feasible.
Previous studies have deduced the required magnetic field strength to explain the peak luminosity of FBOTs $(L_{\rm peak}\sim10^{43\rm{-}44}$\,erg\,s$^{-1})$ and reported $\sim10^{15}$\,G \citep{Rest18_FELT,Margutti19,Lyutikov19}, which indeed falls into the typical strengths of magnetars.
Furthermore, the energy deposition rate by magnetar spin-down, which follows $\propto t^{-2}$, into nebulae behind the shock are in line with the observed overall time scales of FBOTs \citep{Drout14} as modeled in the context of NS mergers by, e.g., \cite{Yu13}.
These facts infer that the magnetar model in the context of AICs may explain the emission mechanism of a subclass of rapidly evolving transients.
These facts also motivate us to explore detailed AIC scenarios based on state-of-the-art SN simulations including their plausible formation channel to magnetar.

To date, there are several AIC simulations with various physical inputs.
Early one-dimensional spherically symmetric (1D) studies \citep{Baron87,Woosley&Baron92,Fryer99_AIC} based on a rather simplified neutrino transport and recent ones \citep{Mor&Piran23,Mori24} have commonly reported electron capture collapse of WDs and formation of NSs, followed by a shock expansion, except a short-term model in \cite{Baron87}.
In multi-D models, AICs were studied with various focuses.
The seminal work by \cite{Dessart06_AIC} has reported that the AIC of rotating WDs in 2D axisymmetry results in a weak explosion triggered by neutrino heating.
\cite{Abdikamalov10_AIC} focused on the aspect of gravitational wave (GW) emission from collapse of rotating WDs by performing general relativistic simulations with a simplified microphysics.
The first 3D study in this context was reported in \cite{Longo-Micchi23}, which conducted fully relativistic 3D simulations, albeit with an M1 grey neutrino transport.
They reported loud GW signals associated with non-axisymmetric rotational instabilities and also witnessed explosion, whose diagnostic explosion energies are $\sim5\times10^{50}$\,erg for rotating models.
Later the same group performed 2D GRMHD simulations with an M1 grey neutrino transport, aiming at the jet formation to account for long gamma-ray bursts and kilonovae facilitated by strong magnetic fields \citep{Cheong25_AIC}.
Very recently, \cite{Batziou&Janka24_AIC} reported quite long-term AIC models in 2D with a sophisticated multi-energy neutrino transport.
They followed the evolution of the system for $\sim7$\,s after NS formation and discussed impacts of WD rotation on the proton/neutron richness in the ejecta.

In this paper, we report a new result for 3D numerical relativity simulations of AICs with a multi-energy M1 neutrino transport.
Our primary focus is on discussing the impacts of non-axisymmetric instabilities on the matter ejection from central proto-NS (PNS), which are significant especially for rapidly rotating models, and also on presenting detailed thermodynamic properties of ejecta based on our up-to-date neutrino transport.
Our results are generally consistent with previous studies, except that our 3D models exhibit quantitatively different ejecta profile, especially for electron fraction, in comparison to previous 2D axisymmetric studies.
Afterward, based on our numerical models, we discuss their peculiar GWs, neutrinos, and electromagnetic signals.

The paper is organized as follows.
In Section~\ref{sec:Numerical Setup}, we explain our initial WD models and how they can be constructed.
In addition, our radiation-MHD scheme is concisely summarized.
The main results are presented in Section~\ref{sec:Results}.
Sections~\ref{sec:Gravitational wave emissions}--\ref{sec:Implications for the electromagnetic counterparts} are devoted to a discussion about their multi-messenger signals.
We summarize our results and discussion in Section~\ref{sec:Summary}.
Throughout this paper, cgs unit is used and Greek indices run from 0 to 3. $c$ and $G$ are the speed of light and gravitational constant, respectively. 

\begin{table*}
  \begin{tabular}{lcccccccccc} \hline \hline
    Model & $\beta_0$ & $\rho_{\rm c,0}$ & $J_0$ & $M_{\rm WD,bar}$& $\Omega_0$  & $\beta_{\rm cb}$ & $t_9$ & $E_{\rm exp,9}$ & $t_{\rm fin}$& $E_{\rm exp,fin}$  \\ 

    & [\%] & [g\,cm$^{-3}$]  &  [10$^{49}$\,g\,cm$^2$\,s$^{-1}$] & [M$_\odot]$& [rad\,s$^{-1}$] & [\%] & [ms] & [10$^{50}$\,erg] &  [ms]&  [10$^{50}$\,erg]  \\ \hline
    R1 & 0.113 &$10^{10}$ & 1.05 & 1.73 & 1.38  & 0.56  & 338 & 2.54 & 325 & 1.01\\
    R2 & 0.217  &$10^{10}$ & 2.08 & 1.74 & 1.91 & 1.09 & 302 & 2.54 & 354 & 2.70\\
    R3 & 0.320  &$10^{10}$ & 2.98 & 1.75 & 2.31 & 1.35 & 317 & 3.37 & 381 & 3.63\\
    R6 & 0.647 &$10^{10}$ & 5.86 & 1.76 & 3.24 & 3.56   & 345 & 3.11 & 487 & 4.18\\
    R6oB & 0.647 &$10^{10}$ & 5.86 & 1.76 &3.24 & 3.56 & 390 & 1.76 & 455 & 2.78\\
    R6o & 0.647 &$10^{10}$ & 5.86 & 1.76 &3.24 & 3.56 & -- & --  & 247 & 0.24\\
 \hline
  \end{tabular}
  \caption{summary of our models. From left to right, the columns represent the model name, the initial ratio of the rotational kinetic energy to the gravitational potential energy $\beta$, central density, angular momentum, rest mass, angular velocity along the rotational axis, the value of $\beta$ at the core bounce, the time $t_9$ at which the maximum shock radius reaches $10^9$\,cm, the explosion energy at $t=t_9$, the time at which the simulation is stopped, and final explosion energy.}
  \label{tab:summary}
\end{table*}
\section{Numerical Setup}
\label{sec:Numerical Setup}
In this section, we first explain the present initial WD models including how we construct them.
Afterward we briefly describe our numerical code and setup.

\subsection{Initial WD models}
\label{sec:Initial WD models}
We perform collapse simulations of rigidly rotating WDs with various rotational parameters, assuming equatorial or octant symmetry.
The initial configuration is in hydrostatic equilibrium and can be constructed for a given central rest mass density $\rho_{\rm c,0}$ and ratio of the rotational kinetic energy, $T_\mathrm{rot}$, to the gravitational potential energy $W$, $\beta\equiv|T_\mathrm{rot}/W|$, together with thermodynamic quantities to close the system \citep[see, e.g.,][]{shibata04}.
As one of our aims in this study is to discuss electromagnetic counterparts of AICs, in particular the one associated with a magnetar formation, we further adopt one magnetized model (R6oB), albeit assuming an octant symmetry for the sake of saving computational resource.

In this study, we consider relatively massive rotating WDs, intended to assume Oxygen-Neon-Magnesium (ONeMg) WDs, for which the central density becomes so high that gravitational collapse due to the electron capture is triggered, rather than the detonation branch, i.e. the type Ia SN \citep{Nomoto&Kondo91}.
For this purpose, we always employ high central density as $\rho_{\rm c,0}=10^{10}$\,g\,cm$^{-3}$.
Regarding the initial WD rotational profile, we choose various parameters $\beta_0\sim0.1$, 0.2, 0.3, and 0.6\,\%.
We note that it is still a matter of debate, how such rapidly rotating massive WDs having $\sim1.7\,M_\odot$ can be born in astrophysically plausible binary systems \citep[see][for more detailed discussion and possible branches to these WDs]{Yoon&Langer05}. 

As for the thermodynamic profile, we simply assume constant electron fraction $Y_e(=0.5)$ and low entropy $s(=0.8$\,$k_{\rm B}$\,baryon$^{-1}$) structures, because of the lack of plausible thermodynamic profiles of WD and also to avoid numerical difficulties originated from dealing with too low temperature or entropy.
This assumption is one of differences in comparison to many of previous AIC models \citep{Dessart06_AIC,Abdikamalov10_AIC,Longo-Micchi23,Batziou&Janka24_AIC}, which give the initial parametrized temperature.

For the present study, we use the SFHo nuclear relativistic mean-field equation of state (EOS) of \citet{SFH}, taking into account the electrons/positrons, and photons contributions.
Since the original SFHo EOS table covers the density and temperature regions only of $\rho\gtrsim10^3$\,g\,cm$^{-3}$ and $T\ge0.1$\,MeV, respectively, we extend the table down to $\rho=3\times10^{-3}$\,g\,cm$^{-3}$ and $T\sim10^{-4}$\,MeV using the EOS of \cite{TimmesEOS}, in order to evolve the matter ejected toward the low-density region accurately. 

Outside the WDs, we initially set up an artificial atmosphere of sufficiently low density. We first define the WD surface $r_{\rm WD}(r,\theta)$ at $\rho=10$\,g\,cm$^{-3}$, where $(r,\theta)$ are the normal radius and azimuthal angle defined in the spherical-polar coordinates and $r_\mathrm{WD}$ is in the range of $1900$--2600\,km.
Beyond the surface, we simply assume a power-law distribution for the atmosphere density $\rho_{\rm atom}(r,\theta)$ as follows:
\begin{equation}
    \rho_{\rm atom}(r,\theta)=10\left(\frac{r_{\rm WD}(r,\theta)}{r}\right)^3{\rm \,g\,cm}^{-3}.
\end{equation}
In the present study, $\rho_\mathrm{atom}$ is always larger than $10^{-3}\,\mathrm{g\,cm^{-3}}$ because the outer boundaries of the computational domain is located at $\sim 1.5 \times 10^4$\,km along each axis.
$Y_e$ and entropy per baryon in the atmosphere are set to be identical with the initial stellar matter, i.e., $Y_e=0.5$ and $s=0.8$\,$k_{\rm B}$\,baryon$^{-1}$, respectively.

Magnetic fields $(\bf{B})$ are given through the vector potential $\bf{A}$ as $\bf B=\nabla\times \bf A$, with 
\begin{eqnarray}
    (A_r, A_\theta, A_\phi)=\left(0,\, 0, \,
    \frac{B_0}{2}\frac{R_0^3}{r^3+R_0^3}r\sin{\theta}
    \right),
\end{eqnarray}
where $B_0$ and $R_0$ represent the magnetic field strength at center and the size of central sphere with uniform magnetic fields, respectively.
$(r,\theta,\phi)$ denote the usual spherical polar coordinates.

Tab.~\ref{tab:summary} summarizes the model name, $\beta_0$, $\rho_{\rm c,0}$, initial total angular momentum $J_0$, initial baryonic WD mass $M_{\rm WD,bar}$, initial angular frequency $\Omega_0$, $\beta$ value at core bounce $\beta_{\rm cb}$, post bounce time $t_9$ when the maximum shock reaches $10^9$\,cm, diagnostic explosion energy measured at $t=t_9$, final simulation time $t_{\rm fin}$, and diagnostic explosion energy at $t=t_{\rm fin}$.
The model names R1, R2, R3, and R6 approximately represent $\beta_0\sim0.1$, 0.2, 0.3, and 0.6\,\%, respectively.
To investigate the impacts of non-axisymmetric effects, we simulate additional two octant symmetry models, whose model name contains ``o''.
The magnetized model is labeled as R6oB, for which we employ relatively strong initial magnetic field $B_0=10^{11}$\,G \citep[see,][for observational and theoretical support for such strong magnetic field]{Wickramasinghe00,Schmidt03,Zhu15}.

\begin{figure}
\begin{center}
\includegraphics[width=\columnwidth]{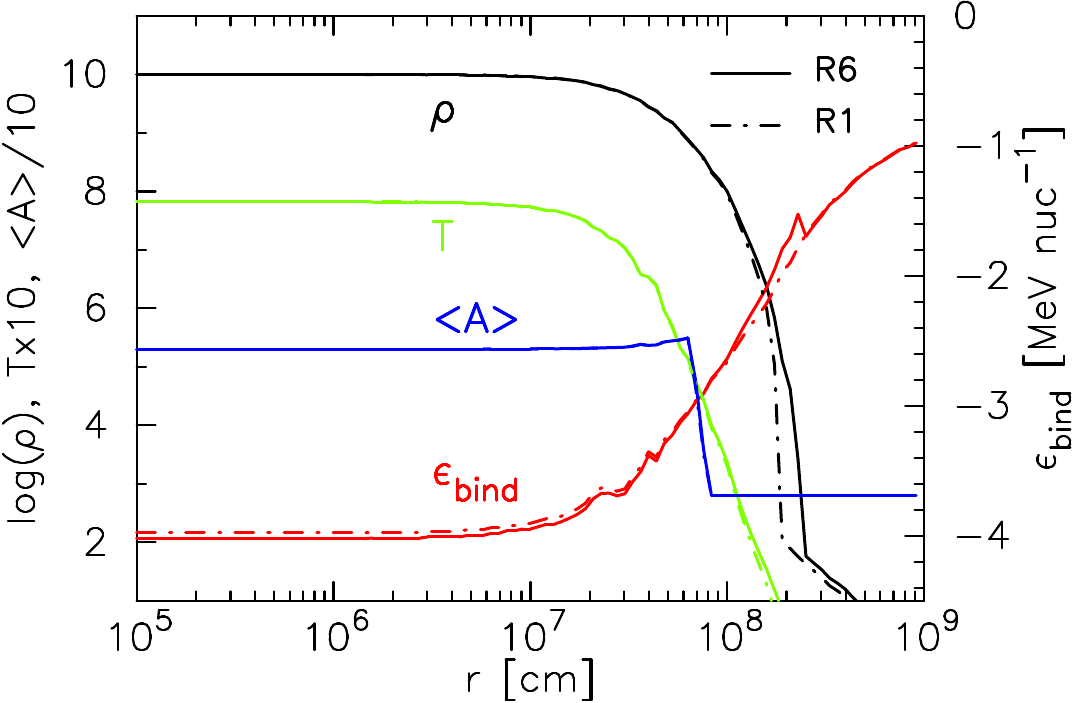}
\caption{Initial configuration of two representative WD models R1 and R6. Shown are the rest mass density in logarithmic scale $\log (\rho\,[{\rm g\,cm^{-3}}])$ (black line), temperature $T$ [MeV] (multiplied by 10; green), mean atomic number $\langle A\rangle$ (multiplied by $1/10$; blue), and binding energy $\varepsilon_{\rm bind}$\,[MeV\,nuc$^{-1}$] (red). 
\label{fig:InitialCondition}}
\end{center}
\end{figure}

Fig.~\ref{fig:InitialCondition} presents the initial spatial profiles of the density, mean atomic number $\langle A\rangle$, and binding energy $\epsilon_{\rm bind}$ along the equator.
Our fastest rotating model R6 has an enlarged equator radius by $\sim30$\,\% compared to $\approx1.92\times10^8$\,cm for the slowest model R1.

\subsection{Radiation GRMHD code}
\label{sec:Radiation GRMHD code}
Numerical computations are carried out using our 3D-GR neutrino-radiation code of \citet{KurodaT21,KurodaT24}.
It solves the Einstein equations based on the BSSN formalism (\citealt{Shibata95, Baumgarte99}) with a constraint violation propagation prescription in terms of a Z4c formalism \citep[cf.][and references therein]{Hilditch13} by which a local constraint violation can be propagated away.
The code is described on a static Cartesian spatial mesh, for which we employ here the same resolution as in \citet{KurodaT21}.
The computational domain extends to $1.5\times10^4$\,km from the center along each axis, in which 2:1 ratio nested boxes from 0 to $L_{\rm max}$ refinement levels are embedded.
In the present study, we set $L_{\rm max}=11$ and each nested box contains $64^3$ cells so that the finest resolution at the center achieves 229~m.
The GR spectral neutrino transport is based on the two-moment scheme with M1 analytical closure \citep[cf.][and references therein]{Shibata11}, with the same neutrino energy resolution of 12 bins in the range of 1--300~MeV, as was used in \citet{KurodaT21}. The weak rates used for the collision integral of the neutrino transport equation are given in \cite{Kotake18}. 

\section{Results}
\label{sec:Results}

In this section, first, we briefly overview the dynamical evolution of the computed models in Section\,\ref{sec:Collapse and explosion dynamics}.
Then in Section\,\ref{sec:Ejecta properties}, we describe the ejecta properties in detail.

\subsection{Collapse and explosion dynamics}
\label{sec:Collapse and explosion dynamics}
%

\begin{figure}
\begin{center}
\includegraphics[width=1.0\columnwidth]{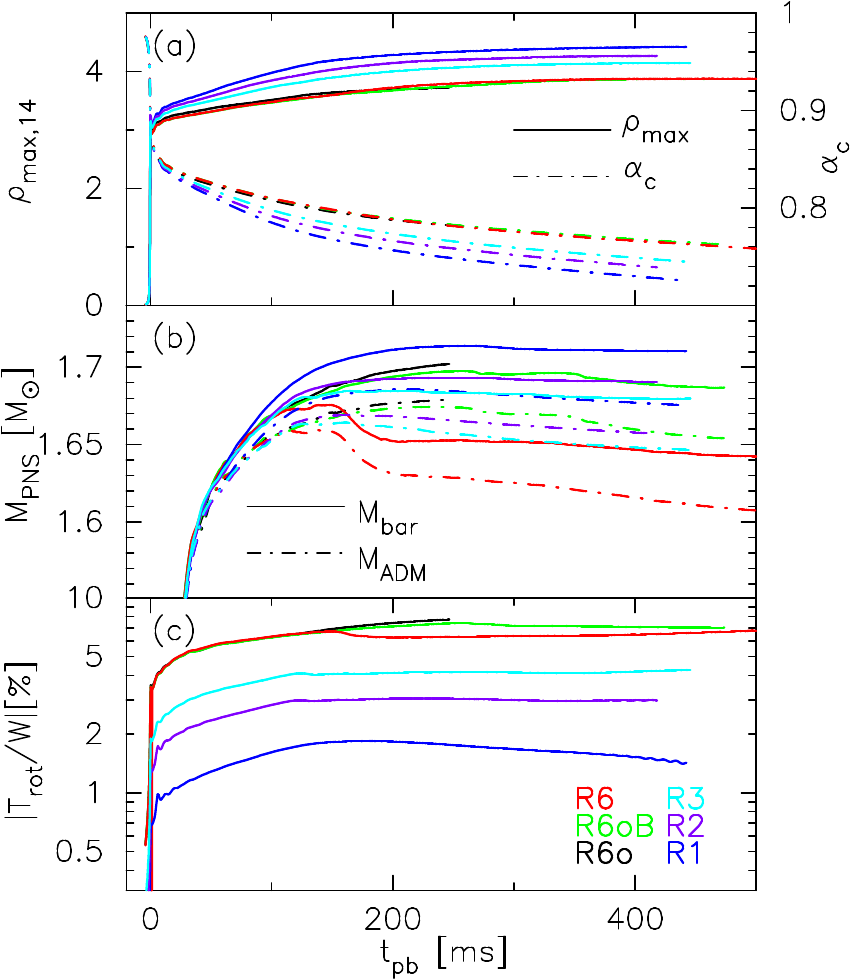}
\caption{Panel (a): Evolution of the maximum rest mass density $\rho_{\rm max,14}$ (solid) in units of $10^{14}$\,g\,cm$^{-3}$ and the central lapse function $\alpha_{\rm c}$ (dash-dotted). (b): The PNS mass in baryonic ($M_{\rm bar}$: solid) and ADM-based value ($M_{\rm ADM}$, see text for definition: dash-dotted). (c): The ratio of rotational kinetic energy $T_\mathrm{rot}$ to gravitational potential energy $W$ of PNS. 
\label{fig:Overall}}
\end{center}
\end{figure}
We begin with describing the overall evolution process.
As explained in Sec.~\ref{sec:Initial WD models}, the inner region of all current WD models are composed of heavy nuclei, and because of the sufficiently high initial central density ($\rho=10^{10}$\,g\,cm$^{-3}$), the electron capture immediately proceeds after the initiation of computation.
Consequently, it deprives the electron degeneracy pressure, which is the main force against WD's self gravity, and induces the gravitational collapse.
At $\sim0.13$\,s after the onset of collapse, all models undergo the core bounce, not by the centrifugal force but by the strong nuclear force, as is evident from the bounce density exceeding a nuclear saturation density $\sim2\times10^{14}$\,g\,cm$^{-3}$.
In the meantime, if we switch off the neutrino matter interactions, i.e., in the absence of electron captures, we confirmed that the core does not initiate collapse and stays nearly its initial state at least for $1$\,s.

There is a slight rotational dependence of the core bounce time on the rotational degree. 
The fastest rotating models R6/R6o show $\sim1$\,ms delay in comparison to the slowest one R1.
Additionally, the faster rotating model exhibits a lower central density and higher central lapse for a give time as seen in panel (a) of Fig.~\ref{fig:Overall}.
All these features are general rotational effects and such a trend is preserved throughout the simulation time.
The proto-neutron star (PNS) mass also shows a clear rotational dependence in both baryonic $M_{\rm PNS,bar}$ and Arnowitt–Deser–Misner (ADM)-based mass $M_{\rm PNS,ADM}$\footnote{See Eq.~(3) in \cite{KurodaT24} for our definition of $M_{\rm PNS,ADM}$. We note that $M_{\rm PNS,ADM}$ is used just as a broad measurement of the gravitational mass of the PNS as it is not yet an isolated system but still surrounded by the original WD envelope, making it difficult to clearly separate the PNS and ejecta.} (panel b).
We define the PNS by the region where the rest mass density exceeds $10^{10}$\,g\,cm$^{-3}$.
Both $M_{\rm PNS,bar}$ and $M_{\rm PNS,ADM}$ become higher for slowly rotating models.
At post bounce times of $t_{\rm pb}\sim100$--150\,ms, the baryonic PNS mass nearly plateaus in models R1, R2, and R3.

At $t_{\rm pb}\sim100$\,ms, the red (R6) and green (R6o) lines start to diverge.
While the PNS mass continuously grows in model R6o, a significant decrease is thereafter observed in the counterpart model R6.
We will later discuss this noteworthy behavior.
The neutrino emission deprives the internal energy from the PNSs and reduces $M_{\rm PNS,ADM}$ with time, which makes the difference between $M_{\rm PNS,bar}$ and $M_{\rm PNS,ADM}$ gradually larger.

\begin{figure}
\begin{center}
\includegraphics[width=\columnwidth]{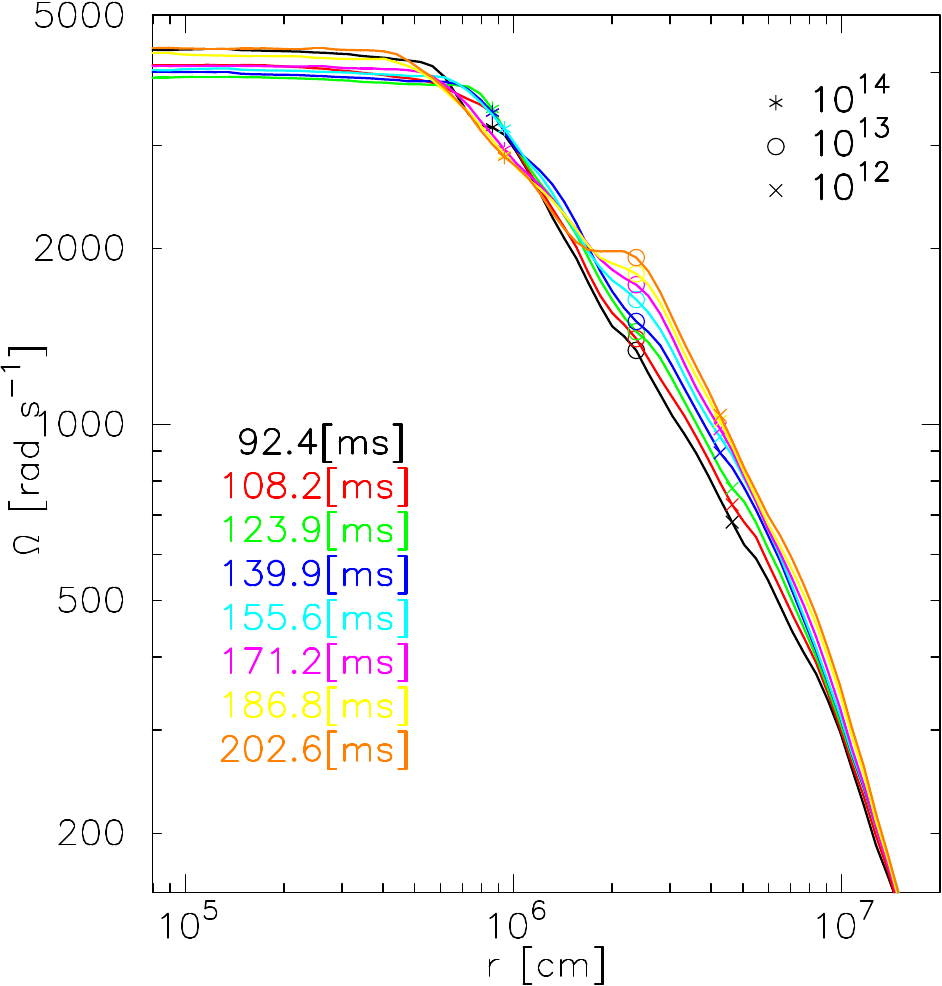}
\caption{Spatial profile of azimuthal angle averaged angular frequency $\Omega(\varpi)$ for model R6 on the equatorial plane and its time evolution as represented by the post bounce time on the lower-left corner.
The radii of angle averaged isodensity surface at $\rho=10^{12,13,14}$\,g\,cm$^{-3}$ are also marked by points shown on the upper-right corner.
\label{fig:Omega}}
\end{center}
\end{figure}
\begin{figure*}
\begin{center}
\includegraphics[angle=-90.,width=\textwidth]{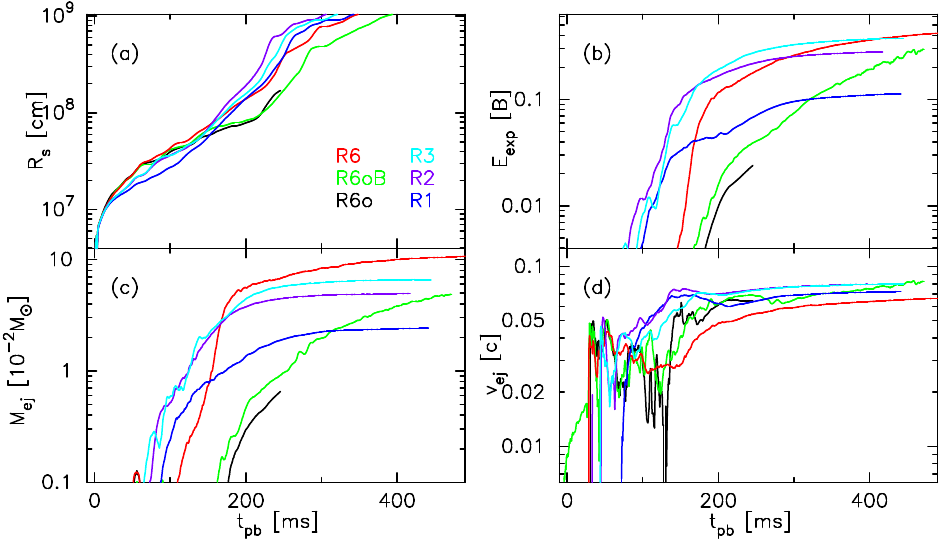}
\caption{Evolution of the maximum shock radius $r_{\rm shock}$ (top-left), diagnostic explosion energy $E_{\rm exp}$ (top-right), ejecta mass $M_\mathrm{ej}$ (bottom-left), and bulk velocity of ejecta  $v_\mathrm{ej}$ (bottom-right) for all models.
\label{fig:RshockEexp}}
\end{center}
\end{figure*}

Panel (c) presents the dimensionless rotational parameter $|T_\mathrm{rot}/W|$. $T_\mathrm{rot}$ and $W$ here are, respectively, the rotational kinetic and gravitational potential energies of the PNS.
From the plot, we see that $|T_\mathrm{rot}/W|$ in all models exceeds 1\,\% after bounce.
In terms of the PNS angular frequency, the frequency goes beyond $1$\,kHz even in the slowest rotating model R1.
One remarkable feature is found in the most rapidly rotating models R6 and R6o and their comparison.
Their $|T_\mathrm{rot}/W|$ reaches $\sim7$\,\% at $t_{\rm pb}\sim150$\,ms and then begin to diverge from one another.
R6 (red line) exhibits a sudden decrease while the octant symmetry model R6o continues increasing.
As mentioned already, the PNS mass in model R6 also shows a similar sudden decrease at the same time.
We attribute these features to the low-$|T_\mathrm{rot}/W|$ instability \citep{Centrella01,Shibata02,Shibata03,Watts05,Ott05,Saijo06,Scheidegger10,Takiwaki18}.
Indeed such a sudden decrease in $|T_\mathrm{rot}/W|$ is reported in previous 3D simulations in the context of rapidly rotating neutron star \citep{Ott05,Shibagaki21}.
Although a similar mass decrease was not explicitly mentioned in these previous 3D studies, the mass decrease can be naively explained by the formation of strong one-armed spiral waves associated with the low-$|T_\mathrm{rot}/W|$ instability, which ejects a fraction of gravitationally-loosely bound PNS envelope through the angular momentum transport by the gravitational torque.

The low-$|T_\mathrm{rot}/W|$ instability is known as an efficient mechanism to redistribute the angular momentum \citep{Saijo06}.
To see whether the observed sudden decrease in both $|T_\mathrm{rot}/W|$ and $M_{\rm PNS}$ can indeed be attributed to the low-$|T_\mathrm{rot}/W|$ instability, we examine how the spatial profile of angular frequency evolves across the corresponding period ($100\le t_{\rm pb}\le200$\,ms).
Fig.~\ref{fig:Omega} plots the spatial profile of azimuthal angle averaged angular frequency $\Omega(\varpi)$ for model R6 on the equatorial plane defined by
\begin{eqnarray}
    \Omega(\varpi) \equiv\left. \frac{1}{2\pi}\int d\phi \frac{v_\phi}{\varpi} \right|_{z=0},
\end{eqnarray}
with $v_\phi$ being the azimuthal component of three-velocity and $\varpi=\sqrt{x^2+y^2}$.
We plot several time snapshots in different colors represented by the post bounce time on the lower-left corner.
From Fig.~\ref{fig:Overall}, we find that $t_{\rm pb}=92$ and 202\,ms correspond to before and after the mass as well as $|T_\mathrm{rot}/W|$ decrease, respectively.
In addition, we overplot the radii of angle averaged isodensity surface at $\rho=10^{12,13,14}$\,g\,cm$^{-3}$ by marks shown on the upper-right corner.
These isodensity surfaces will be used later also in our discussion about GW emissions.
From the figure, we can clearly see that $\Omega(\varpi)$ drastically increases at $r\sim25$\,km, where the density is $\rho\approx 10^{13}$\,g\,cm$^{-3}$, across the corresponding phase.
Such a spin-up at an outer region is consistent with the finding of \cite{Ott05}.
We simultaneously witness that at a slightly inner region $r\sim10$\,km, a spin-down can be observed.
All these features support our interpretation that the low-$|T_\mathrm{rot}/W|$ instability indeed activates during $100\le t_{\rm pb}\le 200$\,ms and redistributes the angular momentum from inner to outer region, which also works to eject a part of spun-up PNS components and reduces the PNS mass.

Next we pay attention to the explosion dynamics.
Fig.~\ref{fig:RshockEexp} plots the time evolution of the maximum shock radius $R_{\rm s}$ in panel (a), diagnostic explosion energy $E_{\rm exp}$ (b), ejecta mass $M_{\rm ej}$ (c), and ejecta velocity $v_{\rm ej}$ (d).
Here we define the maximum shock radius $R_s$ by the outermost radius where the entropy exceeds $s=4\,k_{\rm B}$\,baryon$^{-1}$.
Additionally, $E_{\rm exp}$ and $M_{\rm ej}$ are evaluated by
\begin{eqnarray}
    E_{\rm{exp}}&=&f_\mathrm{s}\int_{e_{\rm{bind}}>0,r>100\,{\rm km}} e_{\rm{bind}} \sqrt{\gamma}\,dx^3 \label{eq:Eexp}, \\
    M_{\rm{ej}}&=&f_\mathrm{s}\int_{e_{\rm{bind}}>0,r>100\,{\rm km}} \rho \sqrt{\gamma}\,dx^3 \label{eq:Mej},
\end{eqnarray}
with the binding energy in each numerical cell $e_{\rm bind}$ being evaluated by \citep{BMuller12a}
\begin{eqnarray}    e_{\rm{bind}}&=&\alpha c^2\left(\tau+\rho \Gamma \right) -\rho \Gamma c^2.\label{eq:ebind}
\end{eqnarray}
Here $\alpha$ is the lapse function, $\tau$ is the relativistic energy density excluding the rest mass energy \citep[see][for the definition]{KurodaT20}, $\rho$ is the rest mass density, $\Gamma$ is the Lorentz factor, and $\gamma$ is the determinant of the three-dimensional spatial metric.
For the volume integral, we apply additional criterion $r>100$\,km on top of $e_{\rm{bind}}>0$ to prevent the matter in the vicinity of the PNS from being identified as unbound.
Note that we assume an equatorial and octant symmetry, where we calculate only a domain at $z>0$ and $(x,y,z)>0$ of the full 3D domain, respectively. Hence, the factor $f_\mathrm{s}=2$ or 8 in Eqs.~(\ref{eq:Eexp}) and (\ref{eq:Mej}) is added, which stems from the assumed equatorial or octant symmetry, to account for the contribution from outside the computational domain.
Furthermore, from $E_{\rm exp}$ and $M_{\rm ej}$, we approximately estimate the bulk velocity of ejecta $v_{\rm ej}$ as follows:
\begin{eqnarray}
    v_{\rm ej}=\sqrt{\frac{2E_{\rm exp}}{M_{\rm ej}}}.
\end{eqnarray}
\begin{figure*}

  \begin{minipage}{0.48\textwidth}
    \centering
    \includegraphics[width=\textwidth]{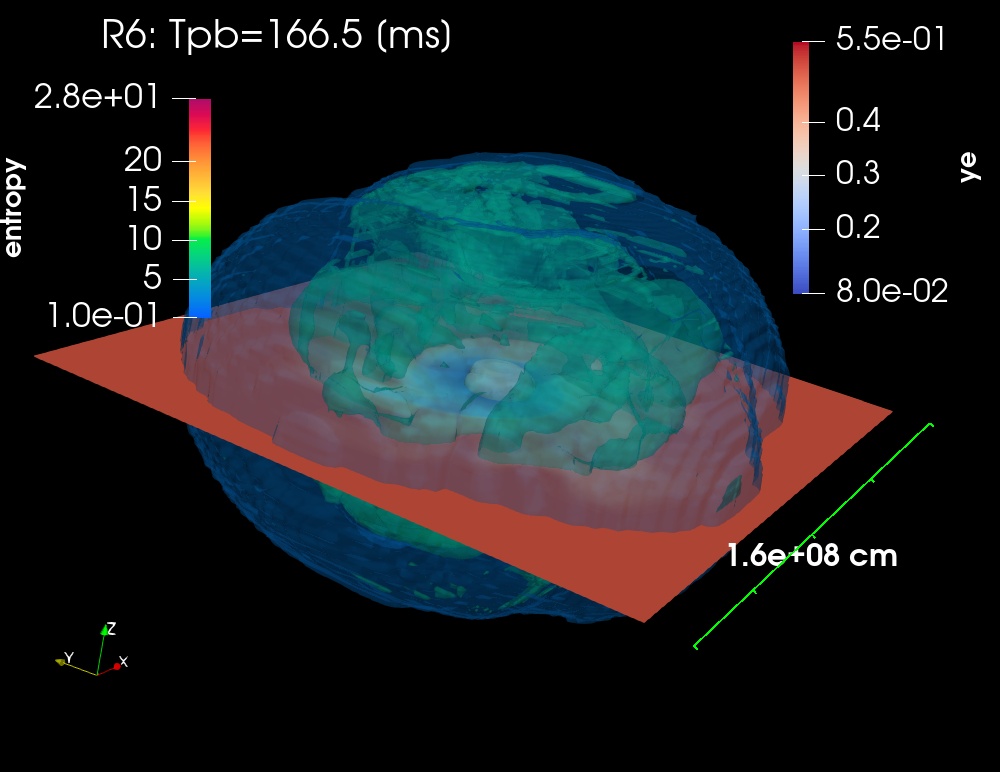}
    \label{fig:3D_1}
  \end{minipage}
  \begin{minipage}{0.48\textwidth}
    \centering
    \includegraphics[width=\textwidth]{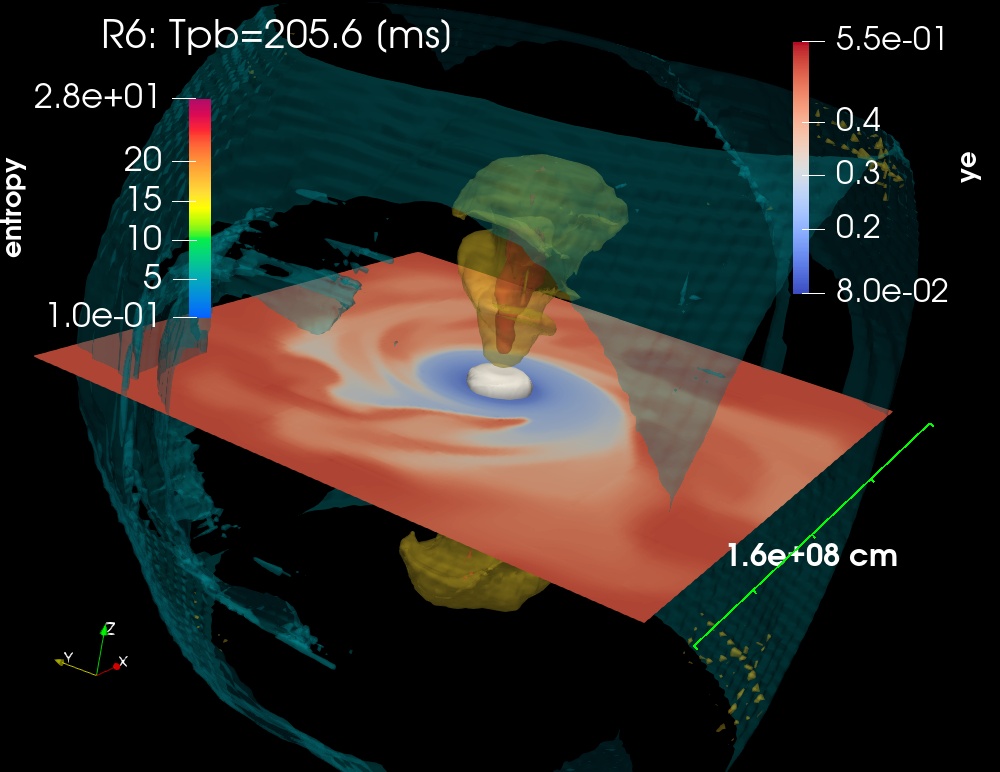}
    \label{fig:3D_2}
  \end{minipage}

  \vspace{-2mm}

  \begin{minipage}{0.48\textwidth}
    \centering
    \includegraphics[width=\textwidth]{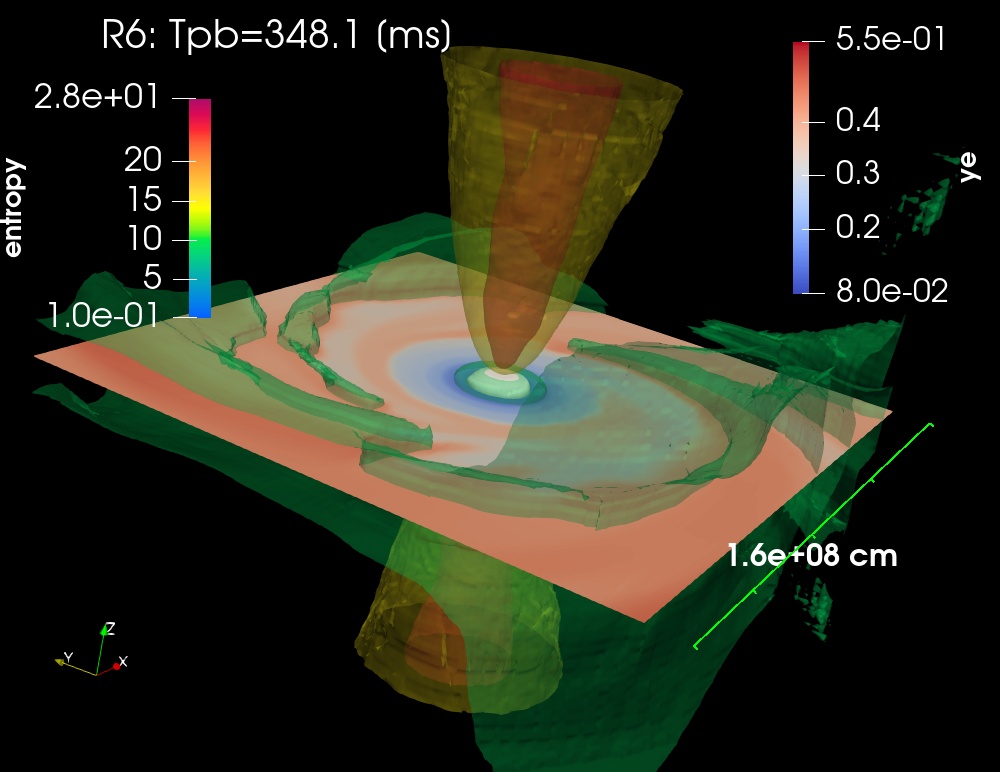}
    \label{fig:3D_3}
  \end{minipage}
  \begin{minipage}{0.48\textwidth}
    \centering
    \includegraphics[width=\textwidth]{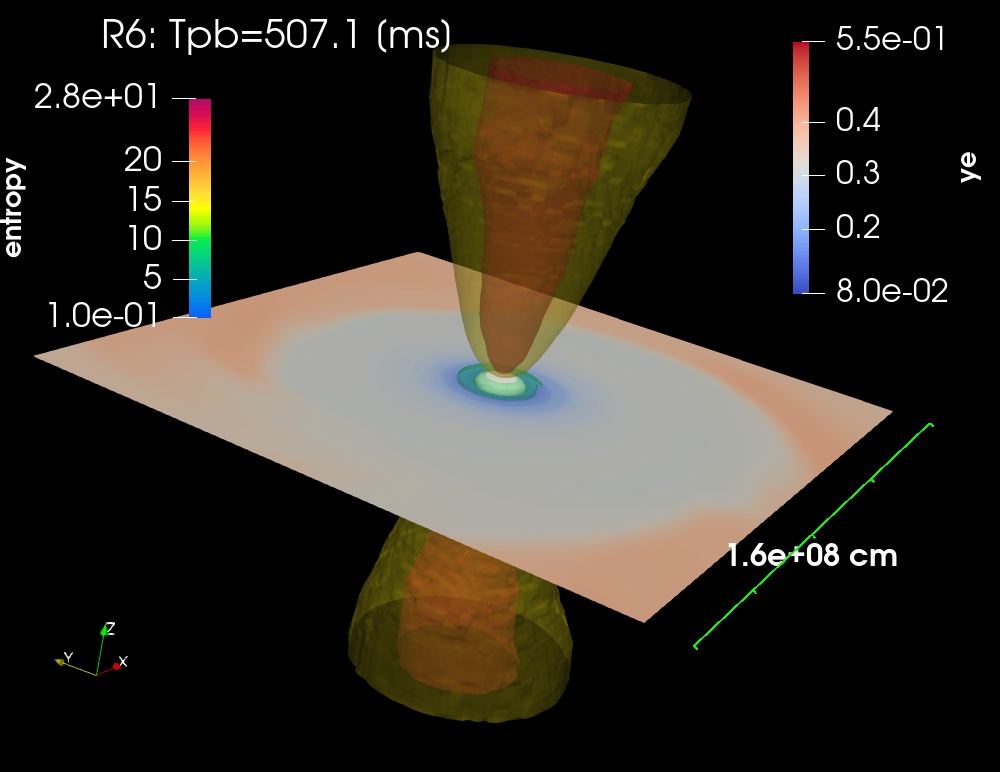}
    \label{fig:3D_4}
  \end{minipage} 
\center    
\caption{The isoentropy surface and $Y_e$ contour for model R6 within $\sim800$\,km from the center.
We depict them for four different post bounce times $t_{\rm pb}=166$ (top-left), 205 (top-right), 348 (bottom-left), and 507\,ms (bottom-right).
Regarding the $Y_e$ contour, we show its slice on the equatorial plane. \label{fig:3D}}
\end{figure*}

The maximum shock evolution indicates that the bounce shock directly leads to the runaway explosion shock wave in all models.
At initial post bounce phase of $t_{\rm pb}\lesssim150$\,ms, the faster rotating models show more extended shock positions, i.e., model R6 shows the largest shock radii.
This results simply from the enhanced centrifugal force, which drives the shock front further out.
During the initial post bounce phase of $t_{\rm pb}\lesssim150$\,ms, however, the more extended shock position implies neither more energetic explosions nor more unbound matters.
Indeed, models R1, R2, and R3 show larger unbound mass and diagnostic explosion energy than those of R6 at $t_{\rm pb}\lesssim150$\,ms.
A drastic increase in the explosion energy continues till $t_{\rm pb}\sim200$\,ms in all models (middle panel).
Thereafter the increase becomes slightly milder.
For $t_{\rm pb}\gtrsim200$\,ms, their values settle to a few times of 0.1\,B$({\rm Bethe}\equiv10^{51}\,{\rm erg})$.
The ejecta mass $M_{\rm ej}$ broadly follows a similar evolutionary path to $E_{\rm exp}$.
For $t_{\rm pb}\gtrsim200$\,ms, $M_{\rm ej}$ has in general a correlation with the WD rotation, and models R1 and R6 exhibit the smallest ($M_{\rm ej}\sim0.02$\,$M_\odot$) and largest ($\sim0.1$\,$M_\odot$) mass ejection, respectively. This correlation indicates that the bulk ejecta velocity is always $\sim 0.1\,c$ regardless of the ejecta mass.

The magnetized model R6oB indicates a rather similar evolution process with its counterpart non-magnetized model R6o at least during our simulation time of R6o ($t_{\rm pb}\sim250$\,ms).
Except a slightly energetic explosion in R6oB, i.e., higher $E_{\rm exp}$ and larger $M_{\rm ej}$, all behaviors shown in Fig.~\ref{fig:RshockEexp} are quantitatively similar between R6oB (green lines) and R6o (black).
Indeed for R6oB we do not observe an emergence of ``magnetically driven jet'', inside which the magnetic pressure dominates over the gas pressure typically by $1$--$2$ orders of magnitude larger \citep[e.g.][]{Obergaulinger20,Bugli20,KurodaT24}.
In these previous SN models, the initial magnetic field imposed is often one order of magnitude stronger ($\sim10^{12}$\,G) than the value chosen in this work, which results in the magnetically driven jet formation relatively soon after core bounce at $t_{\rm pb}\lesssim100$\,ms.
By contrast in model R6oB, the gas pressure always dominates over the magnetic one by approximately one order of magnitude inside the bipolar outflow.
This is because the initial poloidal magnetic fields $B_0=10^{11}$\,G, which is motivated by observational and theoretical studies of strongly magnetized WDs \citep{Wickramasinghe00,Schmidt03,Zhu15}, cannot be amplified to the dynamically relevant strength by linear amplification mechanisms within such a short time scale and/or our numerical resolution (e.g. $\sim450$\,m at just above the PNS surface $r\sim30$\,km) is not sufficient to resolve potential non-linear amplification mechanisms such as MRI \citep{Balbus91}.
From these, we conclude that the explosion witnessed in model R6oB is facilitated mainly by the neutrino heating and the overall explosion dynamics is essentially the same with its counterpart non-magnetized model R6o within our limited simulation times ($t_{\rm pb}\le500$\,ms).
This is more or less consistent with recent AIC models with magnetic fields \citep{Cheong25_AIC}, in which they reported no significant difference during the first $500$\,ms after bounce among the models using initial magnetic fields of $10^{11}$\,G or weaker.

We, however, mention that there are some non-negligible contributions from magnetic fields.
For instance, we find regions outside the outflow opening angle (i.e. $30^\circ\lesssim\theta\lesssim50^\circ$), where the magnetic pressure dominates over the gas pressure by approximately one order of magnitude.
Such highly magnetized regions may possibly influence the mass ejection along with the further magnetic field amplifications.
Additionally, the remnant is strongly magnetized, achieving poloidal ($B_{\rm p}$) and toroidal ($B_\phi$) components of $B_{\rm p}\sim10^{14-15}$\,G and $B_{\phi}\sim10^{16}$\,G inside PNS, respectively.
Such strong magnetic fields together with rapid rotation for formed NSs can indeed be a key ingredient to explain rapidly evolving bright transients as will be discussed in Sec.~\ref{sec:Implications for the electromagnetic counterparts}.

\cite{Longo-Micchi23} have recently performed 3DGR AIC simulations, albeit with a rather simplified neutrino transport and shorter simulations times ($t_{\rm pb}\lesssim160$\,ms) than ours.
Their explosion energy and ejecta mass reach $\sim1$\,B and $\sim3\times10^{-2}$\,$M_\odot$ in rotating models, while these values remain essentially zero in non-rotating model, i.e. no explosion.
Another latest 2D long-time simulations by \cite{Batziou&Janka24_AIC} show $E_{\rm exp}\sim0.01$--0.25\,B and $M_{\rm ej}\sim0.2$--$5\times10^{-2}$\,$M_\odot$ at $t_{\rm pb}\gtrsim4$\,s, which are evaluated at quite a later phase than ours.
Admitting that there are several differences in numerical methods and initial WD models between ours and theirs, such as mass and rotational profiles, if we simply compare our explosion energies and ejecta masses with these previous studies, they broadly show a good agreement except our rapidly rotating model R6, which exhibits a considerably larger ejecta mass, $M_{\rm ej}\sim0.1$\,$M_\odot$.

The exceptionally large ejecta mass for model R6 can be attributed to the low-$|T_\mathrm{rot}/W|$ instability.
When the low-$|T_\mathrm{rot}/W|$ instability sets in at $t_{\rm pb}\sim150$\,ms (see the red line in panel (c) of Fig.~\ref{fig:Overall}), the one-armed spiral wave appears and ejects matters, which rapidly increases the unbound mass as can be seen from the red line in the right panel of Fig.~\ref{fig:RshockEexp}.
Indeed its counterpart octant symmetry model R6o (black line) exhibits approximately an order of magnitude smaller ejecta-mass evolution, albeit from its limited post bounce evolution, since the octant symmetry completely suppresses the odd-mode non-axisymmetric flow motions including the low-$|T_\mathrm{rot}/W|$ instability.

From $E_{\rm exp}$ and $M_{\rm ej}$, we approximately measure the bulk ejecta velocity $v_{\rm ej}$ (panel d).
As can be seen, all ejecta velocities asymptotically approach $v_{\rm ej}\approx0.08$\,c.
Non-magnetized models, especially R2 and R3, exhibit nearly stagnated velocity evolution at $t_{\rm pb}\gtrsim300$\,ms.
Regarding $v_{\rm ej}$ in magnetized case R6oB, it is still growing at the final simulation time, which might be due to the aforementioned non-negligible contributions from magnetic fields.

To visualize more clearly the global explosion morphology in model R6, in particular its one-armed spiral structure, we depict the isoentropy surface and $Y_e$ contour at four different post bounce times in Fig.~\ref{fig:3D}.
Regarding the $Y_e$ contour, we show its slice on the equatorial plane.
At $t_{\rm pb}\sim166$\,ms (top-left), approximately when the low-$|T_\mathrm{rot}/W|$ sets in, the shock front (outermost bluish surface) shows an oblate structure because of the centrifugal expansion.
At $t_{\rm pb}\sim205$\,ms (top-right), we can clearly see fully developed one armed spiral patterns on the equator.
Furthermore, the bipolar outflows driven by the neutrino heating also start to emanate.
This is because the neutrino luminosity and energy are higher near the rotational axis due to the rotational flattening of the neutrino sphere and increase the heating efficiency there.
This is a commonly observed feature in rotating NSs \citep{Summa18,KurodaT20,Obergaulinger21,Takiwaki21}.
The bipolar outflow remains at least till the final simulation time of $t_{\rm pb}\sim500$\,ms in model R6.
Although not shown, all the rest of our rotating models R1--R3 also exhibit a bipolar structure.

\begin{figure*}
\begin{center}
\includegraphics[width=0.98\textwidth]{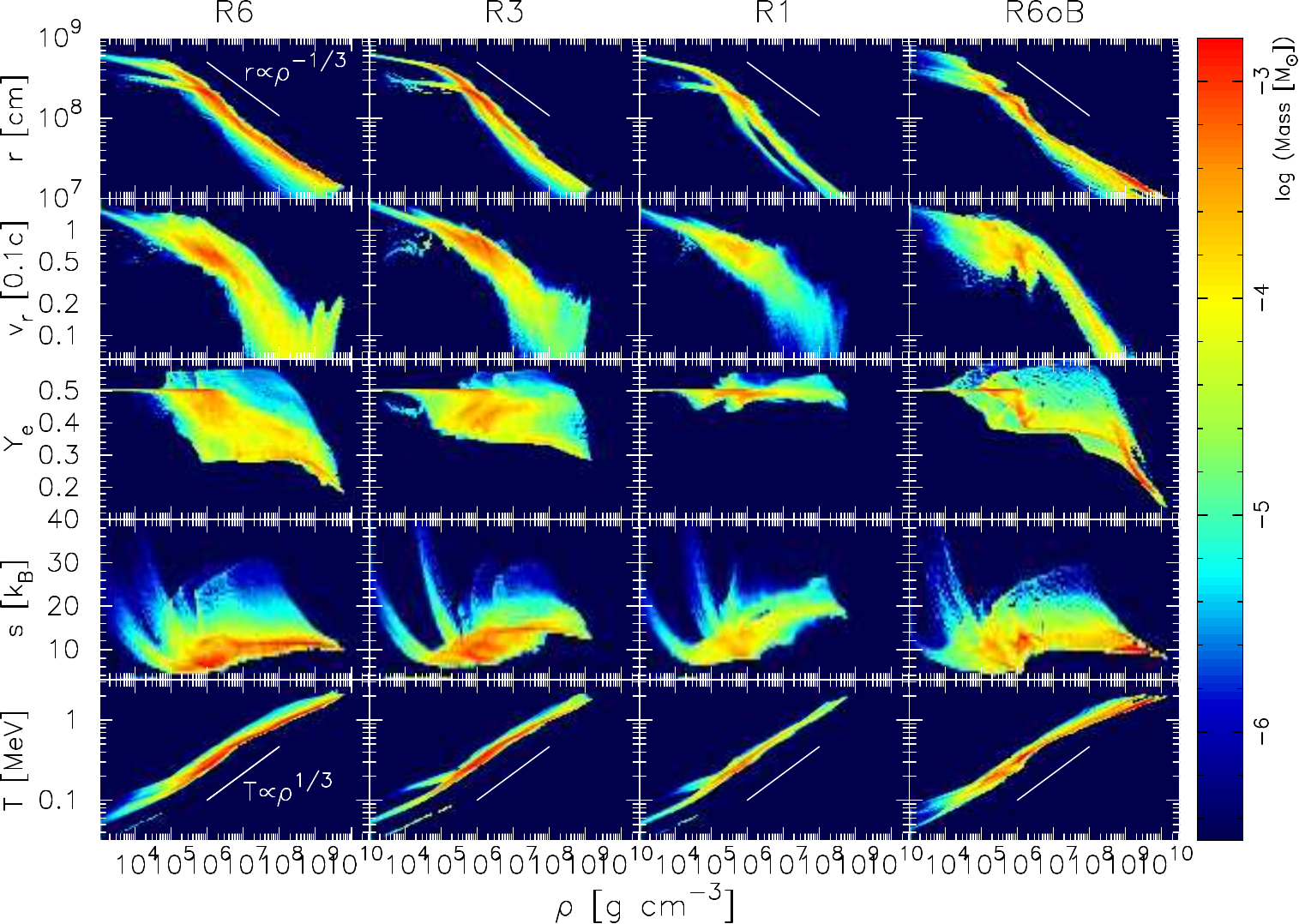}
\caption{Distributions of unbound fluid elements at the final simulation time on the plane of the rest mass density $\rho$ against (from top) the distance $r$ of ejecta from the center, radial velocity $v_r$\,[0.1c], $Y_e$, entropy $s$ [$k_{\rm B}$\,baryon$^{-1}$], and temperature $T$\,[MeV].
The color denotes the integrated mass in units of solar mass in logarithmic scale.
For reference, we also add two white lines in the top and bottom panels, denoting the adiabatic relation $T\propto\rho^{1/3} \propto r^{-1}$.
Note that we exhibit all ejecta profiles regardless of their expanding direction.
\label{fig:ejecta2D}}
\end{center}
\end{figure*}

From the final time snapshot (bottom-right), we observe that the low-$Y_e(\sim0.3)$ matters (whitish region) occupy a large domain on the equator ($r\lesssim800$\,km).
\cite{Longo-Micchi23} reported that the low-$Y_e$ material ($Y_e\lesssim0.3$) is confined only to the central region $r\lesssim200$\,km in rotating models at their final simulation time of $t_{\rm pb}\sim150$\,ms.
At a similar post bounce phase ($t_{\rm pb}\sim166$\,ms), our rotating model shows a consistent $Y_e$ distribution with \cite{Longo-Micchi23}, with those low-$Y_e$ components being confined to $r\lesssim200$--300\,km.
However, our longer time simulations reveal that such low-$Y_e$ components can potentially be ejected at a later phase along with  the development of spiral waves originated from the rotational instability.
Meantime, we observe high-$Y_e$ components ($\gtrsim0.55$) concentrating near the rotational axis (will be explained in the next Sec.~\ref{sec:Ejecta properties}).
This is a consequence of relatively high neutrino luminosity and energy near the rotational axis, which increases the efficiency of neutrino absorption and the $Y_e$ value.
This feature is also consistent with other 2D AIC models accompanying a bipolar-structure \citep{Dessart06_AIC,Batziou&Janka24_AIC}.

\subsection{Ejecta properties}
\label{sec:Ejecta properties}

Next we discuss the ejecta properties. 
In Fig.~\ref{fig:ejecta2D}, we display distributions of unbound fluid elements on the plane of the rest mass density ($x$-axis) versus: (from top) the distance $r$ of ejecta from the center, radial velocity $v_r$\,[0.1c], $Y_e$, entropy $s$ [$k_{\rm B}$\,baryon$^{-1}$], and temperature $T$\,[MeV].
We note again that the ejecta are defined by the unbound material locating for $r \geq 100$\,km (based on Eq.~\ref{eq:Mej}).
The color denotes the integrated mass in units of solar mass in logarithmic scale.
These distributions are evaluated when the maximum shock radius reaches $10^4$\,km, which corresponds to $t_{\rm pb}(=t_9)=338$, $317$, $345$, and $390$\,ms for models R1, R3, R6, and R6oB, respectively (see Tab.~\ref{tab:summary}).

The middle raw of Fig.~\ref{fig:ejecta2D} illustrates the $Y_e$ distribution.
The most noteworthy impact of rotation is the ejection of a more neutron rich matter with the increase of the initial rotation.
This trend can be easily read from the significant amounts of neutron rich ejecta with $0.2\lesssim Y_e\lesssim0.4$ for model R6, while those with only $Y_e\gtrsim0.45$ are ejected for model R1.
Albeit an octant symmetric model, another fastest rotating model R6oB also presents a similar profile to that of non-axisymmetric counterpart model R6. We will discuss the difference between R6 and R6oB later.
As explained in Fig.~\ref{fig:3D}, these low-$Y_e$ ejecta are the outcomes of faster rotation and associated spiral waves, which dredge up the low-$Y_e$ components initially floating in the envelope of rapidly rotating oblate PNS.
At the same time, the initial WD components having $Y_e=0.5$ and initially locating near the WD surface are another major ejecta, which are expelled from its bottom simply along with the explosion without significantly changing their initial $Y_e$ values.

We observe higher entropy ejecta for slower rotating models.
For instance, at $\rho\gtrsim10^7$\,g\,cm$^{-3}$, the dominant ejecta appear at $s\sim20$\,$k_{\rm B}$\,baryon$^{-1}$ for model R1, which decreases to $s\sim14$ and $10$\,$k_{\rm B}$\,baryon$^{-1}$ for models R3 and R6, respectively.
In general, the stellar rotation tends to decrease the neutrino luminosity and mean energy.
This results in the lower entropy of neutrino heated ejecta \citep{Summa18,KurodaT20}.

From the lower edge of the $Y_e$ distribution, we can infer how the ejecta change their $Y_e$ values along with the ejection.
For instance for model R6, a part of ejecta initially locates at $(\rho,Y_e)\sim(10^{10}\,\mathrm{g/cm^3},0.2)$, or in terms of initial location at $r\sim10^7$\,cm (top panel).
Note that these values correspond to those near the surface of the PNS.
These ejecta are afterward expelled by the centrifugal force,  approximately preserving the $Y_e$ values, or by the neutrino heating, which changes the $Y_e$ values.
The increase of $Y_e$ lasts till the neutrino-matter interaction practically ceases at $\rho\sim10^{8-9}$\,g\,cm$^{-3}$.
Thereafter, the lower edge of $Y_e$ remains essentially constant and propagates horizontally leftward on this $(\rho,Y_e)$-plane.
Because of the different degrees of rotation, which can potentially change the initial location and thermodynamic properties of marginally bound matters, the lower edge of constant $Y_e$ evolutionary path can differ from model to model.
Moreover the horizontal, highly populated region in the $(\rho,s)$-plane below $\rho\sim10^{8-9}$\,g\,cm$^{-3}$ (e.g. the red region in the second panel from bottom for model R6) also indicates the cease of active neutrino-matter interactions, which lets the subsequent thermodynamic evolution nearly adiabatic.
Such features are commonly seen also for other models R1--R3.

As has been mentioned, the magnetized model R6oB can be considered as essentially a non-magnetized model during our simulation time ($t_{\rm pb}\lesssim500$\,ms), other than the formation of a strongly magnetized PNS, i.e., magnetar.
Therefore, the overall feature of ejecta properties are more or less similar between R6oB and R6 as presented in Fig.~\ref{fig:ejecta2D}.
However, we can still find non-negligible differences.
For instance, the $Y_e$ distribution extends to a slightly lower side $Y_e \approx 0.14$ for model R6oB than $Y_e \approx 0.18$ for R6.
Having admitted that longer time simulation is certainly essential to assess whether these lowest-$Y_e$ components can be eventually ejected, as they still locate near the PNS ($r\sim100$\,km, from top panel), they are quite loosely unbound.
As they are spatially concentrated at an off-axis region $\theta\sim40^\circ$, where the magnetic pressure dominates over the gas pressure as mentioned previously, we still cannot exclude possibilities of ejection of such low-$Y_e$ components by magnetic field activities at a later phase.
In addition, mass ejection via electromagnetic forces generally suppress the increase of $Y_e$ in comparison to the neutrino heating process, which might result in more visible impacts of magnetic fields on lowering the final $Y_e$ distribution.
To clarify the final properties of the ejecta, it is necessary to study the magnetised model with the longer-term evolution.

The velocity distribution shows that most of ejecta are driven outward at the velocity of $\sim0.01$--0.1\,c, which is consistent with the estimated bulk velocity in Fig.~\ref{fig:RshockEexp}.
The most rapidly rotating model R6 exhibits a broader velocity distribution ($0.01\,{\rm c}\lesssim v_r\lesssim0.1$\,c), while R1 shows a slightly narrower profile ($0.05\,{\rm c}\lesssim v_r\lesssim0.1$\,c).
From a comparison between models R6 and R6oB, we observe a similar profile among them. Namely, R6oB also presents a velocity distribution composed of slow ($\sim0.01\,c$) and fast ($\gtrsim0.1\,c$) ejecta, which is distinct from that of R1.
Temperatures inside ejecta, which have already been accelerated beyond $v_r\gtrsim0.05$\,c, decrease below $T\lesssim0.3$\,MeV due to adiabatic cooling.
Indeed the temperature and rest-mass density approximately satisfy the relation $T\propto\rho^{1/3} \propto r^{-1}$ as can be seen from the two white reference lines in the top and bottom panels, indicating that the ejecta expand approximately adiabatically.
A fraction of these ejecta have relatively low $Y_e$ of $\sim0.3$, which would be of interest from the perspective of nucleosynthesis.
As a comparison with recent 3D study of \cite{Longo-Micchi23} (see their figure 22), we generally observe more scattered distributions, e.g., higher velocities, lower temperatures, and both higher- and lower-$Y_e$ matters.
Beside a difference in the neutrino transport between ours (multi-energy) and theirs (gray single-energy), those differences seen in the distribution can be attributed to our longer simulation times.
The longer-term simulations simply enable us to follow a later phase of the aforementioned nearly adiabatic expansion of ejecta, which generally extends the temperature distribution toward a lower side, and also unbind more matters, which are otherwise still unbound if the simulation time is shorter, by neutrino heating or angular momentum transfer associated with the non-axisymmetric instability, leading to the ejection of both lower- and higher-$Y_e$ components.
Recent long-term models of \cite{Batziou&Janka24_AIC} also report scattered distributions.

We also find that the radial distribution (top-row) is sometimes not aligned along a single line.
This can simply be attributed to the multi component nature of explosion morphology.
For instance, there is an outflow structure along the pole and at the same time an expanding structure is observed toward equator.
Therefore the ejecta having a same density sometimes appear at different locations (or radii) among these distinct explosion components.

\begin{figure*}
\begin{center}
\includegraphics[width=0.99\textwidth]{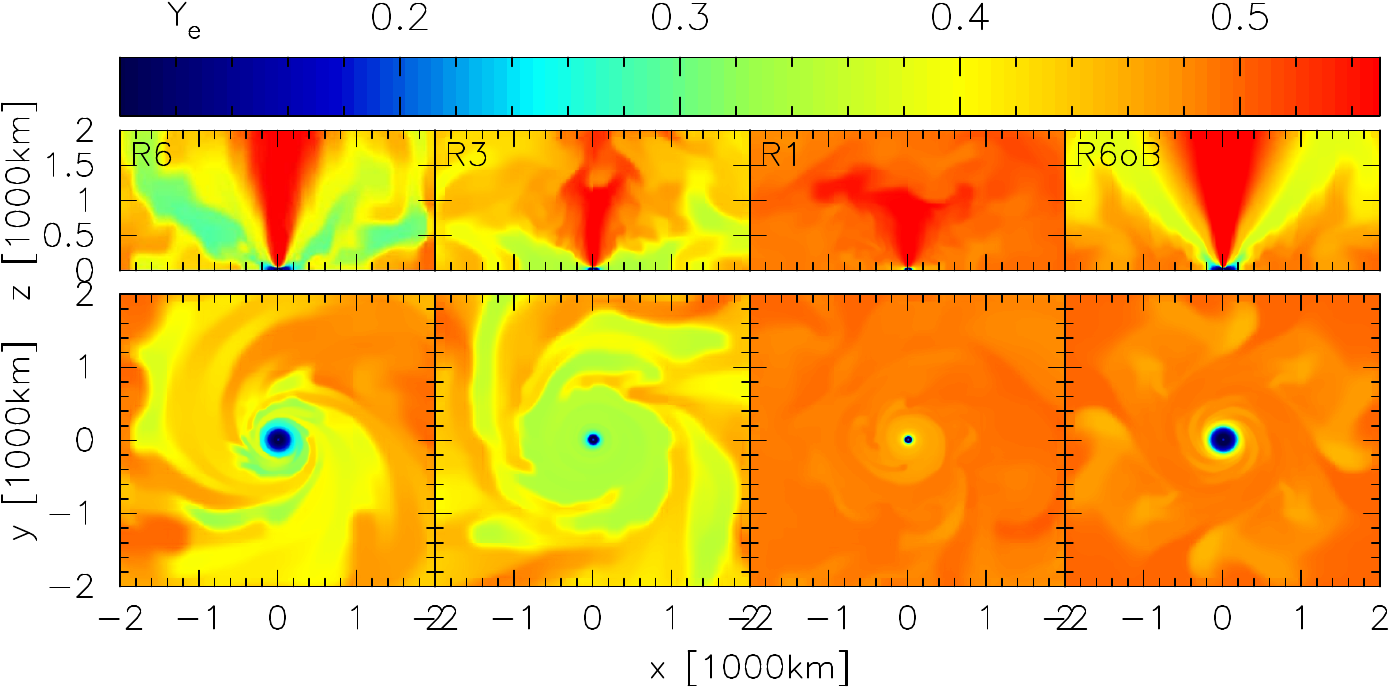}
\caption{Distributions of electron fraction $Y_e$ on the $xz$-plane (top panels) and $xy$-plane (bottom panels), for four selected models R6, R3 , R1, and R6oB at $t=t_9$.
\label{fig:Ye2D}}
\end{center}
\end{figure*}
\begin{figure}
\begin{center}
\includegraphics[width=1.0\columnwidth]{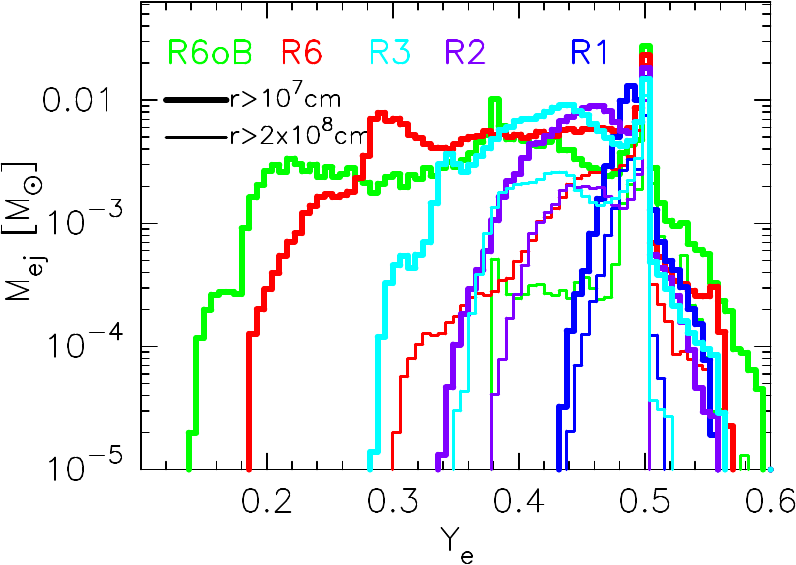}
\caption{Mass histograms as a function of the electron fraction $Y_e$ for all models. The histograms are evaluated when the maximum shock radius reaches $R_{\rm s}=10^9$\,cm. The thick and thin lines denote the ejecta locating at $r\ge10^7$ and $\ge2\times10^8$\,cm, respectively.
\label{fig:Ye_hist}}
\end{center}
\end{figure}

In Fig.~\ref{fig:Ye2D} we show the spatial profile of ejecta $Y_e$ on the $xz$- (top panels) and $xy$-plane (bottom panels) for four selected models R6, R3, R1, and R6oB at $t_{\rm pb}=t_9$.
As already explained, strong spiral waves emanate for model R6, which efficiently transport low-$Y_e$ component from the center to outer envelope. This can be seen in the bottom-left panel.
Such a spiral feature becomes less prominent with the decrease of rotation and is naturally absent in the octant symmetry model R6oB.
As for another remarkable impact of rotation, the central low electron fraction region itself, e.g. the dark blue region having $Y_e\lesssim0.2$, is obviously more extended for model R6 and R6oB than the slower rotating models due to the centrifugal force.
This property also makes the low-$Y_e$ matters, which initially rotate in the vicinity of PNS surface, e.g., at around the edge of central blue region, more easily ejected simply due to their weaker gravitational bound.
On the $xz$-plane, the $Y_e$ distribution exhibits a stronger angle dependence for more rapidly rotating models.
For instance within the outflows ($\theta\lesssim15^\circ$) for model R6, high-$Y_e(\sim0.55)$ matters are the main components, which are surrounded by rather lower electron fraction outflows with $Y_e\lesssim0.3$ along the diagonal direction.
Such an angular distribution is consistent with previous studies \citep{Dessart06_AIC,Batziou&Janka24_AIC,Cheong25_AIC}.
If the initial rotation becomes slower, such features become more moderate and the overall structure approaches more spherically symmetric as can be seen in R1.
From a comparison between R6 and R6oB, the profile on $xz$-plane is more or less similar. We find a clear bipolar structure, which is surrounded by low-$Y_e$ components (yellowish region).
On the whole equatorial plane in R6oB, we find that high-$Y_e$ materials are evidently covering the region, which is a consequence of the absence of spiral waves and is a noticeable difference from the structure seen in the model R6.

To quantitatively discuss the ejecta properties with respect to the WD rotation, we plot the $Y_e$ mass distribution of ejecta for all models in Fig.~\ref{fig:Ye_hist}.
The vertical axis denotes the integrated mass in each $Y_e$ bin.
Since our simulation time is quite limited ($t_{\rm pb}\le500$\,ms), we plot two distributions applying $r>10^7$ or $>2\times10^8$\,cm criterion when evaluating the histogram; the thick and thin lines denote the $Y_e$ histogram for unbound materials beyond $r>10^7$ and $2\times10^8$\,cm, respectively.
We assume that the ejecta which have been already driven away beyond $r>2\times10^8$\,cm, i.e., approximately corresponding to the initial WD radii (see Fig.~\ref{fig:InitialCondition}) could hardly fall back again onto the central PNS and would thus be safely ejected.

As a general feature, the WD rotation broadens the spectrum.
We indeed observe significantly larger ejecta masses at both the lower- and higher-$Y_e$ sides for models R6 and R6oB compared to those for slower rotating models.
In particular, the prolongation of lower side is more sensitive to the rotation.
\cite{Batziou&Janka24_AIC} have also presented that, for a given WD mass, the $Y_e$ distribution becomes broader for more rapidly rotating models.
From the red thin line, AICs of rapidly rotating WDs, like R6, would reasonably eject low-$Y_e$ components down to $0.3$.
We find even lower-$Y_e$ unbound materials with $0.2\le Y_e\le 0.3$ as indicated by the red thick line, i.e. spatially locating between $r=10^7$\,cm and $2\times10^8$\,cm.
However, we expect that their $Y_e$ values would increase afterward, as they are still subject to be exposed to intense neutrino radiation till they reach far enough from the PNS.
Consequently their $Y_e$ distribution concentrates at $Y_e\sim0.28$ as can be seen by a small peak in the thick red line.
The rest-mass density of such ejecta with $Y_e=0.28$ has already been decreased below $10^{8-9}$\,g\,cm$^{-3}$ as explained in Fig.~\ref{fig:ejecta2D}, which is sufficiently low to cease the neutrino-matter interactions and to subsequently preserve their $Y_e$ values.
We can therefore reasonably estimate the final ejecta mass with $Y_e\sim0.3$ to be $\sim0.01\,M_\odot$.
According to previous nucleosynthesis calculations \citep[in a different context, e.g.,][]{Fujibayashi23}, ejecta with $Y_e=0.3$ might produce elements up to the second r-process peak (mass numbers $A\sim130)$, which should be confirmed by our future long-term AIC simulations.
As demonstrated in \cite{Batziou&Janka24_AIC}, some low-$Y_e$ materials with $Y_e=0.3$ which exist in the vicinity of PNSs could be eventually ejected beyond WD surface at a quite later phase ($t_{\rm pb}\gtrsim2$\,s).
This may partially support our statement above.

On proton-rich side $Y_e>0.5$, we observe another broadening effect by rotation.
These proton-rich ejecta are concentrated along the rotation axis as indicated in the upper panels of Fig.~\ref{fig:Ye2D}.
This distribution is explained in the following manner: Near the rotation axis, a neutrino sphere is formed at smaller radii and thus a higher temperature region is formed along the rotation axis compared to that on equator.
This leads to more intense neutrino irradiation with higher emergent neutrino energy and efficiently heats matters, increasing their $Y_e$ values near the rotation axis as reported in previous rotating SN simulations \citep{Summa18,KurodaT20,Obergaulinger21,Takiwaki21}. 
However, we note that there is a caveat that an elongated dumbbell-shaped neutrino sphere due to rapid rotation may artificially increase the efficiency of neutrino absorption for the current M1 neutrino transport as discussed in \cite{Just15} (and references therein).

Before closing this section, we explain the difference seen in the ejecta properties between non-magnetized model R6 and magnetized one R6oB.
In Fig.~\ref{fig:ejecta2D} and also from thick red (R6) and green (R6oB) lines in Fig.~\ref{fig:Ye_hist}, we may apparently interpret that the magnetized model R6oB eject neutron-richer materials ($Y_e\le0.2$) than R6 does.
However, those low-$Y_e$ materials for model R6oB are still locating in the vicinity of the central PNS surface ($r\sim100$\,km, from Fig.~\ref{fig:ejecta2D}).
In the presence of magnetic fields, these low-$Y_e$ materials tend to be defined as unbound according to our current criterion, Eq.~\eqref{eq:Eexp}, because of the additional contribution from magnetic energy, in particular when they locate at a region where the local magnetic energy is quite large, as is the current case.
If we compare ejecta-$Y_e$ values measured at a sufficiently distant region (thin lines in Fig.~\ref{fig:Ye_hist}), we notice that model R6 (thin red line) presents slightly neutron-richer ejecta than R6oB (thin green line).
From these, we interpret that the non-axisymmetric instabilities are the primary factor among the suite of current numerical setups that eject low-$Y_e$ components to outer region.
At the same time we still cannot dismiss the contribution of strong magnetic fields, which certainly decreases the gravitational binding energy of a part of PNS envelope and may facilitate the mass ejection.
We should also add that if we artificially employ stronger magnetic fields, for instance $B_0=10^{12}$\,G as done in \cite{Cheong25_AIC}, the magnetically driven jets may generally eject lower-$Y_e$ components before the ejecta capture a large amount of neutrinos and increase their $Y_e$.
In the meantime, one should pay attention to the compatibility of such strong initial magnetic fields with the values obtained from WD observations and evolution models \citep{Wickramasinghe00,Schmidt03,Zhu15}.

\section{Multimessenger Signals}
\label{sec:Multimessenger Signals}
In this section, we discuss observable multimessenger signals from AICs: GWs, neutrinos, and electromagnetic counterparts.
\subsection{Gravitational wave emissions}
\label{sec:Gravitational wave emissions}
\begin{figure*}
\begin{center}
\includegraphics[angle=0.,width=\textwidth]{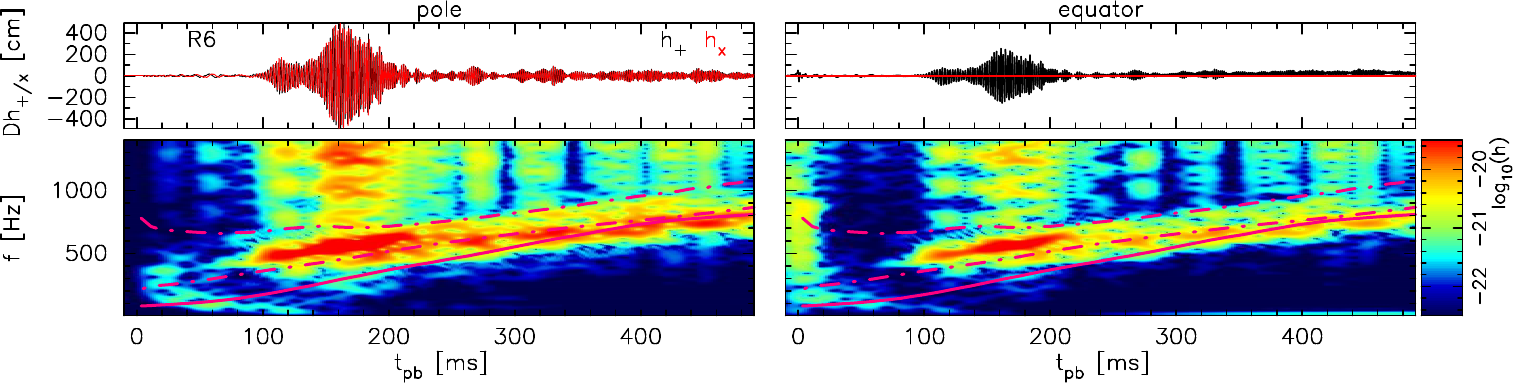}
\includegraphics[angle=0.,width=\textwidth]{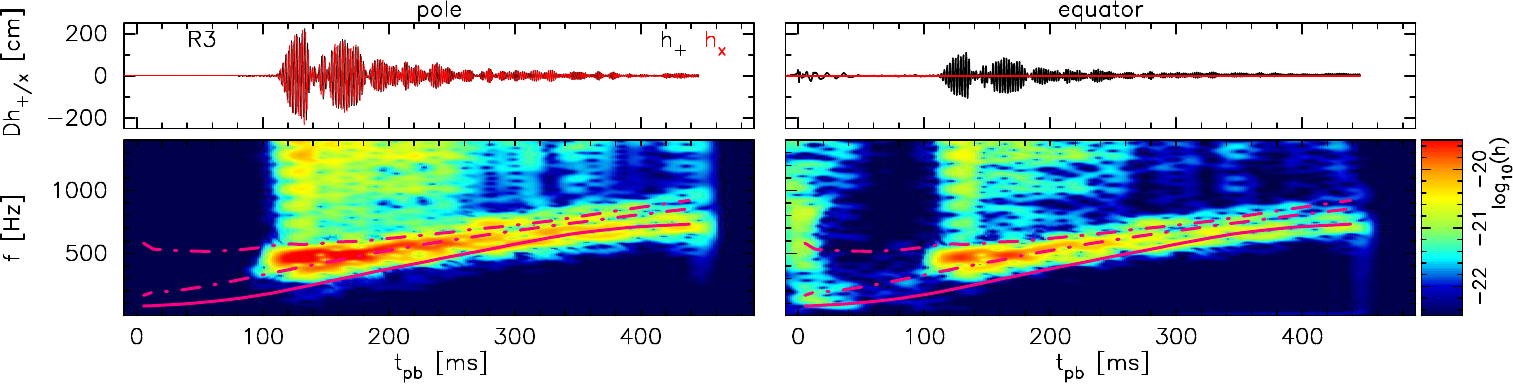}
\includegraphics[angle=0.,width=\textwidth]{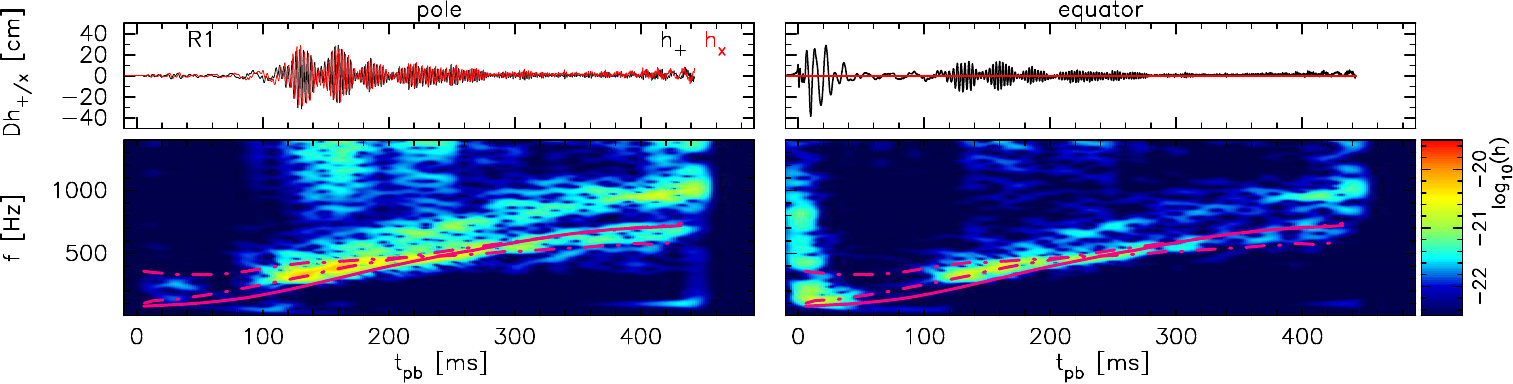}
\caption{GW strains multiplied by the source distance $D$ for plus ($h_+$: black solid lines) and cross ($h_\times$: red lines) modes and spectrograms of their characteristic strains for model R6 seen along the pole (left-hand panels) and the equator (right-hand panels)
assuming a source distance of 10 kpc.
Note that because of the assumed equator symmetry, $h_\times$ along the equator (red line in the top-right panel) completely vanishes.
In the spectral evolution, we overplot the analytical prediction for the peak frequency of $g$-mode oscillations $f_{\rm peak}$ \citep{BMuller13} by the solid line and two angular frequency evolution curves by the dash-dotted lines (see text for their definition).
We change the color of the overplotted lines just for the visualization.
\label{fig:GW}}
\end{center}
\end{figure*}
First we discuss the GW signal.
Fig.~\ref{fig:GW} exhibits $Dh_{+/\times}$ (upper panels) and color coded spectrograms of $|h|$ in logarithmic scale (lower panels), where $D$ and $h_{+/\times}$ are the source distance and GW strain with two polarized modes $+/\times$.
The left and right column are for an observer along the pole and equator (this time, $x$-axis), respectively.
From the top to bottom rows, we display the results for models R6, R3, and R1.
Since R2 shows a consistent trend with the rest of models, we omit its plot this time.
Regarding the magnetized model R6oB, we simply remove it from the discussion below, as its assumed octant symmetry makes its GW signals essentially the same with those of axisymmetric case and the discussion including all non- and axisymmetric models is not so straightforward.
The GW strain $h_{+/\times}$ are evaluated from a standard quadrupole formula \citep{Shibata&Sekiguchi03,KurodaT14}.
To investigate the spectral evolution of GWs, we evaluate the viewing-angle-dependent characteristic strain for an optimally oriented source \citep{Takiwaki21,Shibagaki21}.
In the top panels, two polarized modes are plotted in different colors.
As we assume equator ($z=0$) symmetry, $h_\times$ (red line) is completely suppressed along the equator.
In addition in the spectral evolution, we overplot the analytical prediction for the peak frequency of $g$-mode oscillations $f_{\rm peak}$ \citep{BMuller13} by the solid line and two evolution curves corresponding to the angular frequencies of a rotating object $\omega^n_{\rm rot}$, which is divided by $\pi$, i.e., dash-dotted curves plot $\omega^n_{\rm rot}/\pi(\equiv f^n_{\rm rot};$ explained below).

In the original studies of \cite{Murphy09,BMuller13}, $f_{\rm peak}$ is introduced to explain the Brunt-V\"ais\"ala frequency measured on the neutrino sphere assuming no PNS rotation, which is ultimately translated into the gravitational acceleration and the mean energy of emergent electron antineutrinos \citep{BMuller13}.
Therefore the required quantities to evaluate the gravitational acceleration ($M_{\rm PNS}/R_{\rm PNS}^2$, where $M_{\rm PNS}$ and $R_{\rm PNS}$ refer to the PNS mass and radius, respectively) or emergent neutrino energy are simply applied by spherically averaged values, which is feasible as long as the PNS rotation is significantly slow.
Meanwhile the current models are rotating models and have a significant polar angle dependence due to rotational flattening of the core.
For instance, the PNS radius $R_{\rm PNS}$, which is in a standard definition the isodensity surface at $\rho\sim10^{10}$\,g\,cm$^{-3}$, differs significantly between the polar and equator directions, e.g., more than $10$ times difference in the most rapidly rotating case R6.
Therefore the normally used spherical averaged values may not make sense this time.
We instead measure them along the pole and calculate $f_{\rm peak}$, which can nicely fit the observed spectral feature as exhibited later.
If we calculate $f_{\rm peak}$ using the values measured along equator, it becomes approximately a few orders of magnitude smaller.

The angular frequency $\omega^n_{\rm rot}$ broadly measures the mean angular frequency of a {\it spherical shell}, whose density is in the range from $10^n$ to $10^{n+1}$\,g\,cm$^{-3}$, and is evaluated by
\begin{eqnarray}
    \omega^n_{\rm rot}=J_z^n/I_z^n.
\end{eqnarray}
Here $J^n_z$ and $I^n_z$ denote the angular momentum and moment of inertia of a shell with respect to the $z$-axis and are defined by
\begin{eqnarray}
    J_z^n&=&\int_{10^n<\rho <10^{n+1}\,{\rm g\,cm}^{-3}} dV (x^2+y^2) \frac{S_\phi}{\sqrt{x^2+y^2}},\\
    I_z^n&=&\int_{10^n<\rho <10^{n+1}\,{\rm g\,cm}^{-3}} dV \rho^\ast (x^2+y^2).
\end{eqnarray}
In the above, $dV=dx^3$, $\rho^\ast(=\rho W \sqrt{\gamma})$ is a weighted density, and $S_\phi$ is the azimuthal component of conserved linear momentum of fluid $S_i$.
From a rotating object with an angular frequency $\omega_{\rm rot}$, its characteristic GW frequency $f_{\rm rot}$ becomes \citep[e.g.,][]{Maggiore08}
\begin{eqnarray}
    f_{\rm rot}=2\left(\frac{\omega_{\rm rot}}{2\pi}\right),
\end{eqnarray}
where the factor 2 accounts for two GW emissions during one rotation.
Therefore, the dash-dotted curves, which present $\omega^n_{\rm rot}/\pi=f^n_{\rm rot}$, correspond to the GW frequency from rotating objects, or shells this time, having certain densities.
We plot $f^{13}_{\rm rot}$ and $f^{12}_{\rm rot}$, where $f^{13}_{\rm rot}>f^{12}_{\rm rot}$ during our simulation times for all models.

From Fig.~\ref{fig:GW}, we observe quite loud GW signals for model R6 reaching $Dh\sim500$\,cm between $t_{\rm pb}\sim100$\,ms and $\sim200$\,ms, in particular for a polar observer.
These signals are also found for other models, but with lower amplitude, $Dh\sim 200$\,cm and $30$\,cm for models R3 and R1, respectively.
From Fig.~\ref{fig:Overall}, we can identify that the period, when these signals are emitted, completely overlaps the emergence of strong spiral motions.
Previous 3D rotating NS models \citep{Shibata05c,Ott07,Scheidegger10,KurodaT14,Takiwaki18,Shibagaki21} and also recent 3D AIC models \citep{Longo-Micchi23} have reported such quasi-periodic GW signals with comparable amplitudes.
These signals can be attributed to the emission from a rotating ellipsoid.
Indeed the wave amplitudes become nearly half (e.g. $Dh\sim250$\,cm in R6) for an equator observer, which is consistent with the theoretically expected angle dependence $h_+\propto (1+\cos^2\theta)/2$ \citep{Maggiore08}. 
As already discussed in Fig.~\ref{fig:Omega}, the spiral waves coming from the low-$|T_\mathrm{rot}/W|$ instability redistribute the angular momentum from inner to outer region, and consequently weaken the degree of differential rotation.
Since the low-$|T_\mathrm{rot}/W|$ instability is sustained by strong differential rotation and possibly its associated shear instabilities \citep{Watts05}, the spiral motion is a transient phenomenon and disappears when the degree of the non-axisymmetric deformation becomes small.
Therefore, the associated GWs also subside within a time scale of $\sim10$--100\,ms (see also Fig.~\ref{fig:Omega}), which are also consistent with previous studies \citep{Shibata05c,Saijo06,Ott07,Scheidegger10,KurodaT14,Takiwaki18,Shibagaki21}, while there are inevitable differences originated from the different microphysics input and PNS property.

After the cease of the strong GW emissions associated with the non-axisymmetric deformation ($t_{\rm pb}\gtrsim200$\,ms), the GW amplitudes decrease significantly, by approximately a factor of 5, for all models.
From the spectral evolution, however, ramp-up features are still continuously observed regardless of the models and the observer angle.
For instance for model R6, the feature seems to begin from $f\sim500$\,Hz at $t_{\rm pb}=100$\,ms, which is most likely associated with the aforementioned spiral motion, to $\sim800$\,Hz at $t_{\rm pb}=500$\,ms.
In general, such ramp-up features may be explained by the $g/f$-mode oscillations of the {\it non-rotating} PNS \citep{Morozova18,Torres-Forne19,Sotani20b}, which are indeed observed in many of previous SN simulations with non-rotating progenitor stars \citep{Murphy09,BMuller13,KurodaT16ApJL,Vartanyan20,Mezzacappa20,Shibagaki21}.
However, all models in this study are rotating models with strong spiral waves appearing at $100\,\mathrm{ms}\lesssim t_{\rm pb} \lesssim200$\,ms.
Moreover the peak frequency range during the emergence of strong spiral waves, for instance $f\sim500(300)$--700(500)\,Hz for model R6 (R1), is significantly higher than the typical values ($f\sim100$--200\,Hz) of $g/f$-mode oscillations in the corresponding phase of non-rotating models \citep{Murphy09,BMuller13,KurodaT16ApJL,Vartanyan20,Mezzacappa20,Shibagaki21}.

To understand the origin of ramp-up features, which are quantitatively different from typical values from non-rotating models, we compare the spectral feature and three different frequencies $f_{\rm peak}$ (solid line), $f^{12}_{\rm rot}$ (dash-dotted), and $f^{13}_{\rm rot}$ (dash-dotted).
During the strong GW emission phase ($100\,\mathrm{ms}\lesssim t_{\rm pb} \lesssim200$\,ms), it is obvious that the typical $g/f$-mode frequencies $f_{\rm peak}$ (red solid line) do not match the frequency of background strong features.
The disagreement becomes more pronounced with the increase of the rotation.
We notice that the two frequency curves $f^{12}_{\rm rot}$ and $f^{13}_{\rm rot}$ associated with the fluid rotation appear to be encompassing those strong signals.
This clearly indicates that the initial strong signals are originated from rotating objects, specifically whose density is around $\sim10^{13}$\,g\,cm$^{-3}$.
We note again that one dash-dotted curve $f^n_{\rm rot}$ represents an averaged rotation frequency of fluids, whose density is in the range of $10^n$ and $10^{n+1}$\,g\,cm$^{-3}$.
Our interpretation can be strongly supported by the evolution of angular velocity $\Omega(\varpi)$ in Fig.~\ref{fig:Omega}.
As explained, the angular momentum redistribution and its consequent spin-up are most visible at a region, where the rest-mass density is about $10^{13}$\,g\,cm$^{-3}$.
This indicates that the corresponding region is suddenly accelerated and that the non-axisymmetric flow patterns are rapidly developed, leading to the strong GW emissions.
From these facts, we conclude that the overall ramp-up feature is consisted of two components: the initial rotational origin and the later $g/f$-mode origin mode.

We also find GW signals which are typically seen in rotating SNe.
The first one is the signal at core bounce, namely the Type I signal \citep{Dimmelmeier02B,Dimmelmeier08}.
An observer along the equator would observe burst GW signals, whose amplitude reach $|Dh_+|\sim55$\,cm and 11\,cm for models R6 and R1, respectively, which can be translated into $|h_+|\sim2\times10^{-21}$ and $4 \times 10^{-22}$ assuming a source distance of $D=10$\,kpc.
As models R6 and R1 have a rotational parameter at bounce of $\beta_{\rm cb}\approx 3.56$ and $0.56$\,\% (see, Tab.~\ref{tab:summary}), our result broadly obeys the universal relation between $\beta_{\rm cb}$ and $|h_+|$ at bounce reported by \cite{Dimmelmeier02B,Dimmelmeier08,Abdikamalov10_AIC}.
Regarding another feature, we witness quasi-monotonically growing GWs, only in $h_+$ along the equator, in the explosion phase $t_{\rm pb}\gtrsim200$\,ms (see also Fig.~\ref{fig:RshockEexp}). 
These signals are emitted from anisotropic matter outflows as reported in previous rotating massive stellar collapse simulations exhibiting bipolar explosions \citep{Obergaulinger06,Shibata06,Murphy09,Vartanyan20,Shibagaki24}.
Among the set of our current models, the most rapidly rotating model R6 presents the largest growth rate, as its bipolar structure is most visible.
At $t_{\rm pb}\sim450$\,ms, $Dh_+$ increases to $\sim30$\,cm for model R6, while it does only to $\sim2$\,cm for model R1.
Particularly for model R6, these {\it memory} signals imprint their characteristic frequency into the spectral evolution, appearing at low frequencies $f\lesssim10$\,Hz, which is marginally visible in the spectral evolution along the equator at $t_{\rm pb}\gtrsim300$\,ms and at $f\sim10$\,Hz, i.e. the lower edge of the figure, for model R6.

\begin{figure}
\begin{center}
\includegraphics[width=\columnwidth]{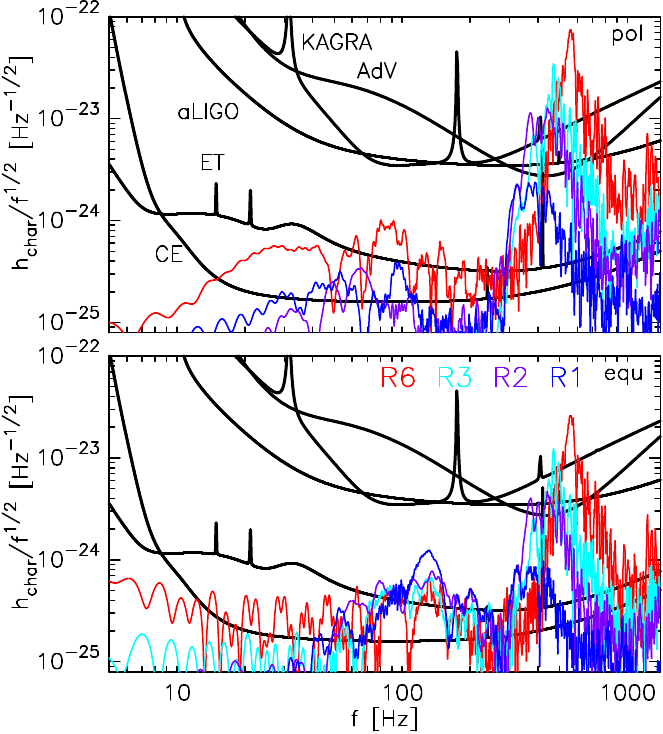}
\caption{Characteristic strain of matter origin GWs (color coded lines) overplotted by the sensitivity curves of the current- and third-generation GW detectors (black): advanced LIGO, advanced VIRGO, KAGRA, Einstein Telescope, and Cosmic Explorer. We assume a source distance of $D=1$\,Mpc.
The upper and lower panels are for an observer along the pole and equator, respectively.
\label{fig:GW_h_ch}}
\end{center}
\end{figure}
Before closing this subsection we discuss the detectability of GWs.
Fig.~\ref{fig:GW_h_ch} plots the detectability of matter origin GWs for models R1, R2, R3, and R6 in different colors.
$h_{\rm char}$ is the GW spectral amplitude assuming a source distance of $D=1$\,Mpc.
The upper and lower panels are for an observer along the pole and equator, respectively.
We overplot the sensitivity curves of the current- and third-generation GW detectors: advanced LIGO (aLIGO), advanced VIRGO (AdV), KAGRA \citep{abbott18det}; Einstein Telescope (ET) \citep{ET_Hild}; and Cosmic Explorer (CE) \citep{CE_Abbott}.
Model R6 exhibits a strong spectral peak at $f=561$\,Hz, whereas slower rotating models have a weaker peak signal appearing at lower frequencies as $f=472$ (R3), 443 (R2), and 392\,Hz (R1).
By comparing the spectral evolution in Fig.~\ref{fig:GW}, these signals peaking at frequencies, $400\lesssim f \lesssim560$\,Hz, are mainly emitted during the emergence of strong spiral waves at $100\le t_{\rm pb}\le 200$\,ms and are thus associated with the non-axisymmetric fluid rotation. Therefore the peak frequency increases with the rotation. 
As has already explained, these peaks are originated from the spiral waves and therefore the peak is approximately halved for an equatorial observer (lower panel).
If we observe rapidly rotating AIC events (e.g., models R6 and R3,) along the pole and $D$ is smaller than $\sim 1$\,Mpc, the peak signal can be well detected even by the current generation detectors.
Using the third generation detectors (ET and CE), we will be able to detect the peak signals from rapidly rotating AIC events even for $D \sim 10$\,Mpc; we may assume a more realistic event rate.
If we observe from equator, rapidly rotating models (R3 and R6) are likely still observable by the thrid generation detectors, while it becomes quite hard for slowly rotating models.

GW signals from the anisotropic matter ejection for model R6 appear at a low frequency range $f\lesssim10$\,Hz for an equatorial observer (red line in the lower panel), which is absent toward the pole.
This low-frequency component, whose energy is integrated over $300\,\mathrm{ms}\le t_{\rm pb}\le500$\,ms according to the top-right panel in Fig.~\ref{fig:GW}, is apparently too weak for the detection.
However, if the current anisotropic matter configuration persists longer time, for instance another 1--2\,s, the GW emission energy of this low-frequency component can be increased by $\sim5$--10 times.
Then we can naively expect that those signals can be detected by the third generation detectors.
In the current study, we do not evaluate GWs originated from non-isotropic neutrino radiation.
As reported in \cite{EMuller12,Vartanyan20,Shibagaki24}, those contributions generally appear at the same low-frequency range ($f\lesssim10$\,Hz). The amplitude can sometimes become by one order of magnitude larger, in particular for rotating models \citep{Shibagaki24}. This makes their detection even more feasible by the third-generation detectors.

\begin{figure*}
\begin{center}
\includegraphics[width=\textwidth]{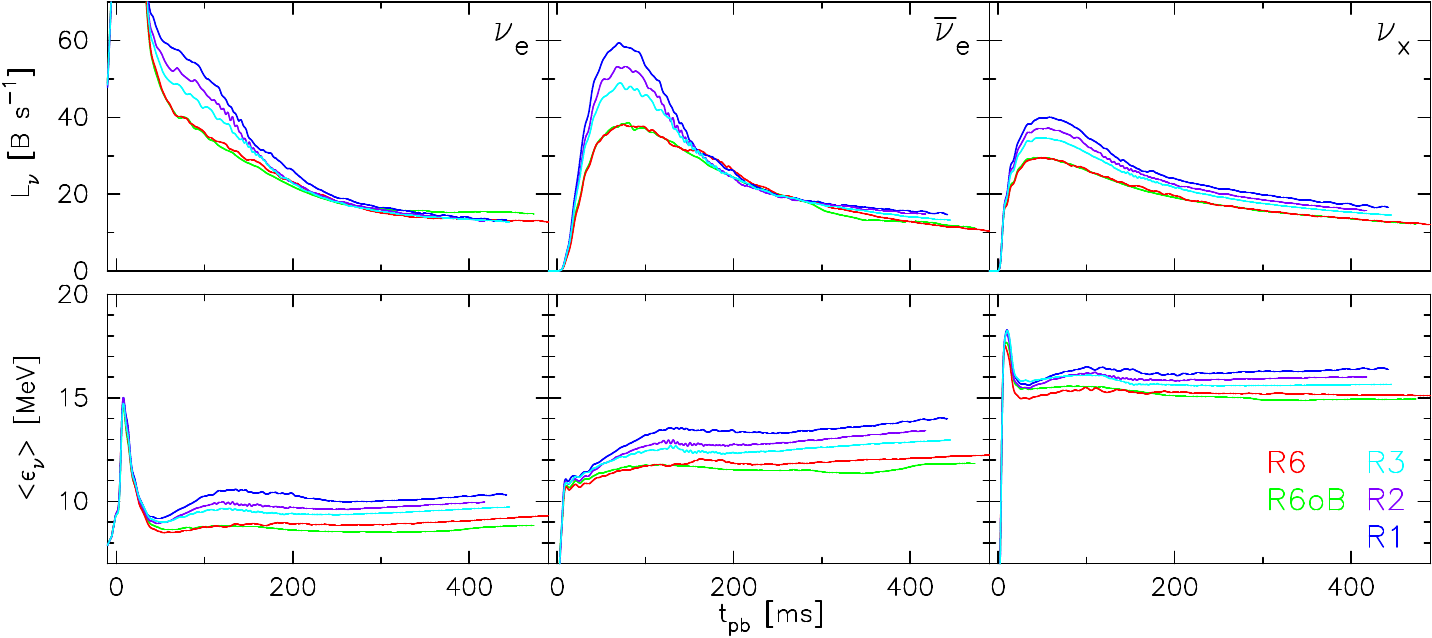}
\caption{Surface integrated neutrino luminosities (upper panels) and mean neutrino energies for all models except R6o: 
Electron neutrinos $\nu_e$ (left), electron antineutrinos $\bar\nu_e$ (middle), and heavy-lepton neutrinos $\nu_x$ (right).
They are measured on a spherical surface at $r=400$\,km.
\label{fig:LnuEnu}}
\end{center}
\end{figure*}
\subsection{Neutrino signals}
\label{sec:Neutrino signals}
In this section, we first discuss the emergent neutrino profiles and then their observability.
Fig.~\ref{fig:LnuEnu} depicts the surface integrated neutrino luminosity $L_\nu$ (upper panels) and mean neutrino energy $\langle \varepsilon_\nu \rangle$ for all models except R6o, whose simulation time is too short for the discussion.
We show profiles for electron neutrinos $\nu_e$ (left), electron antineutrinos $\bar\nu_e$ (middle), and heavy-lepton neutrinos $\nu_x$ (right), measured on a spherical surface at $r=400$\,km.

In the neutrino luminosities, we find a consistent behavior with previous electron-capture SN simulations, namely the absence of high accretion luminosities, which is commonly observed also in the collapse of $less$ massive stars.
For instance, $L_{\nu_e}$ for all models present a smooth decline for $50\,\mathrm{ms} \lesssim t_{\rm pb}\lesssim200$\,ms.
Such a feature has been reported in previous studies, for instance in electron-capture SNe \citep{Hüdepohl10,Radice17} or collapse of low mass stars, like $9.6$\,$M_\odot$ star \citep{BMuller14,KurodaT22}.
It stems from the fact that the progenitors of AICs have essentially no envelope and thus no mass accretion.
While in the collapse of more massive stars with the mass, e.g., $\gtrsim10$\,$M_\odot$, the PNS accretes more mass during this phase, which results in a more prominent, temporal increase of neutrino luminosities at $t_{\rm pb}\sim100$\,ms, in particular for electron neutrinos $L_{\nu_e}$ \citep[e.g.,][]{BMuller12b,Bollig17,Summa18,Vartanyan19,KurodaT22}.
Neutrino mean energies $\langle \varepsilon_\nu \rangle$ for all flavors present a rather flat evolution after bounce ($t_{\rm pb}\gtrsim100$\,ms).
This is a consequence of a nearly saturated PNS mass evolution (see Fig.~\ref{fig:Overall}) and associated no significant PNS contraction (besides the contraction associated with long-term neutrino cooling).
Without the significant PNS contraction, the neutrino spheres for all flavors do not appreciably propagate inward and thus the energy spectra of last scattered neutrinos do not significantly evolve, resulting in a nearly constant and relatively lower mean neutrino energies in comparison to normal massive stellar collapse.

As for the rotational effects, we find a clear inverse trend in the ordering of both $L_{\nu}$ and $\langle \varepsilon_\nu \rangle$ with respect to the WD rotation.
Neutrino luminosities differ the most during the strong spiral wave phase ($50\,\mathrm{ms}\lesssim t_{\rm pb}\lesssim200$\,ms), and afterward we observe a modest rotational dependence.
The mean neutrino energies present $\sim2$\,MeV difference between the slowest (R1) and fastest (R6) rotating models for all flavors.
As a comparison with previous AIC models, our slowest rotating model R1 behaves similarly to an AIC of non-rotating 1.42\,$M_\odot$ WD reported in \cite{Batziou&Janka24_AIC}, albeit during our limited post bounce phase ($t_{\rm pb}\lesssim500$\,ms).
\cite{Radice17}\footnote{In \cite{Radice17}, they present root-mean-squared neutrino energies, which generally leads to $\sim10$\,\% higher energy evaluation than the mean value.} conducted electron-capture SNe using 8.1--8.8\,$M_\odot$ progenitor stars and iron core collapse of low mass ($\sim10$\,$M_\odot$) progenitor stars.
The present neutrino energy evolutions are similar to those of the former electron-capture SNe \citep[see also][]{Hüdepohl10} rather than of the latter iron core collapse of slightly more massive progenitor stars.

\begin{figure*}
\begin{center}
\includegraphics[width=\textwidth]{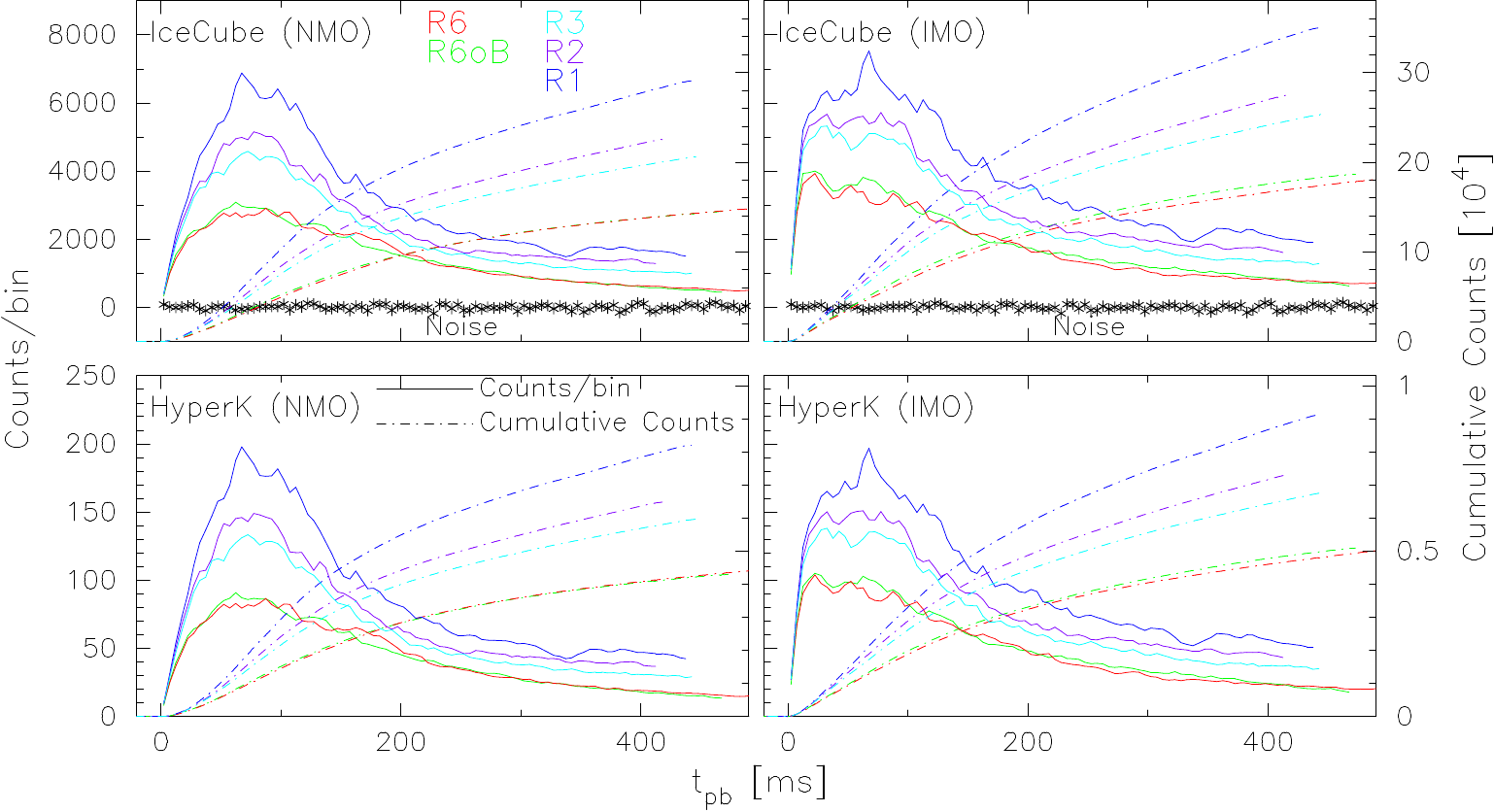}
\caption{Neutrino counts in 5\,ms\,bin (solid line) and cumulative counts (dash-dotted) for IceCube (upper panels) and for Hyper-Kamiokande (lower panels), assuming a source distance of $D=10$\,kpc.
We plot a normal mass ordering (NMO: left column) and inverse mass ordering (IMO: right panels) cases.
In the panel for IceCube, we also add a random shot noise realization (asterisk).
\label{fig:CountRate}}
\end{center}
\end{figure*}
Fig.~\ref{fig:CountRate} plots the expected neutrino counts in 5\,ms\,bins (solid line) and cumulative counts (dash-dotted) for IceCube \citep{abbasi11,salathe12} (upper panels) and Hyper-Kamiokande (HyperK) \citep{abe11,HK18} (lower panels).
The neutrino count is evaluated following \cite{lund10} for IceCube and \cite{Takiwaki18} for HyperK from surface averaged neutrino luminosities and mean energies of emergent neutrinos, thus simply neglecting the observer angle dependence.
We assume a source distance of $D=10$\,kpc.
In addition, we consider two cases: a normal mass ordering (NMO: left column) and inverse mass ordering (IMO: right column) \citep[see, e.g.,][for neutrino mixing parameters used]{Tanabashi18}, without taking into account yet-to-be-clarified collective flavor conversions.
In the upper panels for IceCube, we also add a random shot noise realization (asterisk) with its standard deviation being $\sim\sqrt{R_{\rm bkgd}\times5\,{\rm [ms]}}\sim86$, where $R_{\rm bkgd}=1.48\times10^3$\,ms$^{-1}$ is the assumed background rate of IceCube \citep[e.g.,][]{lund10}.

From the figure, we notice that the faster rotating models exhibit relatively lower event rates.
This trend is simply due to lower neutrino luminosities and mean energies in faster rotating models.
The event rate difference between the slowest and fastest rotating models can be as large as about two to three times for both detectors.
Our simple estimation for the cumulative number of events infers that if such AICs happen at the Galactic center ($D\sim10$\,kpc), we would detect $\sim10^5$ and $\sim10^4$ neutrinos for IceCube and HyperK, respectively. These numbers can be more in longer simulation times.
A less prominent difference between NMO and IMO cases could be another noteworthy feature of AICs.
In typical CCSNe, electron antineutrino luminosity $L_{\bar\nu_e}$ is significantly higher than heavy-lepton neutrino luminosity $L_x$ during the accretion phase, which is more pronounced with the increase of the progenitor mass~\citep[e.g.,][]{BMuller14,Vartanyan19,KurodaT22} and could sometimes result in a factor of two difference.
On the contrary, $L_{\bar\nu_e}$ reported here shows at the largest $\sim50$\,\% difference (R1) and moreover the difference soon nearly diminishes, e.g., at $t_{\rm pb}\gtrsim200$\,ms.
Consequently, the flavor change has a less impact on the final state as Fig.~\ref{fig:CountRate} indicates.

We shortly mention detectability of these AIC events by neutrinos.
From Fig.~\ref{fig:LnuEnu}, we can naively expect significantly high signal to noise ratio ($S/N$) for all models for both detectors, if they happen at a source distance of $D=10$\,kpc.
Neutrino counts for IceCube ($\gtrsim1000$\,in 5\,ms\,bins) well exceed the background noise ($\lesssim90$\,in 5\,ms\,bins) basically throughout all the simulation time and the same is true for HyperK, which has a background-free signal.
From this, we can broadly say that the Galactic events are reasonably observed by neutrinos.
However, for more distant events, there will be a point beyond that only HyperK would feasibly detect neutrino signals.
For instance, for $D=20$\,kpc, neutrino counts for model R6 at the final simulation time $t_{\rm pb}\sim500$\,ms can be estimated as $\sim130$ (IceCube) and $\sim4$ (HyperK) in 5\,ms\,bins.
Thence, from a very crude estimation, the signal to noise ratio becomes $\sim130/86\sim1.5$ for IceCube, while $\sim4/\sqrt{4}\sim2$ for HyperK, where in the latter HyperK case we assume a Poisson error.
Additionally, unlike GWs from AICs (see, Fig.~\ref{fig:GW_h_ch}), the expected horizon distance of AICs via neutrino observations is likely much shorter than $\sim1$\,Mpc.
For instance, the cumulative events for HyperK assuming a source distance of $D=10$\,kpc is $\sim10^4$ at the final simulation time, which becomes $\sim1$ if AIC occurs at $D=1$\,Mpc.

\subsection{Implications for the electromagnetic counterparts}
\label{sec:Implications for the electromagnetic counterparts}

As described in the introduction, the ejecta formed during the AIC as well as the energy injection from the newborn PNS can be the source of various electromagnetic transients, such as FBOTs. Here, we briefly estimate the property of possible electromagnetic signals in various situations based on the results obtained in this work (see also~\citealt{Drout14,Prentice18,Rest18_FELT,Margutti19,Lyutikov19}).

Our simulation results suggest that the ejecta mass, explosion energy, and ejecta velocity are of the order of $M_{\rm ej}\sim0.01$--$0.1\,M_\odot$, $E_{\rm exp}\sim10^{50}\,{\rm erg}$, and $v\sim0.1\,c$, respectively. The lower limit of the opacity of the ejecta can be given by that of the Thomson scattering $\kappa\sim 0.1\,{\rm cm^2}$\,g$^{-1}$. The opacity can be increased up to $1\,{\rm cm^2}$\,g$^{-1}$ for the case that the r-process elements up to the second peak or the first peak are synthesized for $1\%$ ($Y_e\lesssim 0.25$) or $10\%$ ($Y_e\lesssim 0.4$) of the total ejecta mass, respectively~\citep{Wanajo:2014,Tanaka:2019iqp}. Employing these values, the diffusion time scale of photons in the ejecta can be estimated as
\begin{align}
t_{\rm diff}&\sim\sqrt{\frac{3\kappa M_{\rm ej}}{4\pi v c}} \nonumber \\
&\approx3\,{\rm day}\left(\frac{\kappa}{0.2\,{\rm cm^2}\,{\rm g}^{-1}}\right)^{1/2}\left(\frac{M_{\rm ej}}{0.05\,M_\odot}\right)^{1/2}\left(\frac{v}{0.1\,c}\right)^{-1/2}.\label{eq:t_diff}
\end{align}
Here, the fiducial value for the opacity is set by the case that the opacity is dominated by the Thomson scattering of the matter with $Y_e\approx 0.5$.

For the MIC scenario, the collapse can occur deep inside the inflated envelope of the merger remnant. ~\cite{Schwab21} suggests that the envelope extension can vary from $R_*=10^{10}\,{\rm cm}$ to $2\times 10^{12}\,{\rm cm}$ depending on the total mass of the progenitor system. Under the assumption that the envelop mass smaller than the ejecta mass, the interaction of the ejecta with the envelope matter results in the cooling emission with the peak luminosity of
\begin{align}
L_{\rm peak}&\sim\frac{E_{\rm exp}t_{\rm e}}{t_{\rm diff}^2}\nonumber \\
&=3\times 10^{42} {\rm erg}\,{\rm s}^{-1}\left(\frac{R_*}{10^{12}\,{\rm cm}}\right)
\left(\frac{\kappa}{0.2\,{\rm cm^2}\,{\rm g}^{-1}}\right)^{-1}\left(\frac{v}{0.1\,c}\right)^{2}\label{eq:Lp_env}
\end{align}
at the peak time of $t_{\rm diff}$ with the effective temperature of
\begin{align}
T_{\rm peak}&\sim\left(\frac{L_{\rm peak}}{4\pi \sigma_{\rm SB} v^2 t_{\rm diff}^2}\right)^{1/4}\approx 1\times 10^4\,{\rm K}\left(\frac{R_*}{10^{12}\,{\rm cm}}\right)^{1/4}
\nonumber 
\\&~~~~~~~\times\left(\frac{\kappa}{0.2\, {\rm cm^2}\,{\rm g}^{-1}}\right)^{-1/2}\left(\frac{M_{\rm ej}}{0.05\,M_\odot}\right)^{-1/4}\left(\frac{v}{0.1\,c}\right)^{1/4},\label{eq:tempp_env}
\end{align}
where $t_{\rm e}=R_*/v$ and $\sigma_{\rm SB}$ is the Stefan-Boltzmann constant. The bolometric luminosity will rapidly decline after the peak with the time scale of $t_{\rm diff}$, unless there is some energy source that powers the ejecta.

We note that an AIC with an inflated envelope of $R_*\gtrsim 10^{12}\,{\rm cm}$ may occur only for a massive remnant (total mass $\approx 1.9\,M_\odot$)~\citep{Schwab21}. For lower-mass WD cases, the progenitor is likely to be significantly contracted by the time of the collapse~\citep{Schwab16}. In such cases, the interaction between the ejecta and contracted envelope may not produce a bright optical transient, due to the small value of $R_*$~\citep[e.g.,][see also the reference for the absorption and scattering effects on the emission in the presence of thick wind]{Yu19}.

\begin{figure}
    	 \includegraphics[width=\columnwidth]{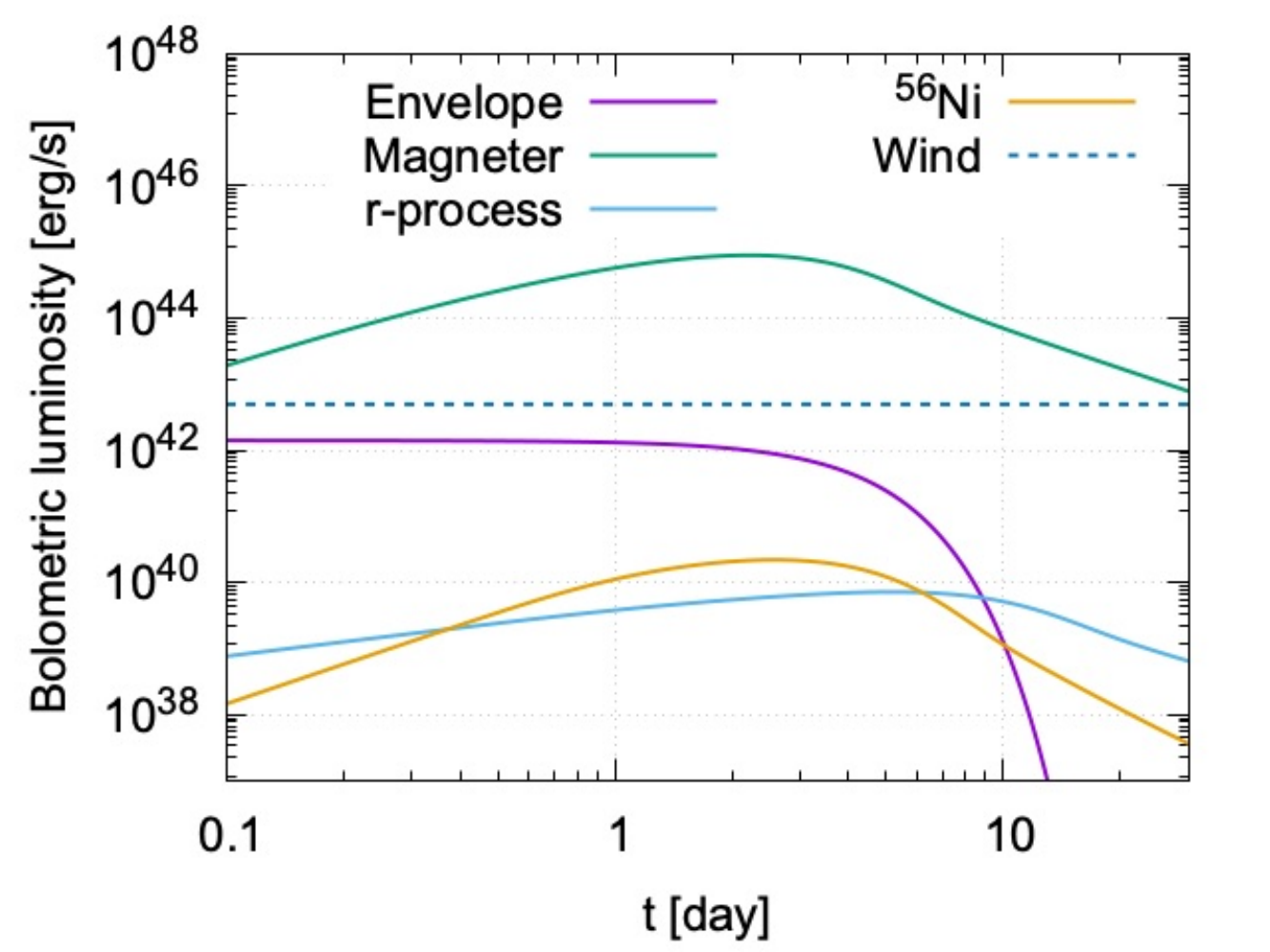}
 	 \caption{Light curves associated with an AIC for various scenarios obtained by numerically solving one-zone models.}
	 \label{fig:LC}
\end{figure}

If a significant amount of the r-process elements or $^{56}{\rm Ni}$ are synthesized in the ejecta, the envelope-ejecta interaction powered emission can be followed by faint long-lasting emission.
Although detailed ejecta mass of these heavy elements should be determined by nucleosynthesis calculation, our approximate estimation gives $M_{\rm Ni}=2.6(2.0)\times10^{-3}$\,$M_\odot$ for R6(R1), where we presume half of highly heated ejecta beyond $5$\,GK with $Y_e\ge0.5$ may become $^{56}$Ni~\citep{Fujibayashi24}.
Regarding the r-process elements, we assume the optimistic case that the r-process elements are synthesized for $10\%$ of the total ejecta mass, and thus $M_r=0.1\,M_{\rm ej}$.
With these assumptions, the emission with the peak luminosity of 
\begin{align}
    L_{\rm peak}&\approx M_{\rm r} {\dot q}_{\rm r,\beta}(t_{\rm diff})\nonumber\\
&\approx 1\times 10^{40}\,{\rm erg}\,{\rm s}^{-1}  \left(\frac{M_{\rm r}}{5\times 10^{-3}\,M_\odot}\right) \left(\frac{t_{\rm diff}}{6 \,{\rm day}}\right)^{-1.3}\label{eq:Lp_r}
\end{align}
and
\begin{align}
L_{\rm peak}&\approx M_{\rm Ni56} f_\gamma(t_{\rm diff}){\dot q}_{\rm Ni56,\gamma}(t_{\rm diff})\nonumber\\
&\approx 3\times 10^{40}\,{\rm erg}\,{\rm s}^{-1}\left(\frac{M_{\rm Ni56}}{3\times 10^{-3}\, M_\odot}\right)\left(\frac{f_{\rm \gamma}(t_{\rm diff})}{0.1} \right)~\left(t_{\rm diff}\ll 9\,{\rm day}\right)\label{eq:Lp_ni}
\end{align}
can occur followed by approximately $\propto t^{-1.3}$ and $\propto t^{-3}$ decline rate, respectively (see Fig.~\ref{fig:LC}). Here, ${\dot q}_{\rm r,\beta}\approx  10^{10}\,{\rm erg}\,{\rm g^{-1}}\,{\rm s}^{-1}\left(\frac{t}{ {\rm day}}\right)^{-1.3}$ and ${\dot q}_{\rm Ni56,\gamma}\approx 3\times 10^{10}\,{\rm erg}\,{\rm g}^{-1}\,{\rm s}^{-1}~(t\ll 9\,{\rm day})$  denote the specific radioactive decay heating rates of the r-process elements ($\beta$ particles)~\citep{Korobkin:2012uy} and $^{56}{\rm Ni}$, respectively. $t_{\rm diff}$ for the r-process powered emission model is modified so as to take the increase in the opacity into account ($\kappa=1\,{\rm cm^2 g^{-1}}$).  
Note that, while the energy deposition with $\beta$ particles of the r-process elements is well thermalized up to 70 days~\citep{Barnes:2016umi}, thermalization of deposited $\gamma$-rays is less efficient after $t_{\rm ineff}\sim 1\,{\rm day}$ for the $\gamma$-ray effective opacity of $0.027\,{\rm cm^2}\,{\rm g}^{-1}$~\citep{Maeda:2005pi} for the typical ejecta parameters. Hence, the thermalization efficiency of $f_{\rm \gamma}=1-e^{-t_{\rm ineff}^2/t^2}$~\citep{Barnes:2016umi} is taken into account for the $^{56}{\rm Ni}$ powered emission. These results here with Eq.~\eqref{eq:Lp_env} suggest that if the radius of the extended envelope is smaller than $\sim10^{10}\,{\rm cm}$, the peak luminosity will be dominated by radioactive decay powered emission with this hypothetical amount of r-process elements or $^{56}$Ni. The photospheric temperature at the peak can be estimated from Eqs.~\eqref{eq:Lp_r} and \eqref{eq:Lp_ni} to be $T_{\rm peak}\approx 2$--$3\times 10^3\,{\rm K}$, which suggests that the emission is bright in the near-infrared wavelengths: The emission from radioactively powered AIC ejecta may appear similar to that of a faint, red kilonova. Distinguishing this emission from a NS merger kilonova will likely require the detailed spectral modeling.

Direct measurements of the radioactive-decay gamma-rays escaping from ejecta may provide the proof for the presence and identification of the nucleosynthesis of radioactive elements~\citep{Patel25}. However, for the plausible amount of synthesized r-process elements and $^{56}$Ni, the gamma-ray luminosity will be too faint to be detected by the current missions ~\citep{Hotokezaka16} unless the event occurs at a very close distance ($\lesssim 1\,{\rm Mpc}$). Nevertheless, the observation by new detectors with the sensitivity improved by an order of magnitude will be a powerful tool for diagnosing nucleosynthesis in AIC ejecta.

The magnetar activity of a newborn NS can also power the ejecta and may be the source of rapidly evolving luminous transients \citep{Yu15}.
The magnetar could be well formed in the aftermath of AIC/MICs, if the collapsing initial WDs are sufficiently magnetized and rapidly rotating.
It is indeed still a matter of debate how such strongly magnetized WDs ($\gtrsim10^{10}$\,G) are feasibly formed in the stellar evolution scenario \citep[see,][for a potential magnetic field amplification process via merger of WDs]{Zhu15}.
Unfortunately, because their focus was on the formation of sub-Chandrasekhar mass WDs, we cannot directly apply their results to AIC/MICs and associated magnetar formation of our current interest.
However, presuming strong magnetic fields ($\gtrsim10^{10}$\,G) inside super-Chandrasekhar mass WDs, the magnetar might be reasonably formed.
This is because those initial fields are first amplified via compression during the collapse by 2--3 orders of magnitude and then additionally by rotational winding and possibly some other non-linear processes, e.g., MRI \citep{Balbus91}.
For instance by rotational winding alone, the compressed magnetic fields are further amplified by 1--2 orders of magnitude during a time scale of $\sim$10--100\,ms, if the nascent NS's rotation period is 1\,kHz.
As a result, the initial magnetic fields could be well amplified in total by 4--5 orders of magnitude, i.e., the magnetar class $\sim10^{14-15}$\,G, within a dynamically relevant time scale (e.g., within $\sim100$\,ms after bounce).
Our rotating magnetized WD model R6oB, which assumes a rather stronger initial poloidal (dipole-like configuration) magnetic field strength of $10^{11}$\,G, achieves $B_{\rm p}\sim10^{14-15}$\,G and toroidal component $B_{\phi}\sim10^{16}$\,G inside PNS with a spin period of $\Omega\ge1000$\,rad\,s$^{-1}$.

In the following discussion, we, therefore, employ the NS dipole magnetic field strength of $B_{\rm p}=10^{15}\,{\rm G}$ and NS rotation angular frequency of $\Omega=1000\,{\rm rad}\,{\rm s}^{-1}$ as fiducial values.
These values are indeed indicated for explaining the peak luminosity of $L_{\rm peak}\sim10^{44}$\,erg\,s$^{-1}$, which is thus associated with particularly luminous events within the observed range ($L_{\rm peak}\sim10^{42-44}$\,erg\,s$^{-1}$, \citealt{Drout14}), and the energy injection from the magnetar to the ejecta can be written in the form of $L_{\rm mag}=L_{\rm mag,0}/(1+t/\tau_{\rm mag})^2$, where the initial spin down luminosity, $L_{\rm mag,0}$, and time scale, $\tau_{\rm mag}$, are given by
\begin{align}
L_{\rm mag,0}&\sim\frac{B_{\rm p}^2 R_{\rm NS}^6 \Omega^4}{6 c^3}\approx2\times 10^{46}\,{\rm erg}\,{\rm s}^{-1}\left(\frac{B_{\rm p}}{10^{15}\,{\rm G}}\right)^{2}\nonumber\\
&~~~~~~~~~~~~~~~~~~~~~~~~
\times\left(\frac{R_{\rm NS}}{12\,{\rm km}}\right)^{6}\left(\frac{\Omega}{1000\,{\rm rad}\,{\rm s}^{-1}}\right)^{4}
\end{align}
and
\begin{align}
\tau_{\rm mag}&\approx 0.6\,{\rm day}\left(\frac{B_{\rm p}}{10^{15}\,{\rm G}}\right)^{-2}\left(\frac{M_{\rm NS}}{1.6\,M_\odot}\right)\nonumber\\
&~~~~~
\times\left(\frac{R_{\rm NS}}{12\,{\rm km}}\right)^{-6}\left(\frac{I_{\rm NS}}{2 \times 10^{45} {\rm g\,cm^2}}\right)\left(\frac{\Omega}{1000\,{\rm rad}\,{\rm s}^{-1}}\right)^{-2},
\end{align}
respectively. Then, the peak luminosity of magnetar powered emission at $t=t_{\rm diff}$ can be estimated as ~\citep{Kasen:2010}
\begin{align}
L_{\rm peak}&\approx 2\times 10^{45}\,{\rm erg}\,{\rm s}^{-1}
\left(\frac{B_{\rm p}}{10^{15}\,{\rm G}}\right)^{-2}\left(\frac{M_{\rm NS}}{1.6\,M_\odot}\right)\nonumber\\
&\times\left(\frac{R_{\rm NS}}{12\,{\rm km}}\right)^{-6}\left(\frac{I_{\rm NS}}{2 \times 10^{45} {\rm g\,cm^2}}\right)^2
\left(\frac{t_{\rm diff}}{2\,{\rm day}}\right)^{-2}~(\tau_{\rm mag}<t_{\rm diff}).\label{eq:Lp_mag}
\end{align}
Here, $M_{\rm NS}$, $R_{\rm NS}$, and $I_{\rm NS}$ denote the NS mass, radius, and moment of inertia and we employ fiducial values of $M_{\rm NS}=1.6\,M_\odot$, $R_{\rm NS}=12$\,km, and $I_{\rm NS}=2\times10^{45}$\,g\,cm$^2$ in the above expression.
We note that, from our numerical models, $M_{\rm NS}(\sim1.6$--$1.7\,M_\odot)$ is nearly saturated even during our simulation time of $\le500$\,ms (Fig.~\ref{fig:Overall}) and is consistent with the value assumed above.
However, regarding $R_{\rm NS}$ and $I_{\rm NS}$, our simulation results give the values relevant only for the PNS phase, which are $R_{\rm NS}\sim20$\,km (on axis) and $\gtrsim100$\,km (equator) and $I_{\rm NS}\sim10^{46}$\,g\,cm$^2$ at the final simulation time and significantly differ from values assumed, though these PNS properties will change along with the subsequent deleptonization of NS ($t_{\rm pb}\gg10$\,s).
We also note that the photon diffusion time scale is modified by updating the ejecta velocity by $v\rightarrow\sqrt{2(E_{\rm exp}+L_{\rm mag,0}\tau_{\rm mag})/M_{\rm ej}}\sim\sqrt{I\Omega^2/M_{\rm ej}}$ considering the ejecta acceleration due to the energy injection from the magnetar~\citep[see ][]{Kasen:2010}. We also note that Eq.~\eqref{eq:Lp_mag} is only valid for the case that the diffusion time scale is longer than the spin-down time scale.

Finally, we consider the emission of the shock breakout in the progenitor stellar wind~\citep{Balberg2011} as well as the emission powered by the wind-ejecta interaction~\citep{Moriya13}. Assuming that the AIC occurs in a wind density profile given by $\rho_{\rm wind}=A/r^2$ with $A$ being an assumed coefficient determined by the mass loss rate and stellar wind velocity and $r$ referring the distance from the event, the peak time and shock luminosity at the peak time of the shock breakout emission can be estimated as~\citep{Balberg2011,Margutti19}
\begin{align}
    t_{\rm peak}&\sim\frac{\kappa A}{c}\approx4\times 10^{-3}\,{\rm day}\left(\frac{\kappa}{0.2\, {\rm cm^2}\,{\rm g}^{-1}}\right)\left(\frac{A}{ 10^2\,A_*}\right)
\end{align}
and
\begin{align}
    L_{\rm sh}&\sim\frac{9\pi}{8}r^2\rho_{\rm wind} v^3=5\times 10^{42}\,{\rm erg}\,{\rm s}^{-1}\left(\frac{A}{10^2 A_*}\right)\left(\frac{v}{0.1\,c}\right)^3,
\end{align}
respectively, with $A_*=5\times 10^{11}\,{\rm g}\,{\rm cm}^{-1}$ which corresponds to the case of the wind mass loss rate of ${\dot M}=10^{-5}\,{M_\odot}\,{\rm yr}^{-1}$ and wind velocity of $v_{\rm wind}=10^3\,{\rm km}\,{\rm s}^{-1}$. The naive estimation of the effective temperature at the peak time is $3\times 10^5\,{\rm K}$. However, since the shock velocity is larger than $0.07\,c$, the actual emission will be bright in $X$-ray ($\gtrsim 1\,{\rm keV}$) due to insufficient thermalization ~\citep{Nakar2010,Katz2010}. For the wind profile $\propto r^{-2}$, the shock luminosity will remain approximately constant and power the emission until the stopping time of the ejecta, which is estimated by $\approx 6\times 10^2\,{\rm day}(M_{\rm ej}/0.05\,M_\odot)(v/0.1\,c)^{-1}(A/10^2\,A_*)$ from $M_{\rm ej}\approx \int_0^{v t} 4\pi r^2 \rho_{\rm wind}dr$. However, the efficiency of converting the matter internal energy into photon energy may become inefficient in the later epoch~\citep{Moriya13}. Hence, the value of $L_{\rm sh}$ should be taken as the upper limit of the photon luminosity after the shock breakout time. 

Figure~\ref{fig:LC} shows the light curves obtained by numerically solving one-zone models of the scenarios discussed above. In the one zone models, the evolution of the internal energy, $E_{\rm int}$, is carried out under the assumption of homogeneous density, radiation dominated pressure, and homologously expanding velocity profile of ejecta. The evolution equation for $E_{\rm int}$ is given by~\citep[e.g., ][]{Kasen:2010} 
\begin{align}
\frac{d E_{\rm int}}{dt}=-\frac{E_{\rm int}}{t}-L_{\rm rad}+H_{\rm heat},
\end{align}
where $L_{\rm rad}\sim E_{\rm int} t/t_{\rm diff}^2$ and $H_{\rm heat}$ denote the bolometric luminosity and total energy injection rate, respectively. The parameters as well as the total energy injection rate used in the model of each scenario are the same as those employed in the discussion above. The values of the peak luminosity estimated above are indeed approximately reproduced by the one-zone models.

With the ejecta property obtained by our work, the energy injection from the newborn magnetar can explain the brighter side of observed FBOTs, whose bolometric luminosities peak at $\gtrsim10^{44}$\,erg\,s$^{-1}$, as being pointed out by previous studies~\citep{Prentice18,Rest18_FELT,Margutti19,Lyutikov19}. However, the brightness and time scale of the emission are highly dependent on the property of the newborn magnetar, and hence, the investigation of the magnetar property based on plausible AIC/MIC progenitor models and considering relevant magnetic field amplification mechanisms will be crucial for the light curve prediction.

Even in the absence of a magnetar formation, there could be various electromagnetic counterparts~\citep[see also, e.g.,][]{Margutti19}. The cooling emission due to the ejecta-envelope interaction can explain the faint FBOT peak luminosities of $\sim10^{42}$\,erg\,s$^{-1}$~\citep{Drout14}, if the envelope radius $R_\ast$ is larger than $10^{12}\,{\rm cm}$. The emission will decay rapidly in the time scale of a few days unless a significant energy injection is present. Hence, in the case of observing the FBOT-like transient, the presence of a power source can be distinguished by the observation on whether the luminosity rapidly drops within a few times the peak time scale or not.

The long-lasting emission followed by such a rapid decline after a few diffusion time scales ($\sim 10\,{\rm day}$) may indicate the production of the r-process elements or $^{56}$Ni in the AIC ejecta. Due to the small ejecta mass and resulting inefficient $\gamma$-ray thermalization, the emission powered by $^{56}$Ni can decline more rapidly than that powered by the r-process heating, while the actual decline rate of the r-process powered emission can also strongly depend on the nucleosynthetic abundances~\citep{Wanajo:2014}. For observing such signals, the deep follow-up observation in the infrared wavelengths, such as by JWST, will be needed also with the nebular spectral modeling~\citep{Hotokezaka:2023aiq,Pognan:2023qhw} since the ejecta are expected to be optically thin after $10\,{\rm day}$.
For example, assuming that the nebular spectra observed in GRB230307A~\citep{JWST:2023jqa} was powered by radioactive decay of r-process elements with ejecta mass $\sim 0.05\,M_\odot$ in the distance of $300\,{\rm Mpc}$, the nebular phase emission powered by radioactive decay may be observed for nearby AIC events with the distance $<100\,{\rm Mpc}$ if the sufficiently large amount of r-process elements ($> 0.005\,M_\odot$) is synthesized.

The AIC ejecta-wind interaction can result in the shock-breakout emission comparably bright to some of FBOTs, as far as the mass loss rate is larger than $10^{-3}\,{M_\odot}\,{\rm yr}^{-1}$. However, as seen above as well as being pointed out in~\cite{Margutti19}, significantly higher wind density (with $0.1$--$1\,{M_\odot}\,{\rm yr}^{-1}$) will be required to explain bright FBOTs with $>10^{44}\,{\rm erg\,s^{-1}}$. Since the emission will be bright in X-rays, wide-field time-domain X-ray telescopes, such as Einstein Probe~\citep{EinsteinProbeTeam:2022fpj,Yuan:2025cbh}, will be useful to observe the shock-breakout emission. For example, the shock breakout emission peaks with the time scale of $1000\,{\rm s}$ with the brightness of $\sim10^{43}\,{\rm erg\,s^{-1}}$ can be detected for the events with the distance $\lesssim 50\,{\rm Mpc}$ by the WXT (~\citealt{Yuan:2025cbh}; see also references therein for the expected detection rate of FBOTs by Einstein Probe), while the detectability should be discussed employing more detailed modeling of the emission~\citep{Nakar2010}. The spectral features of the emission induced by the wind-ejecta interaction can indicate the progenitor of the system. For example, no H/He line feature is expected in the spectra if the progenitor is formed through a merger with its companion WD~\citep{Saio&Nomoto85,Shen12,Schwab21}, whereas such features can be present if the companion is a non-degenerate star with an H/He layer~\citep{Nomoto79,Nomoto&Kondo91}. Note, however, that if the wind shell has already detached far from the progenitor surface at the onset of the AIC, both shock-breakout and shock-powered emission in the early phase may be absent~\citep{Yu19}. Nevertheless, the presence of the dense stellar wind can still be confirmed by observing radio synchrotron emission, which may peak around the ejecta deceleration time scale (100-1000 days for the $A=10^2\,A_*$ wind profile)~\citep[][see also~\citealt{Yu19}]{Margutti19}.

The increase in the ejecta opacity due to the presence of the r-process elements in general leads to the emission with the fainter luminosity, redder spectra, and longer time scale (see, for example, Eqs.~\eqref{eq:Lp_env}, ~\eqref{eq:tempp_env}, and ~\eqref{eq:t_diff}). In fact, the peak luminosity can be fainter for an order of magnitude in the presence of large amount of the r-process elements, and diffusion time scale and effective temperature at the peak can also be longer and lower by a factor of a few, respectively. While the estimation of the ejecta opacity can be a good indicator for the significant production of the r-process elements, its dependence also degenerates with other model parameters, such as the envelope radius, magnetar properties, and wind profiles. Hence, in order to judge the presence of the r-process elements, more detailed modeling of the observations, such as light curve/spectral evolutions, also with the accurate predictions is necessary.

\section{Summary}
\label{sec:Summary}
We presented the results of 3D numerical relativity simulations with multi-energy neutrino transport for core collapse of rotating magnetized WDs, as a modeling for an end stage of single or double degenerate WD evolutions.
The main focuses of this study are on the explosion dynamics and their associated multi-messenger signals: GWs, neutrinos, and electromagnetic counterparts.
AIC of WDs has been recently attracting more attentions because of their unique properties (relatively low explosion energy and small amount of mass ejection), which might be feasible to explain some of the rapidly evolving optical transients.
Our results indeed support such scenarios, albeit depending on yet-to-be-cleared several parameters assumed.

Regarding the explosion dynamics, we witnessed prompt type explosions in all models.
The shock front does not noticeably stall and it reaches $10^9$\,cm at $t_{\rm pb}\sim300$\,ms.
All models including the slowest rotating one R1 show a bipolar explosion, with R6 presenting the most visible jet-like configuration.
At the end of our simulation times ($t_{\rm pb}\sim400$--500\,ms), the diagnostic explosion energy and ejecta mass reach 0.1--0.4\,B and 0.02--0.1\,$M_\odot$, respectively.
These values are in good agreement with latest reports  \citep{Longo-Micchi23,Batziou&Janka24_AIC,Cheong25_AIC} and indeed fall in the range of low explosion energy and small ejecta mass in comparison to typical massive stellar collapse cases.
As for the rotational dependence on the shock dynamics, rapidly rotating models initially exhibit larger shock radii due mainly to centrifugal expansion.
However at a later phase, moderately rotating models show more extended shock radii, which are facilitated by stronger neutrino heating emitted from more compact and hotter PNS because of the less centrifugal support.
As another remarkable feature of rotation, our rapidly rotating model R6 presents the emergence of strong spiral waves dominated by a low (one-armed) mode, which is by definition completely suppressed in the counterpart octant symmetry model R6o.
These spiral waves play a role to increase both the explosion energy and ejecta mass approximately by a factor of 2, from which we can say that full 3D models, i.e. without assuming axisymmetry, are essential.

We also discussed ejecta properties.
As a major trend, the faster rotating model ejects lower $Y_e$ components, which is consistent with \cite{Batziou&Janka24_AIC}.
There are two rotational effects which produce such a trend.
As a leading effect, the rotation simply flattens the central core having low-$Y_e$ matters toward equator.
Thence the edge of such a core, particularly along the equator, becomes gravitationally loosely bound and is consequently more easily ejected.
The appearance of strong spiral waves may also be another factor, as such kind of rotational instability redistributes the angular momentum and the low-$Y_e$ matters in the vicinity of central core tend to be more easily expelled.
Typical ejecta velocities and $Y_e$ extend from $0.05$ to $0.1\,c$ and $\sim0.3$ to well above 0.5, respectively.

The strong spiral waves are indeed the origin of loud GW signals.
The most rapidly rotating model R6 emits GWs, whose amplitudes reach $D|h|\sim500$\,cm for an observer along the rotational axis, during the spiral-wave developing phase.
These signals are not burst type GWs, as represented by the so-called Type I signal from the rotational core bounce, but last for a relatively long time, broadly 100\,ms in all models.
Such long-lasting quasi-periodic strong signals may expand the horizon distance of such events for GWs up to $\sim1$\,Mpc for the current generation GW detectors or even up to $\sim10$\,Mpc for third-generation ground-based detectors, which can be assessed by more sophisticated GW detection analyses.
We also found a new finding that the dominant mode of spectral evolution is consisted of two components: initial spiral wave-origin modes and subsequent ramp-up feature stemming from $g/f$-mode oscillations of ``rotationally flattened'' PNS core.
According to our mode analysis, the initial peak frequency associated with the spiral waves are nicely explained by the typical rotational frequency of a region (more precisely a shell of rotating spheroid), whose density is around $\sim10^{13}$\,g\,cm$^{-3}$ regardless of the initial rotation.
After the fade-out of the strong spiral waves, the spectral peak is smoothly taken over by those of $g/f$-mode oscillations.
We demonstrated that the PNS radius and emergent electron antineutrino energy $\varepsilon_{\bar\nu_e}$, which are often used to describe the ramp-up feature of $g/f$-mode oscillations \citep{BMuller13,marek09gw}, can be measured along the pole to reproduce the spectral evolution.

We also discussed how we would be able to observe such AIC events via neutrinos.
Because of the nature of AICs with essentially no stellar envelope, the mass accretion onto the nascent NS is largely suppressed compared to typical massive stellar collapse.
As a result, the accretion luminosity phase soon ceases and neutrino luminosities for all species become relatively fainter.
Additionally the evolution of mean energy of emergent neutrinos does not present any noticeable increasing trend.
All these features are quite similar to those from iron core collapse at low mass end or electron-capture SNe \citep{Hüdepohl10,BMuller14,Radice17,KurodaT22}.
As another feature, we did not observe any significant differences in neutrino detection rates between normal and inverse mass ordering.
This stems again from the low mass accretion rate in AICs, which results in the aforementioned less distinction in neutrino profiles between different neutrino flavours.
Our simple estimation infers that if such AICs happen at the Galactic center, we would detect $\sim10^5$ and $\sim10^4$ neutrinos for IceCube and HyperK, respectively.
For IceCube, the neutrino counts are well above its background signal, if they happen at the Galactic center.
However, unlike the horizon distance of such events for GWs ($\sim10(1)$\,Mpc for the third(current) generation detectors), relatively faint neutrino luminosity makes its horizon distance much shorter than 1\,Mpc.

Based on the simulation results, we broadly estimated the properties of  possible electromagnetic counterparts in various scenarios which can be associated to AICs. With tuned properties of the newborn magnetar, the energy injection from the magnetar can explain the FBOTs with the ejecta property obtained in our work, as being pointed out by previous studies~\citep{Drout14,Prentice18,Rest18_FELT,Margutti19,Lyutikov19}. Even in the absence of the significant magnetar energy injection to the ejecta, the cooling emission due to the ejecta-envelope interaction can be as bright as some of faint FBOTs~\citep{Drout14} with the inflated envelope of $\gtrsim10^{\rm 12}\,{\rm cm}$. The cooling envelope emission can be followed by the nebular emission powered by r-process elements or $^{56}{\rm Ni}$ productions. The observation of these signals will provide important insights into the diversity in AICs as well as their potential roles in the chemical evolution in the universe. Since the light curves are highly dependent on the assumption on the magnetar, the investigation of its property consistently following its formation process taking the magnetic field effects into account is crucial. In addition, more sophisticated modeling based on the detailed structure and neucleosyntheic abundances of the ejecta as well as taking microphysical process~\citep[e.g.,][]{Nakar2010,Hotokezaka:2023aiq,Pognan:2023qhw} and multi-dimensional radiative transfer effects~\citep[e.g.,][]{Margutti19} into account is needed to provide accurate light curve/spectral predictions, which are left as topics for future investigation.

Before closing, we shortly touch on observable differences in multi-messenger signals between AIC and BNS, both of which could be the potential formation channel to a rapidly rotating NS.
Regarding their GWs, BNS emits typically two to three orders of magnitude larger amplitudes than those from collapse of WDs, with significantly different waveforms \citep[e.g.,][]{Shibata05a}.
We also expect noticeable differences in neutrino signals at bounce in AICs or at merger in BNSs.
On one hand the former AIC exhibits the electron type neutrino burst at core-bounce via deleptonization, the BNS merger, on the other hand, emits electron type anti-neutrino burst \citep{Cusinato22} due to leptonization.
Finally as for the EM counterpart, the AIC may exhibit some features of its companion star, for instance H/He line feature is expected in the spectra if the companion is a non-degenerate star with an H/He layer~\citep{Nomoto79,Nomoto&Kondo91}, which is surely absent in the BNS merger.
As we discussed, the emission from radioactively powered AIC ejecta may appear similar to that of a faint, red kilonova, although we certainly need more detailed nucleosynthesis calculation as well as spectral modeling.
Like these, we can expect that there are many distinct features in the multi-messenger signals between AIC and BNS, which would enable us to easily distinguish these two events.

\section*{Acknowledgements}
TK thanks the members of the CRA for stimulated discussion and useful advice.
TK is also grateful to Tobias Fischer for the extended EOS tables and also to Shota Shibagaki for providing a module to reproduce the background shot noise for IceCube.
We also acknowledge Brian Metzger for his valuable
comments on the observed properties of AT~2018cow.
Numerical computations were carried out on Sakura and Raven at Max Planck Computing and Data Facility.
This work was in part supported by Grant-in-Aid for Scientific Research (No. 23H04900) of Japanese MEXT/JSPS.

\section*{Data Availability}
The data underlying this article will be shared on reasonable request to the corresponding author.
GW strains and neutrino profiles used in Figs.~\ref{fig:GW} and \ref{fig:LnuEnu}, respectively, are available in an online repository \url{https://github.com/kurodatk/GWs_Neutrinos_AIC}.



\bibliographystyle{mnras}
\bibliography{main} 

\begin{thebibliography}{}
\makeatletter
\relax
\def\mn@urlcharsother{\let\do\@makeother \do\$\do\&\do\#\do\^\do\_\do\%\do\~}
\def\mn@doi{\begingroup\mn@urlcharsother \@ifnextchar [ {\mn@doi@}
  {\mn@doi@[]}}
\def\mn@doi@[#1]#2{\def\@tempa{#1}\ifx\@tempa\@empty \href
  {http://dx.doi.org/#2} {doi:#2}\else \href {http://dx.doi.org/#2} {#1}\fi
  \endgroup}
\def\mn@eprint#1#2{\mn@eprint@#1:#2::\@nil}
\def\mn@eprint@arXiv#1{\href {http://arxiv.org/abs/#1} {{\tt arXiv:#1}}}
\def\mn@eprint@dblp#1{\href {http://dblp.uni-trier.de/rec/bibtex/#1.xml}
  {dblp:#1}}
\def\mn@eprint@#1:#2:#3:#4\@nil{\def\@tempa {#1}\def\@tempb {#2}\def\@tempc
  {#3}\ifx \@tempc \@empty \let \@tempc \@tempb \let \@tempb \@tempa \fi \ifx
  \@tempb \@empty \def\@tempb {arXiv}\fi \@ifundefined
  {mn@eprint@\@tempb}{\@tempb:\@tempc}{\expandafter \expandafter \csname
  mn@eprint@\@tempb\endcsname \expandafter{\@tempc}}}

\bibitem[\protect\citeauthoryear{{Abbasi} et~al.,}{{Abbasi}
  et~al.}{2011}]{abbasi11}
{Abbasi} R.,  et~al., 2011, \mn@doi [\aap] {10.1051/0004-6361/201117810}, \href
  {https://ui.adsabs.harvard.edu/abs/2011A&A...535A.109A} {535, A109}

\bibitem[\protect\citeauthoryear{{Abbott} et~al.,}{{Abbott}
  et~al.}{2017}]{CE_Abbott}
{Abbott} B.~P.,  et~al., 2017, \mn@doi [Classical and Quantum Gravity]
  {10.1088/1361-6382/aa51f410.48550/arXiv.1607.08697}, \href
  {https://ui.adsabs.harvard.edu/abs/2017CQGra..34d4001A} {34, 044001}

\bibitem[\protect\citeauthoryear{{Abbott} et~al.,}{{Abbott}
  et~al.}{2018}]{abbott18det}
{Abbott} B.~P.,  et~al., 2018, \mn@doi [Living Reviews in Relativity]
  {10.1007/s41114-018-0012-9}, \href
  {https://ui.adsabs.harvard.edu/abs/2018LRR....21....3A} {21, 3}

\bibitem[\protect\citeauthoryear{{Abdikamalov}, {Ott}, {Rezzolla}, {Dessart},
  {Dimmelmeier}, {Marek}  \& {Janka}}{{Abdikamalov}
  et~al.}{2010}]{Abdikamalov10_AIC}
{Abdikamalov} E.~B.,  {Ott} C.~D.,  {Rezzolla} L.,  {Dessart} L.,
  {Dimmelmeier} H.,  {Marek} A.,   {Janka} H.~T.,  2010, \mn@doi [\prd]
  {10.1103/PhysRevD.81.044012}, \href
  {https://ui.adsabs.harvard.edu/abs/2010PhRvD..81d4012A} {81, 044012}

\bibitem[\protect\citeauthoryear{{Abe} et~al.,}{{Abe} et~al.}{2011}]{abe11}
{Abe} K.,  et~al., 2011, arXiv e-prints, \href
  {https://ui.adsabs.harvard.edu/abs/2011arXiv1109.3262A} {p. arXiv:1109.3262}

\bibitem[\protect\citeauthoryear{{Balberg} \& {Loeb}}{{Balberg} \&
  {Loeb}}{2011}]{Balberg2011}
{Balberg} S.,  {Loeb} A.,  2011, \mn@doi [\mnras]
  {10.1111/j.1365-2966.2011.18505.x}, \href
  {https://ui.adsabs.harvard.edu/abs/2011MNRAS.414.1715B} {414, 1715}

\bibitem[\protect\citeauthoryear{{Balbus} \& {Hawley}}{{Balbus} \&
  {Hawley}}{1991}]{Balbus91}
{Balbus} S.~A.,  {Hawley} J.~F.,  1991, \mn@doi [\apj] {10.1086/170270}, \href
  {https://ui.adsabs.harvard.edu/abs/1991ApJ...376..214B} {376, 214}

\bibitem[\protect\citeauthoryear{Barnes, Kasen, Wu  \&
  Mart\'\i{}nez-Pinedo}{Barnes et~al.}{2016}]{Barnes:2016umi}
Barnes J.,  Kasen D.,  Wu M.-R.,   Mart\'\i{}nez-Pinedo G.,  2016, \mn@doi
  [Astrophys. J.] {10.3847/0004-637X/829/2/110}, 829, 110

\bibitem[\protect\citeauthoryear{{Baron}, {Cooperstein}, {Kahana}  \&
  {Nomoto}}{{Baron} et~al.}{1987}]{Baron87}
{Baron} E.,  {Cooperstein} J.,  {Kahana} S.,   {Nomoto} K.,  1987, \mn@doi
  [\apj] {10.1086/165542}, \href
  {https://ui.adsabs.harvard.edu/abs/1987ApJ...320..304B} {320, 304}

\bibitem[\protect\citeauthoryear{{Batziou}, {Glas}, {Janka}, {Ehring},
  {Abdikamalov}  \& {Just}}{{Batziou} et~al.}{2024}]{Batziou&Janka24_AIC}
{Batziou} E.,  {Glas} R.,  {Janka} H.~T.,  {Ehring} J.,  {Abdikamalov} E.,
  {Just} O.,  2024, \mn@doi [arXiv e-prints] {10.48550/arXiv.2412.02756}, \href
  {https://ui.adsabs.harvard.edu/abs/2024arXiv241202756B} {p. arXiv:2412.02756}

\bibitem[\protect\citeauthoryear{{Baumgarte} \& {Shapiro}}{{Baumgarte} \&
  {Shapiro}}{1999}]{Baumgarte99}
{Baumgarte} T.~W.,  {Shapiro} S.~L.,  1999, \mn@doi [\prd]
  {10.1103/PhysRevD.59.024007}, \href
  {http://ads.nao.ac.jp/abs/1999PhRvD..59b4007B} {59, 024007}

\bibitem[\protect\citeauthoryear{{Bollig}, {Janka}, {Lohs},
  {Mart{\'\i}nez-Pinedo}, {Horowitz}  \& {Melson}}{{Bollig}
  et~al.}{2017}]{Bollig17}
{Bollig} R.,  {Janka} H.~T.,  {Lohs} A.,  {Mart{\'\i}nez-Pinedo} G.,
  {Horowitz} C.~J.,   {Melson} T.,  2017, \mn@doi [\prl]
  {10.1103/PhysRevLett.119.242702}, \href
  {https://ui.adsabs.harvard.edu/abs/2017PhRvL.119x2702B} {119, 242702}

\bibitem[\protect\citeauthoryear{{Bugli}, {Guilet}, {Obergaulinger},
  {Cerd{\'a}-Dur{\'a}n}  \& {Aloy}}{{Bugli} et~al.}{2020}]{Bugli20}
{Bugli} M.,  {Guilet} J.,  {Obergaulinger} M.,  {Cerd{\'a}-Dur{\'a}n} P.,
  {Aloy} M.~A.,  2020, \mn@doi [\mnras] {10.1093/mnras/stz3483}, \href
  {https://ui.adsabs.harvard.edu/abs/2020MNRAS.492...58B} {492, 58}

\bibitem[\protect\citeauthoryear{{Centrella}, {New}, {Lowe}  \&
  {Brown}}{{Centrella} et~al.}{2001}]{Centrella01}
{Centrella} J.~M.,  {New} K. C.~B.,  {Lowe} L.~L.,   {Brown} J.~D.,  2001,
  \mn@doi [\apjl] {10.1086/319634}, \href
  {https://ui.adsabs.harvard.edu/abs/2001ApJ...550L.193C} {550, L193}

\bibitem[\protect\citeauthoryear{{Cheong}, {Pitik}, {Longo Micchi}  \&
  {Radice}}{{Cheong} et~al.}{2025}]{Cheong25_AIC}
{Cheong} P. C.-K.,  {Pitik} T.,  {Longo Micchi} L.~F.,   {Radice} D.,  2025,
  \mn@doi [\apjl] {10.3847/2041-8213/ada1cc}, \href
  {https://ui.adsabs.harvard.edu/abs/2025ApJ...978L..38C} {978, L38}

\bibitem[\protect\citeauthoryear{{Cusinato}, {Guercilena}, {Perego},
  {Logoteta}, {Radice}, {Bernuzzi}  \& {Ansoldi}}{{Cusinato}
  et~al.}{2022}]{Cusinato22}
{Cusinato} M.,  {Guercilena} F.~M.,  {Perego} A.,  {Logoteta} D.,  {Radice} D.,
   {Bernuzzi} S.,   {Ansoldi} S.,  2022, \mn@doi [European Physical Journal A]
  {10.1140/epja/s10050-022-00743-5}, \href
  {https://ui.adsabs.harvard.edu/abs/2022EPJA...58...99C} {58, 99}

\bibitem[\protect\citeauthoryear{{Dan}, {Rosswog}, {Br{\"u}ggen}  \&
  {Podsiadlowski}}{{Dan} et~al.}{2014}]{Dan14}
{Dan} M.,  {Rosswog} S.,  {Br{\"u}ggen} M.,   {Podsiadlowski} P.,  2014,
  \mn@doi [\mnras] {10.1093/mnras/stt1766}, \href
  {https://ui.adsabs.harvard.edu/abs/2014MNRAS.438...14D} {438, 14}

\bibitem[\protect\citeauthoryear{{Dessart}, {Burrows}, {Ott}, {Livne}, {Yoon}
  \& {Langer}}{{Dessart} et~al.}{2006}]{Dessart06_AIC}
{Dessart} L.,  {Burrows} A.,  {Ott} C.~D.,  {Livne} E.,  {Yoon} S.~C.,
  {Langer} N.,  2006, \mn@doi [\apj] {10.1086/503626}, \href
  {https://ui.adsabs.harvard.edu/abs/2006ApJ...644.1063D} {644, 1063}

\bibitem[\protect\citeauthoryear{{Di Stefano}, {Voss}  \& {Claeys}}{{Di
  Stefano} et~al.}{2011}]{Stefano11}
{Di Stefano} R.,  {Voss} R.,   {Claeys} J.~S.~W.,  2011, \mn@doi [\apjl]
  {10.1088/2041-8205/738/1/L1}, \href
  {https://ui.adsabs.harvard.edu/abs/2011ApJ...738L...1D} {738, L1}

\bibitem[\protect\citeauthoryear{{Dimmelmeier}, {Font}  \&
  {M{\"u}ller}}{{Dimmelmeier} et~al.}{2002}]{Dimmelmeier02B}
{Dimmelmeier} H.,  {Font} J.~A.,   {M{\"u}ller} E.,  2002, \mn@doi [\aap]
  {10.1051/0004-6361:20021053}, \href
  {http://ads.nao.ac.jp/abs/2002A%26A...393..523D} {393, 523}

\bibitem[\protect\citeauthoryear{{Dimmelmeier}, {Ott}, {Marek}  \&
  {Janka}}{{Dimmelmeier} et~al.}{2008}]{Dimmelmeier08}
{Dimmelmeier} H.,  {Ott} C.~D.,  {Marek} A.,   {Janka} H.-T.,  2008, \mn@doi
  [\prd] {10.1103/PhysRevD.78.064056}, \href
  {http://ads.nao.ac.jp/abs/2008PhRvD..78f4056D} {78, 064056}

\bibitem[\protect\citeauthoryear{{Drout} et~al.,}{{Drout}
  et~al.}{2014}]{Drout14}
{Drout} M.~R.,  et~al., 2014, \mn@doi [\apj] {10.1088/0004-637X/794/1/23},
  \href {https://ui.adsabs.harvard.edu/abs/2014ApJ...794...23D} {794, 23}

\bibitem[\protect\citeauthoryear{{Fryer}, {Benz}, {Herant}  \&
  {Colgate}}{{Fryer} et~al.}{1999}]{Fryer99_AIC}
{Fryer} C.,  {Benz} W.,  {Herant} M.,   {Colgate} S.~A.,  1999, \mn@doi [\apj]
  {10.1086/307119}, \href
  {https://ui.adsabs.harvard.edu/abs/1999ApJ...516..892F} {516, 892}

\bibitem[\protect\citeauthoryear{{Fujibayashi}, {Sekiguchi}, {Shibata}  \&
  {Wanajo}}{{Fujibayashi} et~al.}{2023}]{Fujibayashi23}
{Fujibayashi} S.,  {Sekiguchi} Y.,  {Shibata} M.,   {Wanajo} S.,  2023, \mn@doi
  [\apj] {10.3847/1538-4357/acf5e5}, \href
  {https://ui.adsabs.harvard.edu/abs/2023ApJ...956..100F} {956, 100}

\bibitem[\protect\citeauthoryear{Fujibayashi, Lam, Shibata  \&
  Sekiguchi}{Fujibayashi et~al.}{2024}]{Fujibayashi24}
Fujibayashi S.,  Lam A. T.-L.,  Shibata M.,   Sekiguchi Y.,  2024, \mn@doi
  [Phys. Rev. D] {10.1103/PhysRevD.109.023031}, 109, 023031

\bibitem[\protect\citeauthoryear{{Garc{\'\i}a-Berro}
  et~al.,}{{Garc{\'\i}a-Berro} et~al.}{2012}]{Garcia-Berro12}
{Garc{\'\i}a-Berro} E.,  et~al., 2012, \mn@doi [\apj]
  {10.1088/0004-637X/749/1/25}, \href
  {https://ui.adsabs.harvard.edu/abs/2012ApJ...749...25G} {749, 25}

\bibitem[\protect\citeauthoryear{{Hachisu}, {Eriguchi}  \& {Nomoto}}{{Hachisu}
  et~al.}{1986}]{Hachisu86}
{Hachisu} I.,  {Eriguchi} Y.,   {Nomoto} K.,  1986, \mn@doi [\apj]
  {10.1086/164487}, \href
  {https://ui.adsabs.harvard.edu/abs/1986ApJ...308..161H} {308, 161}

\bibitem[\protect\citeauthoryear{{Hild} et~al.,}{{Hild} et~al.}{2011}]{ET_Hild}
{Hild} S.,  et~al., 2011, \mn@doi [Classical and Quantum Gravity]
  {10.1088/0264-9381/28/9/09401310.48550/arXiv.1012.0908}, \href
  {https://ui.adsabs.harvard.edu/abs/2011CQGra..28i4013H} {28, 094013}

\bibitem[\protect\citeauthoryear{{Hilditch}, {Bernuzzi}, {Thierfelder}, {Cao},
  {Tichy}  \& {Br{\"u}gmann}}{{Hilditch} et~al.}{2013}]{Hilditch13}
{Hilditch} D.,  {Bernuzzi} S.,  {Thierfelder} M.,  {Cao} Z.,  {Tichy} W.,
  {Br{\"u}gmann} B.,  2013, \mn@doi [\prd] {10.1103/PhysRevD.88.084057}, \href
  {https://ui.adsabs.harvard.edu/abs/2013PhRvD..88h4057H} {88, 084057}

\bibitem[\protect\citeauthoryear{Hotokezaka, Wanajo, Tanaka, Bamba, Terada  \&
  Piran}{Hotokezaka et~al.}{2016}]{Hotokezaka16}
Hotokezaka K.,  Wanajo S.,  Tanaka M.,  Bamba A.,  Terada Y.,   Piran T.,
  2016, \mn@doi [Mon. Not. Roy. Astron. Soc.] {10.1093/mnras/stw404}, 459, 35

\bibitem[\protect\citeauthoryear{Hotokezaka, Tanaka, Kato  \&
  Gaigalas}{Hotokezaka et~al.}{2023}]{Hotokezaka:2023aiq}
Hotokezaka K.,  Tanaka M.,  Kato D.,   Gaigalas G.,  2023, \mn@doi [Mon. Not.
  Roy. Astron. Soc.] {10.1093/mnrasl/slad128}, 526, L155

\bibitem[\protect\citeauthoryear{{H{\"u}depohl}, {M{\"u}ller}, {Janka}, {Marek}
   \& {Raffelt}}{{H{\"u}depohl} et~al.}{2010}]{Hüdepohl10}
{H{\"u}depohl} L.,  {M{\"u}ller} B.,  {Janka} H.~T.,  {Marek} A.,   {Raffelt}
  G.~G.,  2010, \mn@doi [\prl] {10.1103/PhysRevLett.104.251101}, \href
  {https://ui.adsabs.harvard.edu/abs/2010PhRvL.104y1101H} {104, 251101}

\bibitem[\protect\citeauthoryear{{Hyper-Kamiokande Proto-Collaboration}
  et~al.,}{{Hyper-Kamiokande Proto-Collaboration} et~al.}{2018}]{HK18}
{Hyper-Kamiokande Proto-Collaboration} et~al., 2018, \mn@doi [Progress of
  Theoretical and Experimental Physics] {10.1093/ptep/pty044}, \href
  {https://ui.adsabs.harvard.edu/abs/2018PTEP.2018f3C01H} {2018, 063C01}

\bibitem[\protect\citeauthoryear{{Inserra}}{{Inserra}}{2019}]{Inserra19}
{Inserra} C.,  2019, \mn@doi [Nature Astronomy] {10.1038/s41550-019-0854-4},
  \href {https://ui.adsabs.harvard.edu/abs/2019NatAs...3..697I} {3, 697}

\bibitem[\protect\citeauthoryear{{Just}, {Bauswein}, {Ardevol Pulpillo},
  {Goriely}  \& {Janka}}{{Just} et~al.}{2015}]{Just15}
{Just} O.,  {Bauswein} A.,  {Ardevol Pulpillo} R.,  {Goriely} S.,   {Janka}
  H.~T.,  2015, \mn@doi [\mnras] {10.1093/mnras/stv009}, \href
  {https://ui.adsabs.harvard.edu/abs/2015MNRAS.448..541J} {448, 541}

\bibitem[\protect\citeauthoryear{{Kasen} \& {Bildsten}}{{Kasen} \&
  {Bildsten}}{2010}]{Kasen:2010}
{Kasen} D.,  {Bildsten} L.,  2010, \mn@doi [\apj]
  {10.1088/0004-637X/717/1/245}, \href
  {https://ui.adsabs.harvard.edu/abs/2010ApJ...717..245K} {717, 245}

\bibitem[\protect\citeauthoryear{{Katz}, {Budnik}  \& {Waxman}}{{Katz}
  et~al.}{2010}]{Katz2010}
{Katz} B.,  {Budnik} R.,   {Waxman} E.,  2010, \mn@doi [\apj]
  {10.1088/0004-637X/716/1/781}, \href
  {https://ui.adsabs.harvard.edu/abs/2010ApJ...716..781K} {716, 781}

\bibitem[\protect\citeauthoryear{Korobkin, Rosswog, Arcones  \&
  Winteler}{Korobkin et~al.}{2012}]{Korobkin:2012uy}
Korobkin O.,  Rosswog S.,  Arcones A.,   Winteler C.,  2012, \mn@doi [Mon. Not.
  Roy. Astron. Soc.] {10.1111/j.1365-2966.2012.21859.x}, 426, 1940

\bibitem[\protect\citeauthoryear{{Kotake}, {Takiwaki}, {Fischer}, {Nakamura}
  \& {Mart{\'\i}nez-Pinedo}}{{Kotake} et~al.}{2018}]{Kotake18}
{Kotake} K.,  {Takiwaki} T.,  {Fischer} T.,  {Nakamura} K.,
  {Mart{\'\i}nez-Pinedo} G.,  2018, \mn@doi [\apj] {10.3847/1538-4357/aaa716},
  \href {https://ui.adsabs.harvard.edu/abs/2018ApJ...853..170K} {853, 170}

\bibitem[\protect\citeauthoryear{{Kuroda}}{{Kuroda}}{2021}]{KurodaT21}
{Kuroda} T.,  2021, \mn@doi [\apj] {10.3847/1538-4357/abce61}, \href
  {https://ui.adsabs.harvard.edu/abs/2021ApJ...906..128K} {906, 128}

\bibitem[\protect\citeauthoryear{{Kuroda} \& {Shibata}}{{Kuroda} \&
  {Shibata}}{2024}]{KurodaT24}
{Kuroda} T.,  {Shibata} M.,  2024, \mn@doi [\mnras] {10.1093/mnrasl/slae069},
  \href {https://ui.adsabs.harvard.edu/abs/2024MNRAS.533L.107K} {533, L107}

\bibitem[\protect\citeauthoryear{{Kuroda}, {Takiwaki}  \& {Kotake}}{{Kuroda}
  et~al.}{2014}]{KurodaT14}
{Kuroda} T.,  {Takiwaki} T.,   {Kotake} K.,  2014, \mn@doi [\prd]
  {10.1103/PhysRevD.89.044011}, \href
  {http://ads.nao.ac.jp/abs/2014PhRvD..89d4011K} {89, 044011}

\bibitem[\protect\citeauthoryear{{Kuroda}, {Kotake}  \& {Takiwaki}}{{Kuroda}
  et~al.}{2016}]{KurodaT16ApJL}
{Kuroda} T.,  {Kotake} K.,   {Takiwaki} T.,  2016, \mn@doi [\apjl]
  {10.3847/2041-8205/829/1/L14}, \href
  {http://ads.nao.ac.jp/abs/2016ApJ...829L..14K} {829, L14}

\bibitem[\protect\citeauthoryear{{Kuroda}, {Arcones}, {Takiwaki}  \&
  {Kotake}}{{Kuroda} et~al.}{2020}]{KurodaT20}
{Kuroda} T.,  {Arcones} A.,  {Takiwaki} T.,   {Kotake} K.,  2020, \mn@doi
  [\apj] {10.3847/1538-4357/ab9308}, \href
  {https://ui.adsabs.harvard.edu/abs/2020ApJ...896..102K} {896, 102}

\bibitem[\protect\citeauthoryear{{Kuroda}, {Fischer}, {Takiwaki}  \&
  {Kotake}}{{Kuroda} et~al.}{2022}]{KurodaT22}
{Kuroda} T.,  {Fischer} T.,  {Takiwaki} T.,   {Kotake} K.,  2022, \mn@doi
  [\apj] {10.3847/1538-4357/ac31a8}, \href
  {https://ui.adsabs.harvard.edu/abs/2022ApJ...924...38K} {924, 38}

\bibitem[\protect\citeauthoryear{Levan et~al.}{Levan
  et~al.}{2024}]{JWST:2023jqa}
Levan A.~J.,  et~al., 2024, \mn@doi [Nature] {10.1038/s41586-023-06759-1}, 626,
  737

\bibitem[\protect\citeauthoryear{{Longo Micchi}, {Radice}  \&
  {Chirenti}}{{Longo Micchi} et~al.}{2023}]{Longo-Micchi23}
{Longo Micchi} L.~F.,  {Radice} D.,   {Chirenti} C.,  2023, \mn@doi [\mnras]
  {10.1093/mnras/stad2420}, \href
  {https://ui.adsabs.harvard.edu/abs/2023MNRAS.525.6359L} {525, 6359}

\bibitem[\protect\citeauthoryear{{Lund}, {Marek}, {Lunardini}, {Janka}  \&
  {Raffelt}}{{Lund} et~al.}{2010}]{lund10}
{Lund} T.,  {Marek} A.,  {Lunardini} C.,  {Janka} H.-T.,   {Raffelt} G.,  2010,
  \mn@doi [\prd] {10.1103/PhysRevD.82.063007}, \href
  {http://ads.nao.ac.jp/abs/2010PhRvD..82f3007L} {82, 063007}

\bibitem[\protect\citeauthoryear{{Lyutikov} \& {Toonen}}{{Lyutikov} \&
  {Toonen}}{2019}]{Lyutikov19}
{Lyutikov} M.,  {Toonen} S.,  2019, \mn@doi [\mnras] {10.1093/mnras/stz1640},
  \href {https://ui.adsabs.harvard.edu/abs/2019MNRAS.487.5618L} {487, 5618}

\bibitem[\protect\citeauthoryear{Maeda}{Maeda}{2006}]{Maeda:2005pi}
Maeda K.,  2006, \mn@doi [Astrophys. J.] {10.1086/503415}, 644, 385

\bibitem[\protect\citeauthoryear{Maggiore}{Maggiore}{2007}]{Maggiore08}
Maggiore M.,  2007, Gravitational Waves: Volume 1: Theory and Experiments.
Oxford University Press, \mn@doi{10.1093/acprof:oso/9780198570745.001.0001},
  \url {https://doi.org/10.1093/acprof:oso/9780198570745.001.0001}

\bibitem[\protect\citeauthoryear{{Marek}, {Janka}  \& {M{\"u}ller}}{{Marek}
  et~al.}{2009}]{marek09gw}
{Marek} A.,  {Janka} H.-T.,   {M{\"u}ller} E.,  2009, \mn@doi [\aap]
  {10.1051/0004-6361/200810883}, \href
  {http://ads.nao.ac.jp/abs/2009A%26A...496..475M} {496, 475}

\bibitem[\protect\citeauthoryear{{Margutti} et~al.,}{{Margutti}
  et~al.}{2019}]{Margutti19}
{Margutti} R.,  et~al., 2019, \mn@doi [\apj] {10.3847/1538-4357/aafa01}, \href
  {https://ui.adsabs.harvard.edu/abs/2019ApJ...872...18M} {872, 18}

\bibitem[\protect\citeauthoryear{{Marsh}, {Nelemans}  \& {Steeghs}}{{Marsh}
  et~al.}{2004}]{Marsh04}
{Marsh} T.~R.,  {Nelemans} G.,   {Steeghs} D.,  2004, \mn@doi [\mnras]
  {10.1111/j.1365-2966.2004.07564.x}, \href
  {https://ui.adsabs.harvard.edu/abs/2004MNRAS.350..113M} {350, 113}

\bibitem[\protect\citeauthoryear{{Mezzacappa} et~al.,}{{Mezzacappa}
  et~al.}{2020}]{Mezzacappa20}
{Mezzacappa} A.,  et~al., 2020, \mn@doi [\prd] {10.1103/PhysRevD.102.023027},
  \href {https://ui.adsabs.harvard.edu/abs/2020PhRvD.102b3027M} {102, 023027}

\bibitem[\protect\citeauthoryear{{Mor}, {Livne}  \& {Piran}}{{Mor}
  et~al.}{2023}]{Mor&Piran23}
{Mor} R.,  {Livne} E.,   {Piran} T.,  2023, \mn@doi [\mnras]
  {10.1093/mnras/stac2775}, \href
  {https://ui.adsabs.harvard.edu/abs/2023MNRAS.518..623M} {518, 623}

\bibitem[\protect\citeauthoryear{{Mori}, {Sawada}, {Suwa}, {Tanikawa},
  {Kashiyama}  \& {Murase}}{{Mori} et~al.}{2024}]{Mori24}
{Mori} M.,  {Sawada} R.,  {Suwa} Y.,  {Tanikawa} A.,  {Kashiyama} K.,
  {Murase} K.,  2024, \mn@doi [\pasj] {10.1093/pasj/psae104}, \href
  {https://ui.adsabs.harvard.edu/abs/2024PASJ..tmp..106M} {}

\bibitem[\protect\citeauthoryear{{Moriya}, {Blinnikov}, {Tominaga}, {Yoshida},
  {Tanaka}, {Maeda}  \& {Nomoto}}{{Moriya} et~al.}{2013}]{Moriya13}
{Moriya} T.~J.,  {Blinnikov} S.~I.,  {Tominaga} N.,  {Yoshida} N.,  {Tanaka}
  M.,  {Maeda} K.,   {Nomoto} K.,  2013, \mn@doi [\mnras]
  {10.1093/mnras/sts075}, \href
  {https://ui.adsabs.harvard.edu/abs/2013MNRAS.428.1020M} {428, 1020}

\bibitem[\protect\citeauthoryear{{Morozova}, {Piro}  \& {Valenti}}{{Morozova}
  et~al.}{2018}]{Morozova18}
{Morozova} V.,  {Piro} A.~L.,   {Valenti} S.,  2018, \mn@doi [\apj]
  {10.3847/1538-4357/aab9a6}, \href
  {https://ui.adsabs.harvard.edu/abs/2018ApJ...858...15M} {858, 15}

\bibitem[\protect\citeauthoryear{{M{\"u}ller} \& {Janka}}{{M{\"u}ller} \&
  {Janka}}{2014}]{BMuller14}
{M{\"u}ller} B.,  {Janka} H.-T.,  2014, \mn@doi [\apj]
  {10.1088/0004-637X/788/1/82}, \href
  {http://ads.nao.ac.jp/abs/2014ApJ...788...82M} {788, 82}

\bibitem[\protect\citeauthoryear{{M{\"u}ller}, {Janka}  \&
  {Wongwathanarat}}{{M{\"u}ller} et~al.}{2012a}]{EMuller12}
{M{\"u}ller} E.,  {Janka} H.-T.,   {Wongwathanarat} A.,  2012a, \mn@doi [\aap]
  {10.1051/0004-6361/201117611}, \href
  {http://ads.nao.ac.jp/abs/2012A%26A...537A..63M} {537, A63}

\bibitem[\protect\citeauthoryear{{M{\"u}ller}, {Janka}  \&
  {Marek}}{{M{\"u}ller} et~al.}{2012b}]{BMuller12a}
{M{\"u}ller} B.,  {Janka} H.-T.,   {Marek} A.,  2012b, \mn@doi [\apj]
  {10.1088/0004-637X/756/1/84}, \href
  {http://adsabs.harvard.edu/abs/2012ApJ...756...84M} {756, 84}

\bibitem[\protect\citeauthoryear{{M{\"u}ller}, {Janka}  \&
  {Heger}}{{M{\"u}ller} et~al.}{2012c}]{BMuller12b}
{M{\"u}ller} B.,  {Janka} H.-T.,   {Heger} A.,  2012c, \mn@doi [\apj]
  {10.1088/0004-637X/761/1/72}, \href
  {http://ads.nao.ac.jp/abs/2012ApJ...761...72M} {761, 72}

\bibitem[\protect\citeauthoryear{{M{\"u}ller}, {Janka}  \&
  {Marek}}{{M{\"u}ller} et~al.}{2013}]{BMuller13}
{M{\"u}ller} B.,  {Janka} H.-T.,   {Marek} A.,  2013, \mn@doi [\apj]
  {10.1088/0004-637X/766/1/43}, \href
  {http://ads.nao.ac.jp/abs/2013ApJ...766...43M} {766, 43}

\bibitem[\protect\citeauthoryear{{Murphy}, {Ott}  \& {Burrows}}{{Murphy}
  et~al.}{2009}]{Murphy09}
{Murphy} J.~W.,  {Ott} C.~D.,   {Burrows} A.,  2009, \mn@doi [\apj]
  {10.1088/0004-637X/707/2/1173}, \href
  {http://ads.nao.ac.jp/abs/2009ApJ...707.1173M} {707, 1173}

\bibitem[\protect\citeauthoryear{{Nakar} \& {Sari}}{{Nakar} \&
  {Sari}}{2010}]{Nakar2010}
{Nakar} E.,  {Sari} R.,  2010, \mn@doi [\apj] {10.1088/0004-637X/725/1/904},
  \href {https://ui.adsabs.harvard.edu/abs/2010ApJ...725..904N} {725, 904}

\bibitem[\protect\citeauthoryear{{Nomoto} \& {Kondo}}{{Nomoto} \&
  {Kondo}}{1991}]{Nomoto&Kondo91}
{Nomoto} K.,  {Kondo} Y.,  1991, \mn@doi [\apjl] {10.1086/185922}, \href
  {https://ui.adsabs.harvard.edu/abs/1991ApJ...367L..19N} {367, L19}

\bibitem[\protect\citeauthoryear{{Nomoto}, {Nariai}  \& {Sugimoto}}{{Nomoto}
  et~al.}{1979}]{Nomoto79}
{Nomoto} K.,  {Nariai} K.,   {Sugimoto} D.,  1979, \pasj, \href
  {https://ui.adsabs.harvard.edu/abs/1979PASJ...31..287N} {31, 287}

\bibitem[\protect\citeauthoryear{{Obergaulinger} \& {Aloy}}{{Obergaulinger} \&
  {Aloy}}{2020}]{Obergaulinger20}
{Obergaulinger} M.,  {Aloy} M.~{\'A}.,  2020, \mn@doi [\mnras]
  {10.1093/mnras/staa096}, \href
  {https://ui.adsabs.harvard.edu/abs/2020MNRAS.tmp...95O} {p.~95}

\bibitem[\protect\citeauthoryear{{Obergaulinger} \& {Aloy}}{{Obergaulinger} \&
  {Aloy}}{2021}]{Obergaulinger21}
{Obergaulinger} M.,  {Aloy} M.~{\'A}.,  2021, \mn@doi [\mnras]
  {10.1093/mnras/stab295}, \href
  {https://ui.adsabs.harvard.edu/abs/2021MNRAS.503.4942O} {503, 4942}

\bibitem[\protect\citeauthoryear{{Obergaulinger}, {Aloy}, {Dimmelmeier}  \&
  {M{\"u}ller}}{{Obergaulinger} et~al.}{2006}]{Obergaulinger06}
{Obergaulinger} M.,  {Aloy} M.~A.,  {Dimmelmeier} H.,   {M{\"u}ller} E.,  2006,
  \mn@doi [Astron. Astrophys.] {10.1051/0004-6361:20064982}, \href
  {http://adsabs.harvard.edu/abs/2006A%26A...457..209O} {457, 209}

\bibitem[\protect\citeauthoryear{{Ott}, {Ou}, {Tohline}  \& {Burrows}}{{Ott}
  et~al.}{2005}]{Ott05}
{Ott} C.~D.,  {Ou} S.,  {Tohline} J.~E.,   {Burrows} A.,  2005, \mn@doi [\apjl]
  {10.1086/431305}, \href {http://ads.nao.ac.jp/abs/2005ApJ...625L.119O} {625,
  L119}

\bibitem[\protect\citeauthoryear{{Ott}, {Dimmelmeier}, {Marek}, {Janka},
  {Hawke}, {Zink}  \& {Schnetter}}{{Ott} et~al.}{2007}]{Ott07}
{Ott} C.~D.,  {Dimmelmeier} H.,  {Marek} A.,  {Janka} H.~T.,  {Hawke} I.,
  {Zink} B.,   {Schnetter} E.,  2007, \mn@doi [\prl]
  {10.1103/PhysRevLett.98.261101}, \href
  {https://ui.adsabs.harvard.edu/abs/2007PhRvL..98z1101O} {98, 261101}

\bibitem[\protect\citeauthoryear{{Pasham} et~al.,}{{Pasham}
  et~al.}{2021}]{Pasham22}
{Pasham} D.~R.,  et~al., 2021, \mn@doi [Nature Astronomy]
  {10.1038/s41550-021-01524-8}, \href
  {https://ui.adsabs.harvard.edu/abs/2022NatAs...6..249P} {6, 249}

\bibitem[\protect\citeauthoryear{Patel, Metzger, Cehula, Burns, Goldberg  \&
  Thompson}{Patel et~al.}{2025}]{Patel25}
Patel A.,  Metzger B.~D.,  Cehula J.,  Burns E.,  Goldberg J.~A.,   Thompson
  T.~A.,  2025, \mn@doi [Astrophys. J. Lett.] {10.3847/2041-8213/adc9b0}, 984,
  L29

\bibitem[\protect\citeauthoryear{Pognan, Grumer, Jerkstrand  \& Wanajo}{Pognan
  et~al.}{2023}]{Pognan:2023qhw}
Pognan Q.,  Grumer J.,  Jerkstrand A.,   Wanajo S.,  2023, \mn@doi [Mon. Not.
  Roy. Astron. Soc.] {10.1093/mnras/stad3106}, 526, 5220

\bibitem[\protect\citeauthoryear{{Prentice} et~al.,}{{Prentice}
  et~al.}{2018}]{Prentice18}
{Prentice} S.~J.,  et~al., 2018, \mn@doi [\apjl] {10.3847/2041-8213/aadd90},
  \href {https://ui.adsabs.harvard.edu/abs/2018ApJ...865L...3P} {865, L3}

\bibitem[\protect\citeauthoryear{{Pursiainen} et~al.,}{{Pursiainen}
  et~al.}{2018}]{Pursiainen18}
{Pursiainen} M.,  et~al., 2018, \mn@doi [\mnras] {10.1093/mnras/sty2309}, \href
  {https://ui.adsabs.harvard.edu/abs/2018MNRAS.481..894P} {481, 894}

\bibitem[\protect\citeauthoryear{{Radice}, {Burrows}, {Vartanyan}, {Skinner}
  \& {Dolence}}{{Radice} et~al.}{2017}]{Radice17}
{Radice} D.,  {Burrows} A.,  {Vartanyan} D.,  {Skinner} M.~A.,   {Dolence}
  J.~C.,  2017, \mn@doi [\apj] {10.3847/1538-4357/aa92c5}, \href
  {https://ui.adsabs.harvard.edu/abs/2017ApJ...850...43R} {850, 43}

\bibitem[\protect\citeauthoryear{{Reboul-Salze}, {Barr{\`e}re}, {Kiuchi},
  {Guilet}, {Raynaud}, {Fujibayashi}  \& {Shibata}}{{Reboul-Salze}
  et~al.}{2024}]{Reboul-Salze24}
{Reboul-Salze} A.,  {Barr{\`e}re} P.,  {Kiuchi} K.,  {Guilet} J.,  {Raynaud}
  R.,  {Fujibayashi} S.,   {Shibata} M.,  2024, \mn@doi [arXiv e-prints]
  {10.48550/arXiv.2411.19328}, \href
  {https://ui.adsabs.harvard.edu/abs/2024arXiv241119328R} {p. arXiv:2411.19328}

\bibitem[\protect\citeauthoryear{{Rest} et~al.,}{{Rest}
  et~al.}{2018}]{Rest18_FELT}
{Rest} A.,  et~al., 2018, \mn@doi [Nature Astronomy]
  {10.1038/s41550-018-0423-2}, \href
  {https://ui.adsabs.harvard.edu/abs/2018NatAs...2..307R} {2, 307}

\bibitem[\protect\citeauthoryear{{Saijo} \& {Yoshida}}{{Saijo} \&
  {Yoshida}}{2006}]{Saijo06}
{Saijo} M.,  {Yoshida} S.,  2006, \mn@doi [\mnras]
  {10.1111/j.1365-2966.2006.10229.x}, \href
  {http://ads.nao.ac.jp/abs/2006MNRAS.368.1429S} {368, 1429}

\bibitem[\protect\citeauthoryear{{Saio} \& {Nomoto}}{{Saio} \&
  {Nomoto}}{1985}]{Saio&Nomoto85}
{Saio} H.,  {Nomoto} K.,  1985, \aap, \href
  {https://ui.adsabs.harvard.edu/abs/1985A&A...150L..21S} {150, L21}

\bibitem[\protect\citeauthoryear{{Saio} \& {Nomoto}}{{Saio} \&
  {Nomoto}}{2004}]{Saio&Nomoto04}
{Saio} H.,  {Nomoto} K.,  2004, \mn@doi [\apj] {10.1086/423976}, \href
  {https://ui.adsabs.harvard.edu/abs/2004ApJ...615..444S} {615, 444}

\bibitem[\protect\citeauthoryear{{Salathe}, {Ribordy}  \&
  {Demir{\"o}rs}}{{Salathe} et~al.}{2012}]{salathe12}
{Salathe} M.,  {Ribordy} M.,   {Demir{\"o}rs} L.,  2012, \mn@doi [Astroparticle
  Physics] {10.1016/j.astropartphys.2011.10.012}, \href
  {https://ui.adsabs.harvard.edu/abs/2012APh....35..485S} {35, 485}

\bibitem[\protect\citeauthoryear{{Scheidegger}, {K{\"a}ppeli}, {Whitehouse},
  {Fischer}  \& {Liebend{\"o}rfer}}{{Scheidegger} et~al.}{2010}]{Scheidegger10}
{Scheidegger} S.,  {K{\"a}ppeli} R.,  {Whitehouse} S.~C.,  {Fischer} T.,
  {Liebend{\"o}rfer} M.,  2010, \mn@doi [\aap] {10.1051/0004-6361/200913220},
  \href {http://cdsads.u-strasbg.fr/abs/2010A%26A...514A..51S} {514, A51}

\bibitem[\protect\citeauthoryear{{Schmidt} et~al.,}{{Schmidt}
  et~al.}{2003}]{Schmidt03}
{Schmidt} G.~D.,  et~al., 2003, \mn@doi [\apj] {10.1086/377476}, \href
  {https://ui.adsabs.harvard.edu/abs/2003ApJ...595.1101S} {595, 1101}

\bibitem[\protect\citeauthoryear{{Schwab}}{{Schwab}}{2021}]{Schwab21}
{Schwab} J.,  2021, \mn@doi [\apj] {10.3847/1538-4357/abc87e}, \href
  {https://ui.adsabs.harvard.edu/abs/2021ApJ...906...53S} {906, 53}

\bibitem[\protect\citeauthoryear{{Schwab}, {Quataert}  \& {Kasen}}{{Schwab}
  et~al.}{2016}]{Schwab16}
{Schwab} J.,  {Quataert} E.,   {Kasen} D.,  2016, \mn@doi [\mnras]
  {10.1093/mnras/stw2249}, \href
  {https://ui.adsabs.harvard.edu/abs/2016MNRAS.463.3461S} {463, 3461}

\bibitem[\protect\citeauthoryear{{Segretain}, {Chabrier}  \&
  {Mochkovitch}}{{Segretain} et~al.}{1997}]{Segretain97}
{Segretain} L.,  {Chabrier} G.,   {Mochkovitch} R.,  1997, \mn@doi [\apj]
  {10.1086/304015}, \href
  {https://ui.adsabs.harvard.edu/abs/1997ApJ...481..355S} {481, 355}

\bibitem[\protect\citeauthoryear{{Shen}, {Bildsten}, {Kasen}  \&
  {Quataert}}{{Shen} et~al.}{2012}]{Shen12}
{Shen} K.~J.,  {Bildsten} L.,  {Kasen} D.,   {Quataert} E.,  2012, \mn@doi
  [\apj] {10.1088/0004-637X/748/1/35}, \href
  {https://ui.adsabs.harvard.edu/abs/2012ApJ...748...35S} {748, 35}

\bibitem[\protect\citeauthoryear{{Shibagaki}, {Kuroda}, {Kotake}  \&
  {Takiwaki}}{{Shibagaki} et~al.}{2021}]{Shibagaki21}
{Shibagaki} S.,  {Kuroda} T.,  {Kotake} K.,   {Takiwaki} T.,  2021, \mn@doi
  [\mnras] {10.1093/mnras/stab228}, \href
  {https://ui.adsabs.harvard.edu/abs/2021MNRAS.502.3066S} {502, 3066}

\bibitem[\protect\citeauthoryear{{Shibagaki}, {Kuroda}, {Kotake}, {Takiwaki}
  \& {Fischer}}{{Shibagaki} et~al.}{2024}]{Shibagaki24}
{Shibagaki} S.,  {Kuroda} T.,  {Kotake} K.,  {Takiwaki} T.,   {Fischer} T.,
  2024, \mn@doi [\mnras] {10.1093/mnras/stae1361}, \href
  {https://ui.adsabs.harvard.edu/abs/2024MNRAS.531.3732S} {531, 3732}

\bibitem[\protect\citeauthoryear{{Shibata} \& {Nakamura}}{{Shibata} \&
  {Nakamura}}{1995}]{Shibata95}
{Shibata} M.,  {Nakamura} T.,  1995, \mn@doi [\prd] {10.1103/PhysRevD.52.5428},
  \href {http://ads.nao.ac.jp/abs/1995PhRvD..52.5428S} {52, 5428}

\bibitem[\protect\citeauthoryear{{Shibata} \& {Sekiguchi}}{{Shibata} \&
  {Sekiguchi}}{2003}]{Shibata&Sekiguchi03}
{Shibata} M.,  {Sekiguchi} Y.-I.,  2003, \mn@doi [\prd]
  {10.1103/PhysRevD.68.104020}, \href
  {http://cdsads.u-strasbg.fr/abs/2003PhRvD..68j4020S} {68, 104020}

\bibitem[\protect\citeauthoryear{{Shibata} \& {Sekiguchi}}{{Shibata} \&
  {Sekiguchi}}{2004}]{shibata04}
{Shibata} M.,  {Sekiguchi} Y.-I.,  2004, \mn@doi [\prd]
  {10.1103/PhysRevD.69.084024}, \href
  {http://ads.nao.ac.jp/abs/2004PhRvD..69h4024S} {69, 084024}

\bibitem[\protect\citeauthoryear{{Shibata} \& {Sekiguchi}}{{Shibata} \&
  {Sekiguchi}}{2005}]{Shibata05c}
{Shibata} M.,  {Sekiguchi} Y.-I.,  2005, \mn@doi [\prd]
  {10.1103/PhysRevD.71.024014}, \href
  {https://ui.adsabs.harvard.edu/abs/2005PhRvD..71b4014S} {71, 024014}

\bibitem[\protect\citeauthoryear{{Shibata}, {Karino}  \& {Eriguchi}}{{Shibata}
  et~al.}{2002}]{Shibata02}
{Shibata} M.,  {Karino} S.,   {Eriguchi} Y.,  2002, \mn@doi [\mnras]
  {10.1046/j.1365-8711.2002.05724.x}, \href
  {https://ui.adsabs.harvard.edu/abs/2002MNRAS.334L..27S} {334, L27}

\bibitem[\protect\citeauthoryear{{Shibata}, {Karino}  \& {Eriguchi}}{{Shibata}
  et~al.}{2003}]{Shibata03}
{Shibata} M.,  {Karino} S.,   {Eriguchi} Y.,  2003, \mn@doi [\mnras]
  {10.1046/j.1365-8711.2003.06699.x}, \href
  {http://ads.nao.ac.jp/abs/2003MNRAS.343..619S} {343, 619}

\bibitem[\protect\citeauthoryear{{Shibata}, {Taniguchi}  \& {Ury{\=
  u}}}{{Shibata} et~al.}{2005}]{Shibata05a}
{Shibata} M.,  {Taniguchi} K.,   {Ury{\= u}} K.,  2005, \mn@doi [\prd]
  {10.1103/PhysRevD.71.084021}, \href
  {http://ads.nao.ac.jp/abs/2005PhRvD..71h4021S} {71, 084021}

\bibitem[\protect\citeauthoryear{{Shibata}, {Liu}, {Shapiro}  \&
  {Stephens}}{{Shibata} et~al.}{2006}]{Shibata06}
{Shibata} M.,  {Liu} Y.~T.,  {Shapiro} S.~L.,   {Stephens} B.~C.,  2006,
  \mn@doi [\prd] {10.1103/PhysRevD.74.104026}, \href
  {http://ads.nao.ac.jp/abs/2006PhRvD..74j4026S} {74, 104026}

\bibitem[\protect\citeauthoryear{{Shibata}, {Kiuchi}, {Sekiguchi}  \&
  {Suwa}}{{Shibata} et~al.}{2011}]{Shibata11}
{Shibata} M.,  {Kiuchi} K.,  {Sekiguchi} Y.,   {Suwa} Y.,  2011, Progress of
  Theoretical Physics, \href {http://ads.nao.ac.jp/abs/2011PThPh.125.1255S}
  {125, 1255}

\bibitem[\protect\citeauthoryear{{Smartt} et~al.,}{{Smartt}
  et~al.}{2018}]{Smartt18}
{Smartt} S.~J.,  et~al., 2018, The Astronomer's Telegram, \href
  {https://ui.adsabs.harvard.edu/abs/2018ATel11727....1S} {11727, 1}

\bibitem[\protect\citeauthoryear{{Sotani} \& {Takiwaki}}{{Sotani} \&
  {Takiwaki}}{2020}]{Sotani20b}
{Sotani} H.,  {Takiwaki} T.,  2020, \mn@doi [\mnras] {10.1093/mnras/staa2597},
  \href {https://ui.adsabs.harvard.edu/abs/2020MNRAS.498.3503S} {498, 3503}

\bibitem[\protect\citeauthoryear{{Spruit}}{{Spruit}}{2002}]{Spruit02}
{Spruit} H.~C.,  2002, \mn@doi [\aap] {10.1051/0004-6361:20011465}, \href
  {https://ui.adsabs.harvard.edu/abs/2002A&A...381..923S} {381, 923}

\bibitem[\protect\citeauthoryear{{Steiner}, {Hempel}  \& {Fischer}}{{Steiner}
  et~al.}{2013}]{SFH}
{Steiner} A.~W.,  {Hempel} M.,   {Fischer} T.,  2013, \mn@doi [\apj]
  {10.1088/0004-637X/774/1/17}, \href
  {http://adsabs.harvard.edu/abs/2013ApJ...774...17S} {774, 17}

\bibitem[\protect\citeauthoryear{{Summa}, {Janka}, {Melson}  \&
  {Marek}}{{Summa} et~al.}{2018}]{Summa18}
{Summa} A.,  {Janka} H.-T.,  {Melson} T.,   {Marek} A.,  2018, \mn@doi [\apj]
  {10.3847/1538-4357/aa9ce8}, \href
  {https://ui.adsabs.harvard.edu/abs/2018ApJ...852...28S} {852, 28}

\bibitem[\protect\citeauthoryear{{Takiwaki} \& {Kotake}}{{Takiwaki} \&
  {Kotake}}{2018}]{Takiwaki18}
{Takiwaki} T.,  {Kotake} K.,  2018, \mn@doi [\mnras] {10.1093/mnrasl/sly008},
  \href {https://ui.adsabs.harvard.edu/abs/2018MNRAS.475L..91T} {475, L91}

\bibitem[\protect\citeauthoryear{{Takiwaki}, {Kotake}  \&
  {Foglizzo}}{{Takiwaki} et~al.}{2021}]{Takiwaki21}
{Takiwaki} T.,  {Kotake} K.,   {Foglizzo} T.,  2021, \mn@doi [\mnras]
  {10.1093/mnras/stab2607}, \href
  {https://ui.adsabs.harvard.edu/abs/2021MNRAS.508..966T} {508, 966}

\bibitem[\protect\citeauthoryear{{Tanabashi} et~al.,}{{Tanabashi}
  et~al.}{2018}]{Tanabashi18}
{Tanabashi} M.,  et~al., 2018, \mn@doi [\prd] {10.1103/PhysRevD.98.030001},
  \href {https://ui.adsabs.harvard.edu/abs/2018PhRvD..98c0001T} {98, 030001}

\bibitem[\protect\citeauthoryear{Tanaka, Kato, Gaigalas  \& Kawaguchi}{Tanaka
  et~al.}{2020}]{Tanaka:2019iqp}
Tanaka M.,  Kato D.,  Gaigalas G.,   Kawaguchi K.,  2020, \mn@doi [Mon. Not.
  Roy. Astron. Soc.] {10.1093/mnras/staa1576}, 496, 1369

\bibitem[\protect\citeauthoryear{{Timmes} \& {Swesty}}{{Timmes} \&
  {Swesty}}{2000}]{TimmesEOS}
{Timmes} F.~X.,  {Swesty} F.~D.,  2000, \mn@doi [\apjs] {10.1086/313304}, \href
  {https://ui.adsabs.harvard.edu/abs/2000ApJS..126..501T} {126, 501}

\bibitem[\protect\citeauthoryear{{Torres-Forn{\'e}}, {Cerd{\'a}-Dur{\'a}n},
  {Passamonti}, {Obergaulinger}  \& {Font}}{{Torres-Forn{\'e}}
  et~al.}{2019}]{Torres-Forne19}
{Torres-Forn{\'e}} A.,  {Cerd{\'a}-Dur{\'a}n} P.,  {Passamonti} A.,
  {Obergaulinger} M.,   {Font} J.~A.,  2019, \mn@doi [\mnras]
  {10.1093/mnras/sty2854}, \href
  {https://ui.adsabs.harvard.edu/abs/2019MNRAS.482.3967T} {482, 3967}

\bibitem[\protect\citeauthoryear{{Tout}, {Wickramasinghe}, {Liebert},
  {Ferrario}  \& {Pringle}}{{Tout} et~al.}{2008}]{Tout08}
{Tout} C.~A.,  {Wickramasinghe} D.~T.,  {Liebert} J.,  {Ferrario} L.,
  {Pringle} J.~E.,  2008, \mn@doi [\mnras] {10.1111/j.1365-2966.2008.13291.x},
  \href {https://ui.adsabs.harvard.edu/abs/2008MNRAS.387..897T} {387, 897}

\bibitem[\protect\citeauthoryear{{Usov}}{{Usov}}{1992}]{Usov92}
{Usov} V.~V.,  1992, \mn@doi [\nat] {10.1038/357472a0}, \href
  {https://ui.adsabs.harvard.edu/abs/1992Natur.357..472U} {357, 472}

\bibitem[\protect\citeauthoryear{{Vartanyan} \& {Burrows}}{{Vartanyan} \&
  {Burrows}}{2020}]{Vartanyan20}
{Vartanyan} D.,  {Burrows} A.,  2020, \mn@doi [\apj]
  {10.3847/1538-4357/abafac}, \href
  {https://ui.adsabs.harvard.edu/abs/2020ApJ...901..108V} {901, 108}

\bibitem[\protect\citeauthoryear{{Vartanyan}, {Burrows}, {Radice}, {Skinner}
  \& {Dolence}}{{Vartanyan} et~al.}{2019}]{Vartanyan19}
{Vartanyan} D.,  {Burrows} A.,  {Radice} D.,  {Skinner} M.~A.,   {Dolence} J.,
  2019, \mn@doi [\mnras] {10.1093/mnras/sty2585}, \href
  {https://ui.adsabs.harvard.edu/abs/2019MNRAS.482..351V} {482, 351}

\bibitem[\protect\citeauthoryear{{Wanajo}, {Sekiguchi}, {Nishimura}, {Kiuchi},
  {Kyutoku}  \& {Shibata}}{{Wanajo} et~al.}{2014}]{Wanajo:2014}
{Wanajo} S.,  {Sekiguchi} Y.,  {Nishimura} N.,  {Kiuchi} K.,  {Kyutoku} K.,
  {Shibata} M.,  2014, \mn@doi [\apjl] {10.1088/2041-8205/789/2/L39}, \href
  {https://ui.adsabs.harvard.edu/abs/2014ApJ...789L..39W} {789, L39}

\bibitem[\protect\citeauthoryear{{Watts}, {Andersson}  \& {Jones}}{{Watts}
  et~al.}{2005}]{Watts05}
{Watts} A.~L.,  {Andersson} N.,   {Jones} D.~I.,  2005, \mn@doi [\apjl]
  {10.1086/427653}, \href {http://ads.nao.ac.jp/abs/2005ApJ...618L..37W} {618,
  L37}

\bibitem[\protect\citeauthoryear{{Wickramasinghe} \&
  {Ferrario}}{{Wickramasinghe} \& {Ferrario}}{2000}]{Wickramasinghe00}
{Wickramasinghe} D.~T.,  {Ferrario} L.,  2000, \mn@doi [\pasp]
  {10.1086/316593}, \href
  {https://ui.adsabs.harvard.edu/abs/2000PASP..112..873W} {112, 873}

\bibitem[\protect\citeauthoryear{{Woosley} \& {Baron}}{{Woosley} \&
  {Baron}}{1992}]{Woosley&Baron92}
{Woosley} S.~E.,  {Baron} E.,  1992, \mn@doi [\apj] {10.1086/171338}, \href
  {https://ui.adsabs.harvard.edu/abs/1992ApJ...391..228W} {391, 228}

\bibitem[\protect\citeauthoryear{{Yoon, S.-C.} \& {Langer, N.}}{{Yoon, S.-C.}
  \& {Langer, N.}}{2005}]{Yoon&Langer05}
{Yoon, S.-C.} {Langer, N.} 2005, \mn@doi [A&A] {10.1051/0004-6361:20042542},
  435, 967

\bibitem[\protect\citeauthoryear{Yu, Zhang  \& Gao}{Yu et~al.}{2013}]{Yu13}
Yu Y.-W.,  Zhang B.,   Gao H.,  2013, \mn@doi [The Astrophysical Journal
  Letters] {10.1088/2041-8205/776/2/L40}, 776, L40

\bibitem[\protect\citeauthoryear{{Yu}, {Li}  \& {Dai}}{{Yu}
  et~al.}{2015}]{Yu15}
{Yu} Y.-W.,  {Li} S.-Z.,   {Dai} Z.-G.,  2015, \mn@doi [\apjl]
  {10.1088/2041-8205/806/1/L6}, \href
  {https://ui.adsabs.harvard.edu/abs/2015ApJ...806L...6Y} {806, L6}

\bibitem[\protect\citeauthoryear{Yu, Chen  \& Wang}{Yu et~al.}{2019}]{Yu19}
Yu Y.-W.,  Chen A.,   Wang B.,  2019, \mn@doi [Astrophys. J. Lett.]
  {10.3847/2041-8213/aaf960}, 870, L23

\bibitem[\protect\citeauthoryear{{Yuan}, {Zhang}, {Chen}  \& {Ling}}{{Yuan}
  et~al.}{2022}]{EinsteinProbeTeam:2022fpj}
{Yuan} W.,  {Zhang} C.,  {Chen} Y.,   {Ling} Z.,  2022, in {Bambi} C.,
  {Sangangelo} A.,  eds, , Handbook of X-ray and Gamma-ray Astrophysics.
p.~86, \mn@doi{10.1007/978-981-16-4544-0_151-1}

\bibitem[\protect\citeauthoryear{Yuan et~al.}{Yuan et~al.}{2025}]{Yuan:2025cbh}
Yuan W.,  et~al., 2025, \mn@doi [Sci. China Phys. Mech. Astron.]
  {10.1007/s11433-024-2600-3}, 68, 239501

\bibitem[\protect\citeauthoryear{{Zhu}, {Pakmor}, {van Kerkwijk}  \&
  {Chang}}{{Zhu} et~al.}{2015}]{Zhu15}
{Zhu} C.,  {Pakmor} R.,  {van Kerkwijk} M.~H.,   {Chang} P.,  2015, \mn@doi
  [\apjl] {10.1088/2041-8205/806/1/L1}, \href
  {https://ui.adsabs.harvard.edu/abs/2015ApJ...806L...1Z} {806, L1}

\makeatother
\end{thebibliography}








\bsp	
\label{lastpage}
\end{document}